\documentclass[11pt,a4paper]{article}
\pdfoutput=1
\usepackage{jheppub}

\usepackage[normalem]{ulem}
\usepackage{amsmath}
\usepackage{amssymb}
\usepackage{amscd}
\usepackage{enumerate}
\usepackage{amsfonts}
\usepackage{epsfig}
\usepackage{mathtools}
\usepackage{yfonts}
\usepackage{bbold}
\usepackage{breqn}
\usepackage{dsfont}

\usepackage{graphicx}
\usepackage{bm}

\definecolor{davecolor}{rgb}{0.95,  0.5,  0.2}

\definecolor{darkgreen}{rgb}{0,0.5,0}
\definecolor{darkblue}{rgb}{0,0,0.6}
\definecolor{purple}{rgb}{0.4,0.15,0.21}
\definecolor{black}{rgb}{.2,.2,.2}

\newcommand{\beq}{\begin{equation}}
\newcommand{\eeq}{\end{equation}}
\newcommand\eeqn{\end{eqnarray}}
\newcommand\beqn{\begin{eqnarray}}

\makeatletter
\def\l@subsubsection#1#2{}
\let\cat@comma@active\@empty
\makeatother

\begin{document}
\title{
Probing beyond ETH at large $c$}
\author{Thomas Faulkner,}
\author{Huajia Wang}
\affiliation{Department of Physics, University of Illinois, 1110 W. Green St., Urbana IL 61801-3080, U.S.A. \vspace{5mm}}

\emailAdd{tomf@illinois.edu}
\emailAdd{rockwhj@illinois.edu}

\abstract{
We study probe corrections to the Eigenstate Thermalization Hypothesis (ETH) in the context of 2D CFTs with large central charge and a sparse spectrum of low dimension operators. In particular, we focus on observables in the form of non-local composite operators $\mathcal{O}_{obs}(x)=\mathcal{O}_L(x)\mathcal{O}_L(0)$ with $h_L\ll c$. As a light probe, $\mathcal{O}_{obs}(x)$ is constrained by ETH and satisfies $\langle \mathcal{O}_{obs}(x)\rangle_{h_H}\approx \langle \mathcal{O}_{obs}(x)\rangle_{\text{micro}}$ for a high energy energy eigenstate $| h_H\rangle$. In the CFTs of interests, $\langle \mathcal{O}_{obs}(x)\rangle_{h_H}$ is related to a Heavy-Heavy-Light-Light (HL) correlator, and can be approximated by the vacuum Virasoro block, which we focus on computing. A sharp consequence of ETH for $\mathcal{O}_{obs}(x)$ is the so called ``forbidden singularities", arising from the emergent thermal periodicity in imaginary time. Using the monodromy method, we show that finite probe corrections of the form $\mathcal{O}(h_L/c)$ drastically alter both sides of the ETH equality, replacing each thermal singularity with a pair of branch-cuts. Via the branch-cuts, the vacuum blocks are connected to infinitely many additional ``saddles". We discuss and verify how such violent modification in analytic structure leads to a natural guess for the blocks at finite $c$: a series of zeros that condense into branch cuts as $c\to\infty$. We also discuss some interesting evidences connecting these to the Stoke's phenomena, which are non-perturbative $e^{-c}$ effects. As a related aspect of these probe modifications, we also compute the Renyi-entropy $S_n$ in high energy eigenstates on a circle. For subsystems much larger than the thermal length, we obtain a WKB solution to the monodromy problem, and deduce from this the entanglement spectrum. }
\keywords{ETH, two dimensional CFT, virasoro block, forbidden singularities, Stoke's phenomena, Renyi entropy}
\arxivnumber{1712.03464}
\maketitle

\section{Introduction}
The question of characterizing and identifying chaos in quantum systems has drawn interest from many interdisciplinary directions in theoretical physics, ranging from quantum information to black hole physics \cite{Schack, Zurek, Shenker, Sachdev, Kitaev}. Classically, the notion of chaos is defined by exponential sensitivity to initial perturbations, and is closely related to non-linear dynamics. An isolated quantum system, on the other hand, always evolves unitarily. Despite this, an isolated quantum system can still exhibits chaotic behavior by acting as its own thermal bath and thermalizing small subsystems. Systems of such nature are expected to arise from generic non-integrable dynamics. While a precise understanding of the underlying mechanism has been lacking, there exits one concrete conjecture regarding such systems, namely the Eigenstate Thermalization Hypothesis (ETH) \cite{Srednicki, Deutsch, Rigol, D'Alessio}. 

ETH states that in finitely excited energy eigenstates, a class of few-body observables have expectation values that are close to those of micro-canonical ensembles. The difference is negligible in the thermodynamic limit. More recently, a proposal \cite{Tolya2} that complements ETH (dubbed canonical universality) was made, which states that all states with sufficiently small energy fluctuation are approximately thermal. On the other hand, an alternative  formulation of ETH (dubbed subsystem ETH) was put forward in \cite{Tolya1}, which proposed a universal form for the subsystem reduced density matrix of finitely excited eigenstates, and discussed its relation to the canonical/micro-canonical results. Numerical support of ETH in terms of reduced density matrices was also found in \cite{Garrison}.

Analytically, ETH is difficult to track in generic non-integrable systems. In special cases however, ETH can arise from universal dynamics in a certain class of conformal field theories (CFTs). One such example is two-dimensional CFTs with a large central charge $c$ and a sparse spectrum of low-lying operators. These theories are also believed to have a weakly coupled gravity dual in $AdS_3$ \cite{Maldacena, Witten, Gubser}. Via the state-operator correspondence, observables evaluated in high energy eigenstate on a circle are conformally related to correlation functions involving operators of high conformal dimensions. In two dimensional CFTs correlation functions can be decomposed into atomic ingredients called Virasoro conformal blocks or conformal partial waves. They are completely fixed by kinematics (i.e. infinite dimensional Virasoro symmetry) and play crucial roles in the progress of constraining CFTs \cite{Heemskerk,Showk,Liendo,Liam2013,Komargodski,Liam2013_2,Liam2014,Alday}, for example via the bootstrap program \cite{Ferrara,Polyakov,Rattazzi}. In the case of interest, the vacuum Virasoro block corresponding to the identity operator dominates the sum over partial waves, and thus encodes universal features in CFT dynamics such as entanglement entropy \cite{Ryu2006,Ryu2006_2,Hartman2013,Faulkner2013,Barrella,Asplund} and chaotic properties \cite{Roberts,Liam2016_3}. 

A general class of objects studied in CFTs take the form of 4-point correlation functions. The cases most relevant for probing ETH involve the heavy-heavy-light-light (HL) limit \cite{Hartman2013, Asplund, Liam2015, Balasubramanian}: 
\beqn
f\left(x,\bar{x}\right)&=& \langle \mathcal{O}_H (0)\mathcal{O}_L\left(x,\bar{x}\right)\mathcal{O}_L(1,1)\mathcal{O}_H(\infty)\rangle\nonumber\\
& \propto & \langle H |\mathcal{O}_L\mathcal{O}_L| H\rangle \approx  \langle \mathcal{O}_L\mathcal{O}_L\rangle_{\beta_H}\nonumber
\eeqn
where the conformal dimensions are set such that $h_H/c \gg 1, h_L/c\ll 1$. A sharp signature of ETH in this case is the ``forbidden singularities" on the complex $x$ plane. They arise as the images of the OPE singularity due to the emergent thermal periodicity along the imaginary time direction. They are said to be forbidden because the only true singularities in Euclidean correlation functions are OPE singularities. These can be precisely reproduced from the large $c$ vacuum Virasoro block contribution to $f\left(x,\bar{x}\right)$ in the HL limit. In this case, we can view ETH as a consequence of the infinitely powerful kinematics in 2D CFTs, whereas generically it is an extremely complex dynamical phenomena. 

In this paper, we extend previous studies of ETH in the context HL correlators at large $c$, focusing on the vacuum Virasoro block contributions. In particular, we move away from the ``probe limit" $h_L/c\ll 1$ and focus on effects characterized by small but finite probe strength $h_L/c$, especially how they interplay with the conformal ratio $x,\bar{x}$ of the blocks (or the separation between the $\mathcal{O}_L$'s). Similar limit was studied for correlators in Liouville theory in \cite{Balasubramanian}. There are a few motivations for studying such extensions. The original statement of ETH restricts the class of observables to few-body operators that do not cause substantial energy fluctuations, namely operators in the probe limit. Operators away from the probe limit are not expected to observe ETH in general and specifics about the heavy micro-states could in principle be encoded in the way they are ``back-reacted" on by the non-probe observables. However, in the cases of interest, moving away from the probe limit of the HL correlator does not affect the dominance by the vacuum Virasoro block. We therefore expect some universal modifications to ETH due to finite probe effects, which are again fixed by kinematics. The general goal of this paper is to extract such universal modifications. 

On the other hand, all quantum mechanical systems (including black holes) are expected to have a discrete spectrum with finitely many degrees of freedom. While ETH characterizes universal behaviors of chaotic systems in the thermodynamic limit, how they exit from ETH characterizes the underlying finiteness of the systems. The most interesting question of such nature is the black hole information paradox \cite{Hawking1976,tHooft1995,Mathur2009TheIP,Maldacena2001,Almheiri}, in which unitary evolution of a pure micro-state, upon forming black holes, exhibits thermal features and thus loses information. Information loss in $AdS_3$ black holes is directly related to the large $c$ HL correlators, where the heavy operator $\mathcal{O}_H$ creates a black hole in the bulk, and for late enough time the observable $\mathcal{O}_L(t)\mathcal{O}_L(0)\sim e^{-\pi t/\beta_H}$ experiences an exponential decay by probing the black hole background. Naively such exponential decay is related to the forbidden singularities via analytic continuation. Resolving the information paradox in this context amounts to finding out how does finite $c$ effects stop the exponential decay at later time, or smoothen out the forbidden singularities \cite{Liam2016, Liam2016_2, Hongbin}. For HL correlators, one can organize the finite $c$ corrections into two types, $\mathcal{O}\left(h_L/c\right)$ or $\mathcal{O}\left(1/c\right)$. While the two are indistinguishable in the probe limit, as we move away from it, there is a natural separation between the two types. Studying the finite probe effects therefore serves as an intermediate step, which as it turns out is also an important step towards finite $c$. As we shall demonstrate, it strongly constrains the form of the resolution at finite $c$.     

A more specific reason for moving away from the probe limit is for studying Renyi-entropies in 2D CFTs, which directly probe the entanglement data in terms of the reduced density matrices. For a single interval on the circle in micro-state $| H\rangle$ they can be calculated as 4-point functions \cite{Hartman2013, Asplund,Faulkner2013,Headrick2010,Chen2013,Perlmutter2014}:
\beq
S_n(\theta)\equiv \frac{1}{1-n}\ln{\text{Tr}\rho_H(\theta)^n},\;\;\text{Tr}\rho_H(\theta)^n\propto \langle H | \sigma_n(\theta)\sigma_n(0)| H\rangle_{\text{CFT}^n/\mathbb{Z}_n} 
\eeq
where $\sigma_n$ is the twist operator defined in the orbifolded CFT (with central charge $nc$) and carry a well-defined conformal dimension $h_n = \frac{nc}{24}\left(1-\frac{1}{n^2}\right)$. The probe limit $n\to 1$ is related to the entanglement entropy $S_{EE}\propto \lim_{n\to 1} S_n$, and satisfies ETH in the sense that $S^H_{EE}\approx S^{\beta_H}_{EE}$ \cite{Hartman2013,Asplund,Liam2014}. This is consistent with subsystem ETH since the equality of entanglements implies the relative entropy, comparing the two reduced density matrices, must be small in the large $c$ limit.
However, for $n>1$ it is no longer true \cite{Asplund,Tolya1,Garrison,Lin2016,He2017,Basu:2017kzo,He:2017txy} 
that the Renyi entropies can be matched to the naive thermal Renyi entropies. 
The discrepancies is thus 
sensitive to more fine-grained informations about the reduced density matrix of the pure state.
Here we will view them as encoded in the non-probe effects from the now heavy twist operators $\sigma_n$. 

The plan of the discussion is as follows. In section~\ref{sec: review}, we review the monodromy method that computes Virasoro blocks in the limit of infinite $c$, and use it to recall in the probe limit the emergence of ETH from the HL correlators. In section~\ref{sec: semi-resol}, we propose a drastic change in the analytic properties of the correlators by re-summing probe effects, and check numerically using the monodromy method. In section~\ref{sec: inverse_laplace}, we study the other side of ETH and show that by re-summing probe effects, a similar alteration in the analytic structure of the correlator arises in the micro-canonical ensemble. In section~\ref{sec: Renyi}, we switch gears and compute the excited state Renyi-entropy for a finite arc, and extract from this features of the entanglement spectrum of the subsystems. In section~\ref{sec: finite_c}, using results obtained from re-summing probe corrections, we discuss the blocks at finite $c$. Interesting connections to the Stoke's phenomena will be revealed.

\section{ETH at the leading order}\label{sec: review}
Before we begin, let us make explicit the thermodynamic limit taken in our context, and the notion of ETH related to it. For doing this, let us specify the relevant scales and express the limit in terms of dimensionless ratios. We consider 2D CFTs defined on a circle $\mathbb{S}^1$ of radius $L$. Via radial quantization, an energy eigen-state $| H\rangle$ on $\mathbb{S}^1$ can be obtained from inserting an operator $\mathcal{O}_H(0)$ with scaling dimension $h_H$. The energy and energy density are given by: 
\beq
E\propto \frac{2 h_H}{L},\;\;\mathcal{E}\propto \frac{2 h_H}{L^2} 
\eeq
One can define an effective temperature by relating it to the average energy density in a canonical ensemble: 
\beq
\mathcal{E}_T \propto c T^{2}\;\rightarrow T_H L \propto \sqrt{\frac{2h_H}{c}}
\eeq
where $c$ is the central charge. At this point there are two choices for thermodynamic limit, the one we focus on in this paper corresponds to sending  $c\to \infty$ while holding the ratio $h_H/c$ finite. The observables we are interested in consist of non-local composite operators $\mathcal{O}_{obs}\sim\mathcal{O}_L(x)\mathcal{O}_L(0)$. They come with a length scale $x$, which we take to be fixed in the thermodynamic limit we are taking. In this case, both $x/L$ and $\beta_H/x$ are finite. As we shall review shortly, the corresponding ETH statement:
\beq\label{eq: eth}
\langle \mathcal{O}_{obs} \rangle_H \approx \langle \mathcal{O}_{obs}\rangle_{micro} 
\eeq
was established in the probe limit $h_L/c\to 0$ \cite{Liam2014}. Modifications to (\ref{eq: eth}) by finite $h_L/c$ corrections will be the key focus of this paper. 

There is a different thermodynamic limit one can take, namely by sending $h_H\to\infty$ but keeping $c$ finite. In this limit, at least one of the ratios $\left(\beta_H/x, x/L\right)$ needs to be vanishing. ETH in this limit has not been established. It has been proposed that  the generalized Gibbs ensembles augmented by infinitely many KdV charges are required to capture ETH in this case \cite{Vidmar,Sasaki,Bazhanov,Boer,Lashkari:2017hwq}. 

\subsection{The monodromy method}
Let us now review the monodromy method that is useful for computing the infinite $c$ limit of Virasoro blocks. Details of the method can be referred to in \cite{Harlow, Liam2014, Hartman2013}. 

For general 4-point functions, we can decompose them into Virasoro blocks: 
\begin{equation}
\langle \mathcal{O}_1 (z_1)\mathcal{O}_2 (z_2)\mathcal{O}_3 (z_3)\mathcal{O}_4 (z_4)\rangle =\sum_{h, \bar{h}} C^{12}_{h,\bar{h}} C^{34}_{h,\bar{h}} \mathcal{V}^{12,34}_h(z_1,z_2,z_3,z_4)  \bar{\mathcal{V}}^{12,34}_{\bar{h}}(\bar{z}_1,\bar{z}_2,\bar{z}_3,\bar{z}_4) 
\end{equation}
where $\mathcal{O}_i$ are primary operators with dimensions $\lbrace h_i, \bar{h}_i\rbrace$, and $\lbrace h, \bar{h}\rbrace$ label the dimensions of the internal family. From now on we focus only on the holomorphic part $\mathcal{V}^{12,34}_h(z_1,z_2,z_3,z_4)$. The monodromy method allows one to compute the block in the ``semi-classical" limit: $h_i = \frac{c}{6}\epsilon_i$, $c\to \infty$, while holding fixed $\epsilon_i$. To proceed, one first solves the following second order differential equation: 
\begin{eqnarray}
\Psi''(z)+T(z)\Psi(z)=0,\;\;T(z)=\sum_i\left\lbrace\frac{\epsilon_i}{(z-z_i)^2}-\frac{6}{c}\frac{p_i}{z-z_i}\right\rbrace
\end{eqnarray}
By conformal transformations we can always place the 4 insertions at $(z_1,z_2,z_3,z_4)=(0,x,1,\infty)$, so that the block is only a function of the conformal invariant ``moduli" $x$. The $p_i$ are called ``accessory parameters". They are not independent, but should be arranged to make $T(z)$ vanish as $z^{-4}$ at infinity, so that the $z=\infty$ is a regular point for the differential equation (before sending $z_4$ to $\infty$). This imposes three constraints among $p_i$: 
\begin{equation}
\sum_i p_i =0,\; \sum_i\left(p_i z_i -\epsilon_i\right)=0,\;\sum_i\left(p_i z_i^2-2\epsilon_i z_i\right)=0
\end{equation}
After solving these constraints, the system depends only on one accessary parameter $p(x)$:
\begin{equation}
T(z)=\frac{\epsilon_1}{z^2}+\frac{\epsilon_2}{(z-x)^2}+\frac{\epsilon_3}{(1-z)^2}+\frac{\sum_i\epsilon_i-2\epsilon_4}{z(1-z)}-\frac{p(x) x(1-x)}{z(z-x)(1-z)}
\end{equation}
There are two linearly independent solutions $\lbrace\Psi^+,\Psi^-\rbrace$, each with 4 regular singularities at $z_i$. For any given $x$, they give rise to a monodromy structure that depends on $p(x)$. To compute the block $\mathcal{V}^{12,34}_h(z_i)$, $p(x)$ should be tuned such that the monodromy matrix $M_{12}$ for any contour encircling only $z_1,z_2$ satisfies the following: 
\begin{equation}
\text{Tr}M_{12} = -2\cos{\left(\pi \Lambda_h\right)},\;h=\frac{c}{24}\left(1-\Lambda_h^2\right)
\end{equation}
In particular, for the Virasoro vacuum block, the monodromy is trivial: $h=0\to \text{Tr}M_{12}=2$. The above monodromy problem defines the accessory parameter as a function $p(x)$, the block is then given by integrating this accessory parameter: 
\begin{equation}
\lim_{c\to \infty}\mathcal{V}^{12,34}_h(x)=e^{-\frac{c}{6} f(x)},\;\;\frac{\partial f(x)}{\partial x}=p(x)
\end{equation}
\subsection{Forbidden singularities}
Now we observe ETH by taking the HL limit: 
\begin{equation}
\epsilon_1=\epsilon_2=\epsilon_L\ll 1, \epsilon_3=\epsilon_4=\epsilon_H\gg 1
\end{equation}
Via the state operator correspondence, $\mathcal{O}_{3,4}$ create the high energy ``background" state, and $\mathcal{O}_{1,2}$ are used to probe such a background state. One can then organize the solution to the monodromy problem in expansion of the probe operator's conformal dimension $\epsilon_L$ \cite{Liam2014}: 
\begin{eqnarray}\label{eq: monodromy_expansion}
\Psi^{\pm}(z)&=& \Psi^{\pm}_0(z)+\epsilon_L \Psi^{\pm}_1(z) + \epsilon_L^2\Psi^{\pm}_2(z)+...,\;\;p(x)= \epsilon_L\; p_0(x)+\epsilon_L^2\; p_1(x)+...\nonumber\\
T(z)&=& T_0(z) + \epsilon_L T_1(z)+\epsilon_L^2 T_2(z)+...\nonumber\\
T_0(z)&=&\frac{\epsilon_H}{(1-z)^2},\;\;T_1(z)=\frac{1}{z^2}+\frac{1}{(z-x)^2}+\frac{2}{z(1-z)}-\frac{p_0(x)x(1-x)}{z(z-x)(1-z)}\nonumber\\
T_n(z)&=&-\frac{p_n(x)x(1-x)}{z(z-x)(1-z)}, n\geq 1
\end{eqnarray}
From the point of view of $\text{AdS}_3/\text{CFT}_2$, this expansion corresponds to the gravitational back-reaction of the probe operator to the bulk geometry. The quantum corrections $\mathcal{O}(1/c^n)$ are not captured by the monodromy method.

In the s-channel, the correlator is dominated by the Virasoro vacuum block. The series (\ref{eq: monodromy_expansion}) can be obtained order by order by sustaining trivial monodromy around a contour containing $z=0$ and $z=x$. 
The leading order solution can be obtained straight-forwardly: 
\begin{equation}\label{eq:monodromy_lead}
\Psi_0''(z)+T_0(z)\Psi_0(z)=0\rightarrow \Psi^{\pm}_0(z)=(1-z)^{\frac{1\pm \sqrt{1-4\epsilon_H}}{2}}
\end{equation}
which automatically satisfy the trivial monodromy condition.  

\begin{figure}[h!]
\centering
\includegraphics[width=0.45\textwidth]{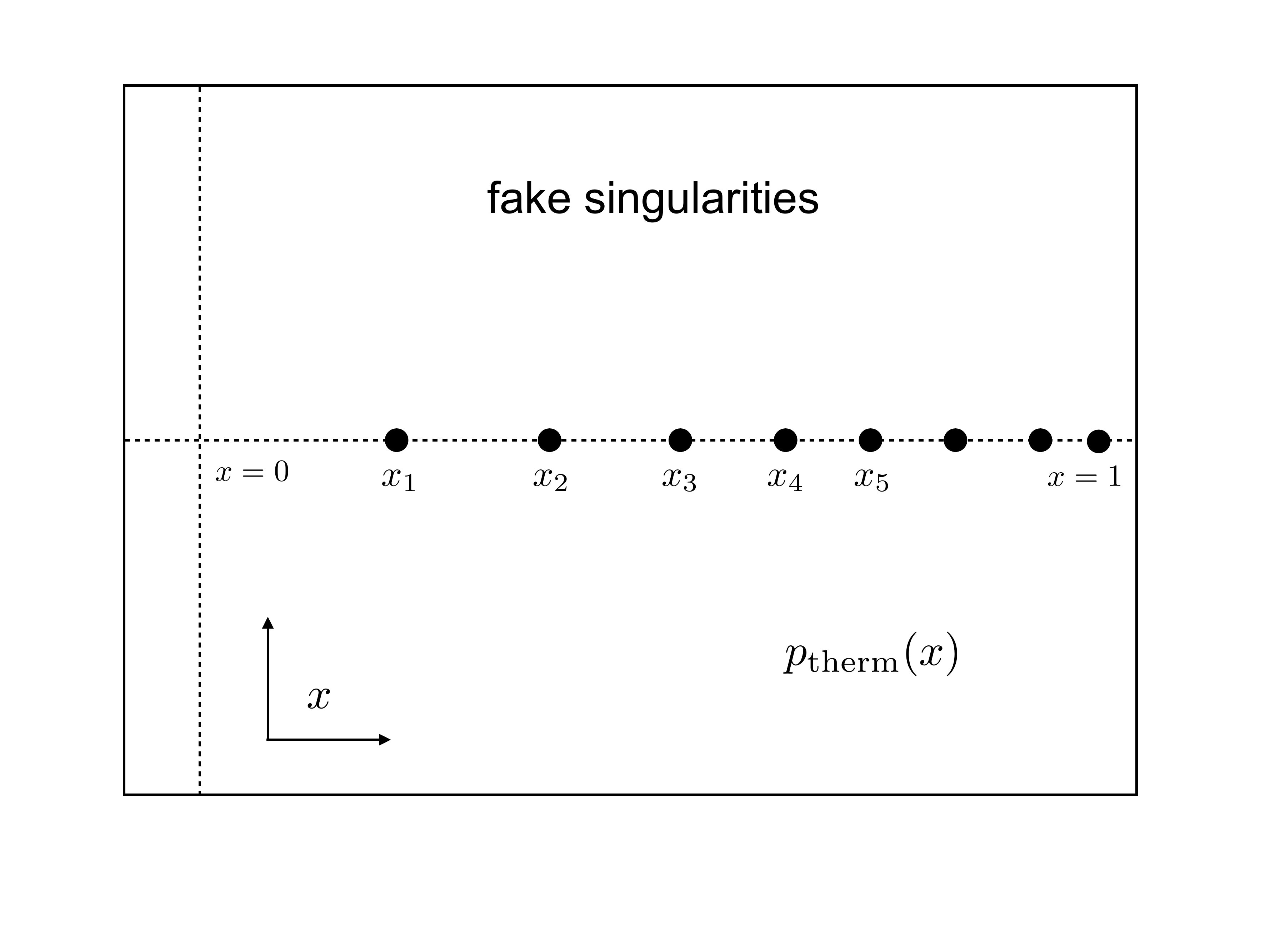}
\includegraphics[width=0.45\textwidth]{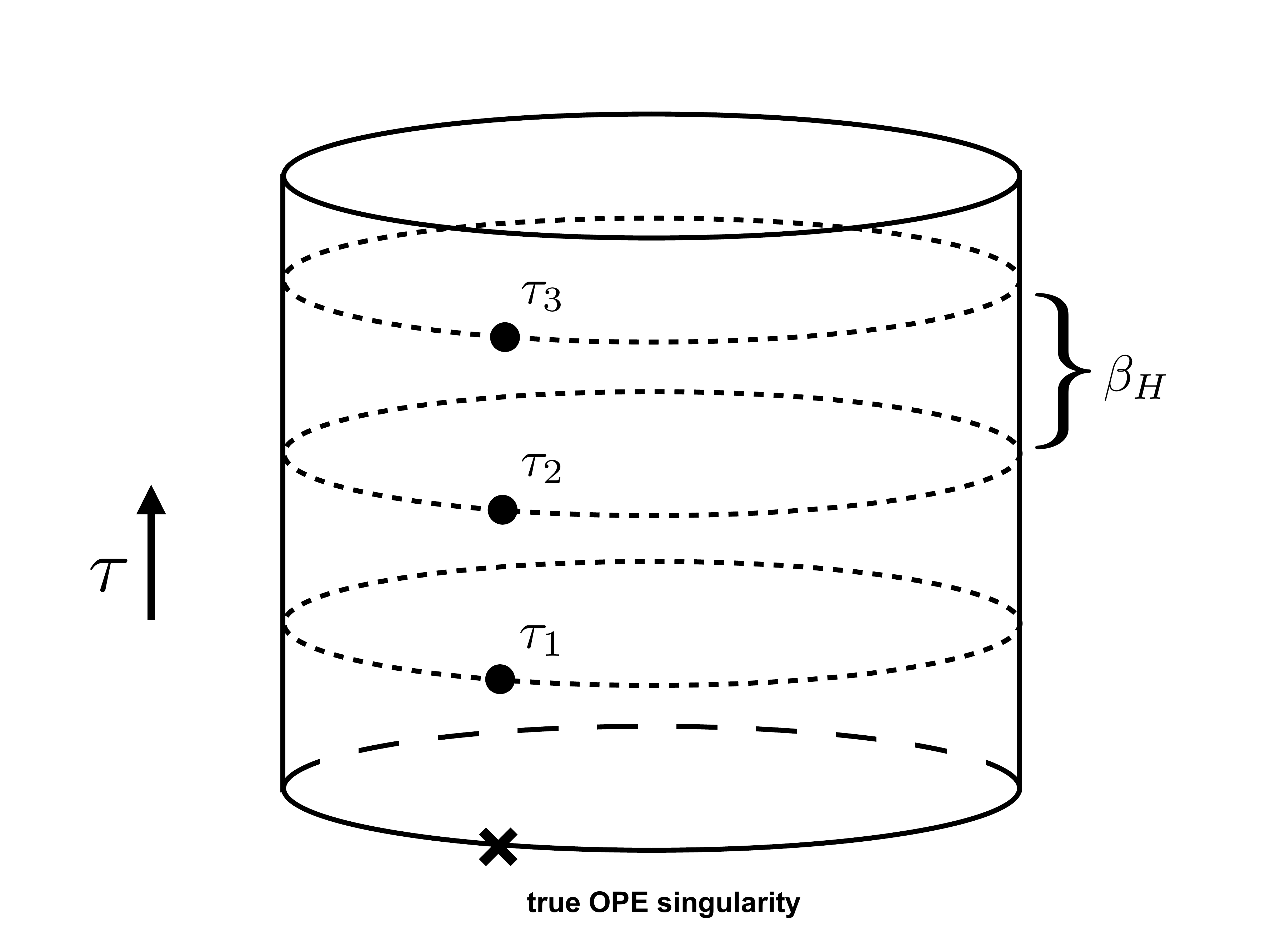}
\caption{\small{Forbidden singularites in the leading order results (left), via a conformal mapping, are related to the thermal images of the OPE singularity (right), a sign of emergent thermality.}}\label{fig: fake-singularities}
\end{figure}

The next order solution will determine the leading order accessory parameter $p_0(x)$: 
\begin{equation}
\Psi^{''\pm}_1(z)+T_0(z)\Psi^{\pm}_1(z)=-T_1(z)\Psi^{\pm}_0(z)
\end{equation}
This is the same differential equation as (\ref{eq:monodromy_lead}), but with a known inhomogeneous source on the right hand side. Using standard methods such as variation of parameters, one can solve for the next order solution $\Psi^{\pm}_1(z)$, and computes the corresponding correction to the monodromy: 
\begin{equation}
\delta \text{Tr}{M_{0x}}\sim 1-(1-x)^{i\alpha_H}+p_0(x)\left[(1-x)^{i\alpha_H}-1\right](x-1)-i\alpha_H\left[1+(1-x)^{i\alpha_H}\right]
\end{equation}
where $\alpha_H=\sqrt{4\epsilon_H-1}$. Sustaining trivial monodromy at this order solves $p_0(x)$:
\begin{equation}\label{eq:accessory_lead}
p_0(x)=\frac{-1+i\alpha_H+(1-x)^{i\alpha_H}(1+i\alpha_H)}{(x-1)\left[(1-x)^{i\alpha_H}-1\right]}
\end{equation}
Integrating over $p_0(x)$ and fixing the integration constant by requiring the vacuum block to agree with the short distance expansion $f(x)\sim 2\epsilon_L\log{(x)}$ gives the leading order $\epsilon_L$ results: 
\begin{equation}\label{eq: leading_mono}
\mathcal{V}^{12,34}_{\text{vac(x)}}=e^{-\frac{c}{6}f(x)},\;\;f(x)=2\epsilon_L\ln{\left(\frac{1-(1-x)^{i\alpha_H}}{i\alpha_H}\right)}+\epsilon_L(1-i\alpha_H)\ln{(1-x)}+\mathcal{O}(\epsilon_L^2)
\end{equation}
To check ETH, we recall the thermal two-point functions on a circle. In the high temperature limit they can be approximated by those on infinite lines:
\begin{equation}
\langle \mathcal{O}_L(\tau)\mathcal{O}_L(0) \rangle_\beta = \left[\frac{\beta}{\pi}\sin{\left(\frac{\pi \tau}{\beta}\right)}\right]^{-2h_L}
\end{equation}
Mapping from the cylinder (with circumference $2\pi $) to the complex plane by $x=1-e^{-\tau}$ and comparing with (\ref{eq: leading_mono}), we identify an ``effective temperature" $\beta_H = \frac{2\pi }{\alpha_H}$. In other words:
\begin{equation}
\langle h_H |\mathcal{O}_L(\tau)\mathcal{O}_L(0)|h_H\rangle \approx \langle \mathcal{O}_L(\tau)\mathcal{O}_L(0)\rangle_{\beta_H}
\end{equation}
which is a manifestation of ETH. As a consequence, additional singularities emerge at $x_n=1-e^{-\frac{2\pi n}{\alpha_H}}, n\in \mathbb{N}$, which correspond to the thermal images of the OPE singularity at $\tau=0$ (see figure \ref{fig: fake-singularities}). Eventually, the exact block at finite $c$ should only have OPE singularities. These additional singularities are artifacts of the particular limit that is taken, and should disappear as all corrections are considered.

\section{``Resolution" by probe effects}\label{sec: semi-resol}
In this section we zoom into the forbidden singularities. They emerge as a consequence of ETH at leading order in $\epsilon_L$ characterizing the probe limit, and should be resolved by finite $c$ effects. The resolution encodes data about how ETH is modified and eventually breaks down away from the probe limit.  In principle, one can compute corrections from both the probe corrections $\mathcal{O}(h_L/c)$ as well as bulk loops $\mathcal{O}(1/c)$ order by order, and look for clues about the resolution. However, perturbative calculations only reveal higher orders of divergences \cite{Liam2016_4,Perlmutter2015,Beccaria2016,Liam2017}. This is usually an indication for re-summation. By moving away from the probe limit and taking $h_L/c$ to be small but finite, there is a hiearchy between probe and bulk loop corrections. It is natural to re-sum the former first. By doing this we find that they drastically change the form of the singularities. 

To understand the re-summation, recall that the accessory parameter is solved from the monodromy equation: 
\begin{equation}
\text{Tr}M_{0x}\left(p, \epsilon_L, x\right)=2
\end{equation}
We have taken a step back from the series expansion ansatz (\ref{eq: monodromy_expansion}) and restored $\text{Tr}M_{0x}$ as a function of both $p$ and $\epsilon_L$. Notice that the series expansion for $p_0(x)$ begins at the linear order in $\epsilon_L$, so we can take itself as an independent small parameter and re-write the monodromy equation in a double series expansion: 
\begin{equation}\label{eq:monodromy general}
\delta \text{Tr}M_{0x}=\text{Tr}M_{0x}-2 = \sum_{m,n\geq 0} G_{mn}(x) p^m \epsilon_L^n=0
\end{equation}
The leading order solution (\ref{eq:accessory_lead}) is obtained by approximating (\ref{eq:monodromy general}) with the linear equation in $p$, and at the leading order in $\epsilon_L$:   
\begin{equation}\label{eq:lead}
\delta\text{Tr}M_{0x}\sim p(x)\left[(1-x)^{i\alpha_H}-1\right](x-1)+\epsilon_L\left\lbrace 1-i\alpha_H -(1+i\alpha_H)(1-x)^{i\alpha_H}\right\rbrace=0
\end{equation}
Away from the singularities we have $p(x)\sim \epsilon_L$, higher power terms of $p$ are thus more suppressed. Including them together with higher power terms of $\epsilon_L$ in the monodromy equation fixes corrections to the solution of (\ref{eq:lead}), and perturbation theory is valid. However, as one approaches the singularities $x\approx x_n$, the coefficient of $p(x)$ vanishes:
\begin{equation}
\delta \text{Tr}M_{0x}\sim -\alpha_H(x-x_n)p(x)+2\epsilon_L\alpha_H=0
\end{equation}
The approximating linear equation (\ref{eq:lead}) is then degenerate, naive series expansion results in powers of $\epsilon_L(x-x_n)^{-1}$, which is more divergent at higher orders. Perturbation theory breaks down, and we need to re-sum the $\epsilon_L$ corrections. At the level of solving the monodromy equation, there is a very simple mechanism for re-summation: now that the linear equation approximation becomes degenerate near $x\approx x_n$, one simply supplements it with the next order term $\mathcal{O}(p^2)$:
\beq
\delta\text{Tr} M_{0x} \sim -b_n p^2 - \alpha_H (x-x_n) p + 2\epsilon_L \alpha_H = 0
\eeq 
From the point of view of the full monodromy equation, $\lbrace x_n\rbrace$ play no special roles, they are just roots of the particular coefficient $G_{10}(x)$. Therefore, we expect that $b_n\approx G_{20}(x_n)\neq 0$ for generic cases, and the leading order poles at $x=x_n$ are resolved into a pair of branches: 
\beqn\label{eq: quadratic}
p_0(x) &=& \frac{2\epsilon_L}{x-x_n}\approx \begin{cases} 
p^-(x),\; & x<x_n\\
p^+(x),\;& x>x_n
\end{cases}\nonumber\\
p^{\pm}(x) &=& \frac{1}{2b_n}\left(-\alpha_H(x-x_n) \pm \sqrt{\alpha_H^2(x-x_n)^2+8b_n \alpha_H\epsilon_L}\right)
\eeqn
For the non-generic cases where $G_{20}(x_n)=0$, the resolutions instead take the form of higher order radicals. Though potentially interesting, we will not consider them in this paper. For generic cases the divergences at the leading order poles $x=x_n$ are regularized by $p^{\pm}(x_n)= \pm \sqrt{\frac{2\alpha_H}{b_n}}\epsilon_L^{1/2}$, which is non-analytic in $\epsilon_L$ and can only arise from an infinite re-summation in the original $\epsilon_L$ expansion. At this step, instead of forbidden poles we have ``forbidden branch-point" singularities at 
\begin{equation}
x^{\pm}_n \approx x_n \pm i\sqrt{\frac{8 b_n \epsilon_L}{\alpha_H}}\approx x_n\pm\frac{4i\epsilon_L}{|p(x_n)|}
\end{equation}

\subsection{Additional saddles}
We have seen that locally, the partial re-summation transforms the leading order poles into branch cuts: $p_0(x)$ near $x_n$ splits into two branches $p^{\pm}(x)$. Globally, the branching structure raises the following question: what do the infinitely many additional branches (one for each forbidden singularity) correspond to? To answer this, recall the monodromy condition that computes the Virasoro block of internal dimension $h$: 
\begin{equation}
\text{Tr}M_{12} = -2\cos{\left(\pi \Lambda_h\right)},\;h=\frac{c}{24}\left(1-\Lambda_h^2\right)
\end{equation}
For each $h$, there are infinitely many other choices of $h'$ that share the same monodromy problem, related by 
\begin{equation}
\Lambda_h = \Lambda_{h'}+2n,\;\; n\in\mathbb{Z}
\end{equation}
For the vacuum block $h=0$, we have $h_n = -\frac{c}{6}n(n+1)$. The monodromies of these solutions wind around $n$ times before going back to trivial. Speculatively, they can be interpreted as additional ``saddles" in some path-integral formulation of computing the block, from which the monodromy problem arises as the equation of motion. We will come back to this point in section \ref{sec: finite_c} when we consider what happens at finite $c$. The roles of these additional ``saddles" have been discussed in \cite{Liam2016,Liam2016_2} for the late time behavior of correlation functions and the information paradox. Here we find that these additional ``saddles" are also important for resolving the forbidden singularities. In fact, together they form a much more elaborate object: an infinitely sheeted Riemann surface, whose details we will describe shortly. 

\subsection{Numerical results}
In this section, we present results for the solving the monodromy problem numerically at small but finite $\epsilon_L$. By doing this we are effectively summing over all $\epsilon_L$ corrections. 

\begin{figure}[h!]
\centering
\includegraphics[width=0.6\textwidth]{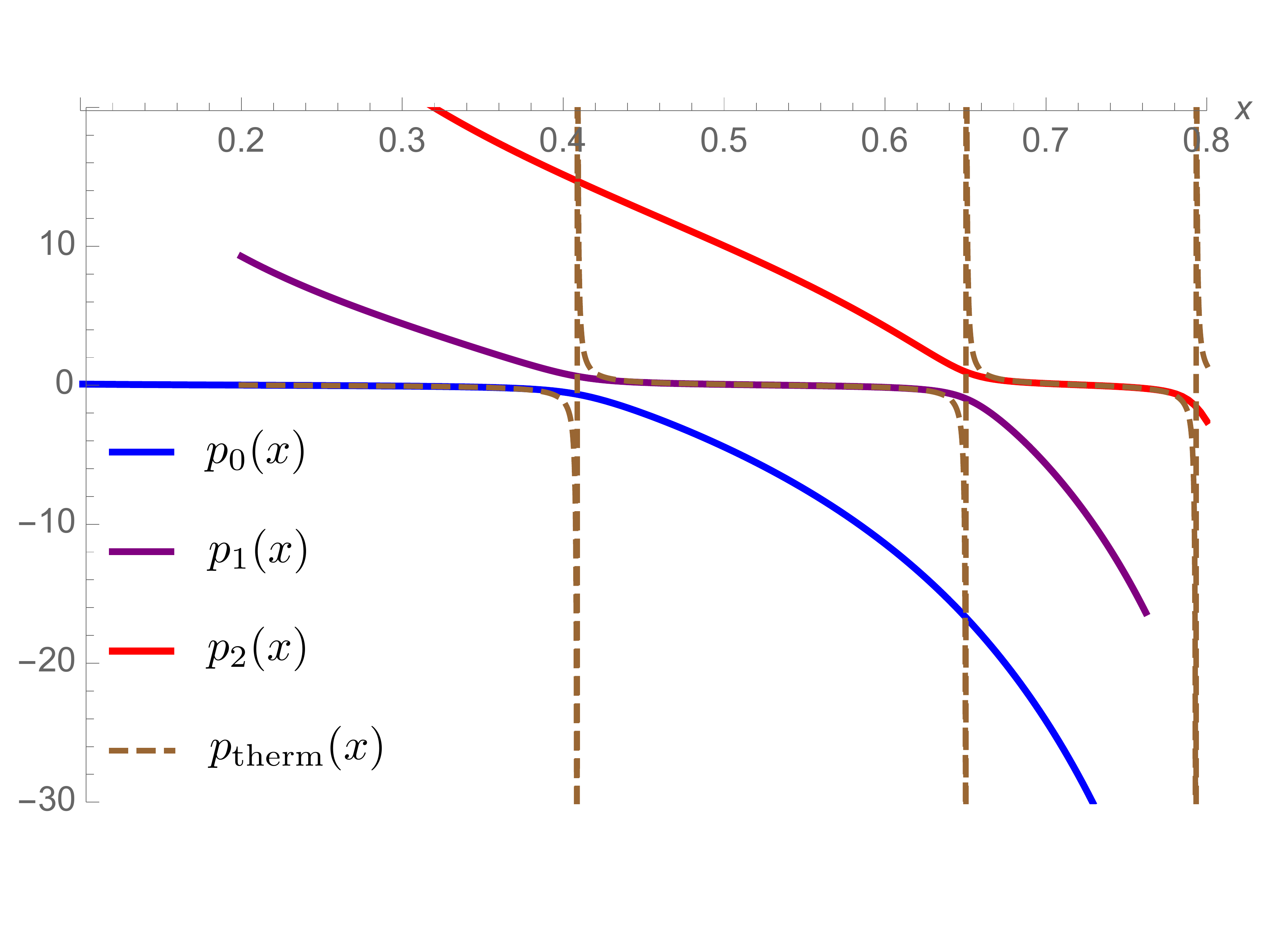}
\caption{\small{Solutions ($\epsilon_H=36, \epsilon_L=5*10^{-3}$) for the accessory parameter $p_n(x)$ for $n=0,1,2$ (solid), compared against the leading order in $\mathcal{O}(\epsilon_L)$ result $p_0(x)=p_{\text{therm}}(x)$ that exhibits thermal singularities.}}\label{fig:accessory_real}
\end{figure}

In figure \ref{fig:accessory_real} we generate solutions for $p(x)$ along the real-axis $0<x<1$. In particular, we start near $x=0$ and extend to finite $x$. The initial values of $p(x)$ at $x\approx 0$ are determined by the known small $x$ expansion of the classical conformal blocks:
\beq
p_n(x) = \frac{1}{2}n(1+n)+\frac{n(1+n)+2\epsilon_L}{x}+\mathcal{O}(x) 
\eeq
The index $n$ labels the additional saddles related to the vacuum block: $h_n = -\frac{c}{6}n(n+1)$, with $n=0$ being the vacuum block. For $n\geq 1, h_n<0$, and thus they should not be taken as physical intermediate states. 
We see that the leading order result $p_{\text{therm}}(x)$ splits into infinitely many branches $p_n(x)$ corresponding to the additional saddles. One can check that near each forbidden singularity $x_n$, $p_{n-1}(x)$ and $p_{n}(x)$ behave exactly like the two square-root branches predicted by the naive quadratic solutions in (\ref{eq: quadratic}). 

\begin{figure}[h!]
\centering
\includegraphics[width=0.3\textwidth]{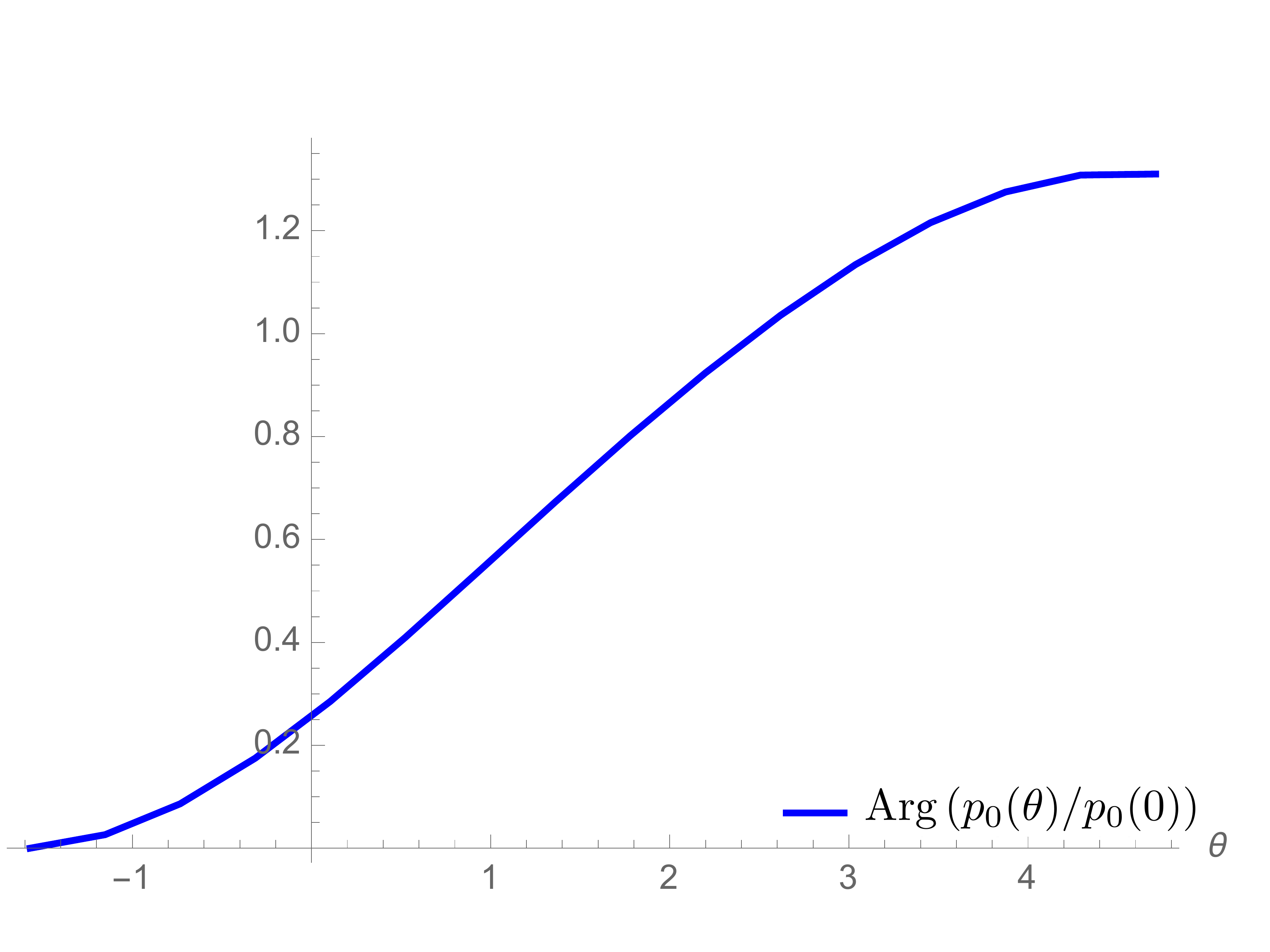}
\includegraphics[width=0.3\textwidth]{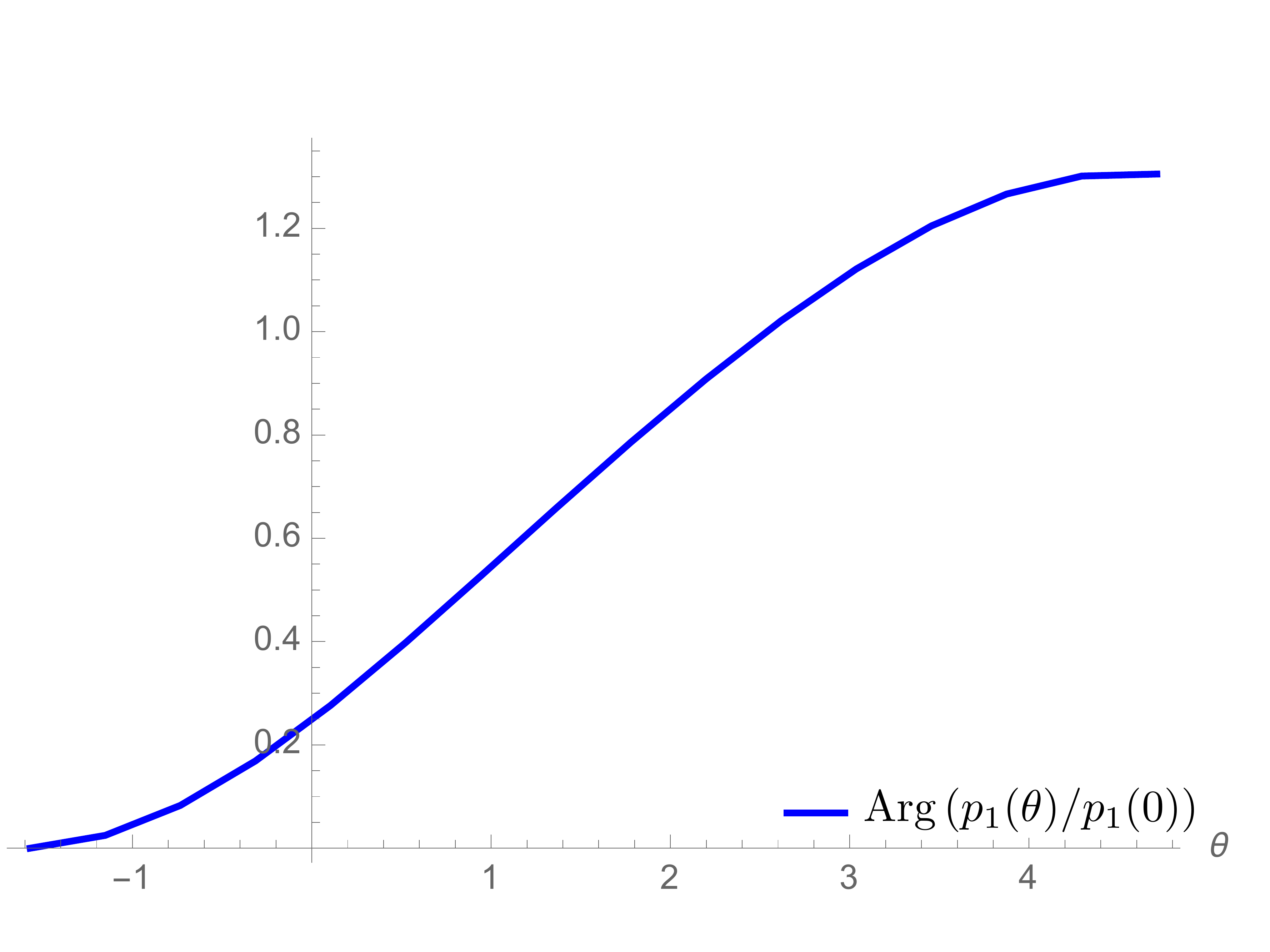}
\includegraphics[width=0.3\textwidth]{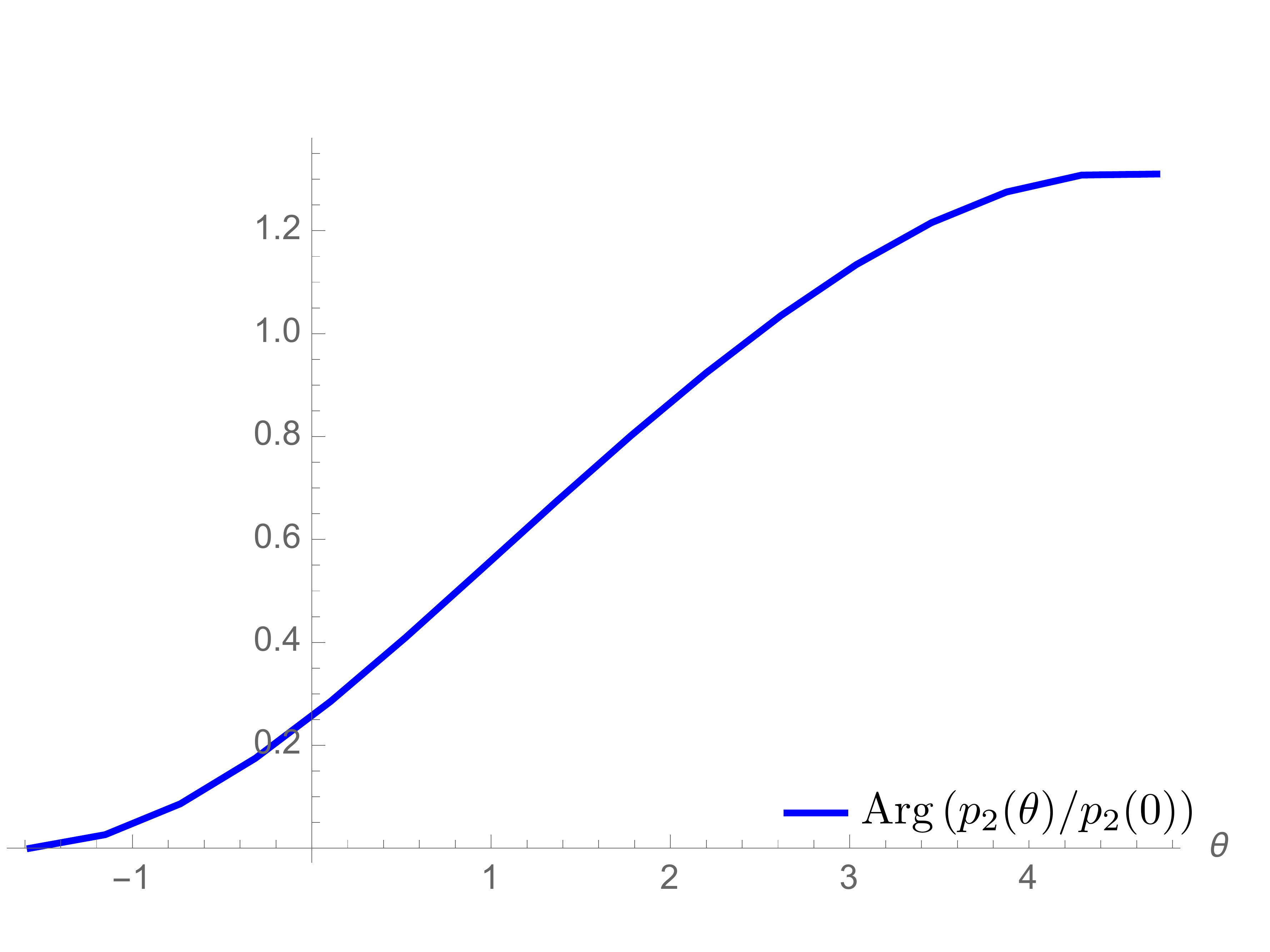}
\caption{\small{Examples of the monodromy of $p_{n-1}(x)$ around the branch point above the forbidden singularities: $x^+_n = x_n + i\frac{4\epsilon_L}{|p_{n-1}(x_n)|}$ for $n=1,2,3$}}\label{fig: branch-points}
\end{figure}

We verify the existence of the branch-points suggested in (\ref{eq: quadratic}) by tracking $p_{n-1}(x)$ around some tiny circles $x=x^{\pm}_n + \epsilon e^{i\theta}, \theta \in (0,2\pi)$ centered about the predicted branch points $x^{\pm}_n=x_n\pm \frac{4i\epsilon_L}{|p_{n-1}(x_n)|}$. Non-trivial monodromies (figure \ref{fig: branch-points}) are detected around these branch points. 
\subsection{Global structure}\label{sec: semi_global}
\begin{figure}[h!]
\centering
\includegraphics[width=0.6\textwidth]{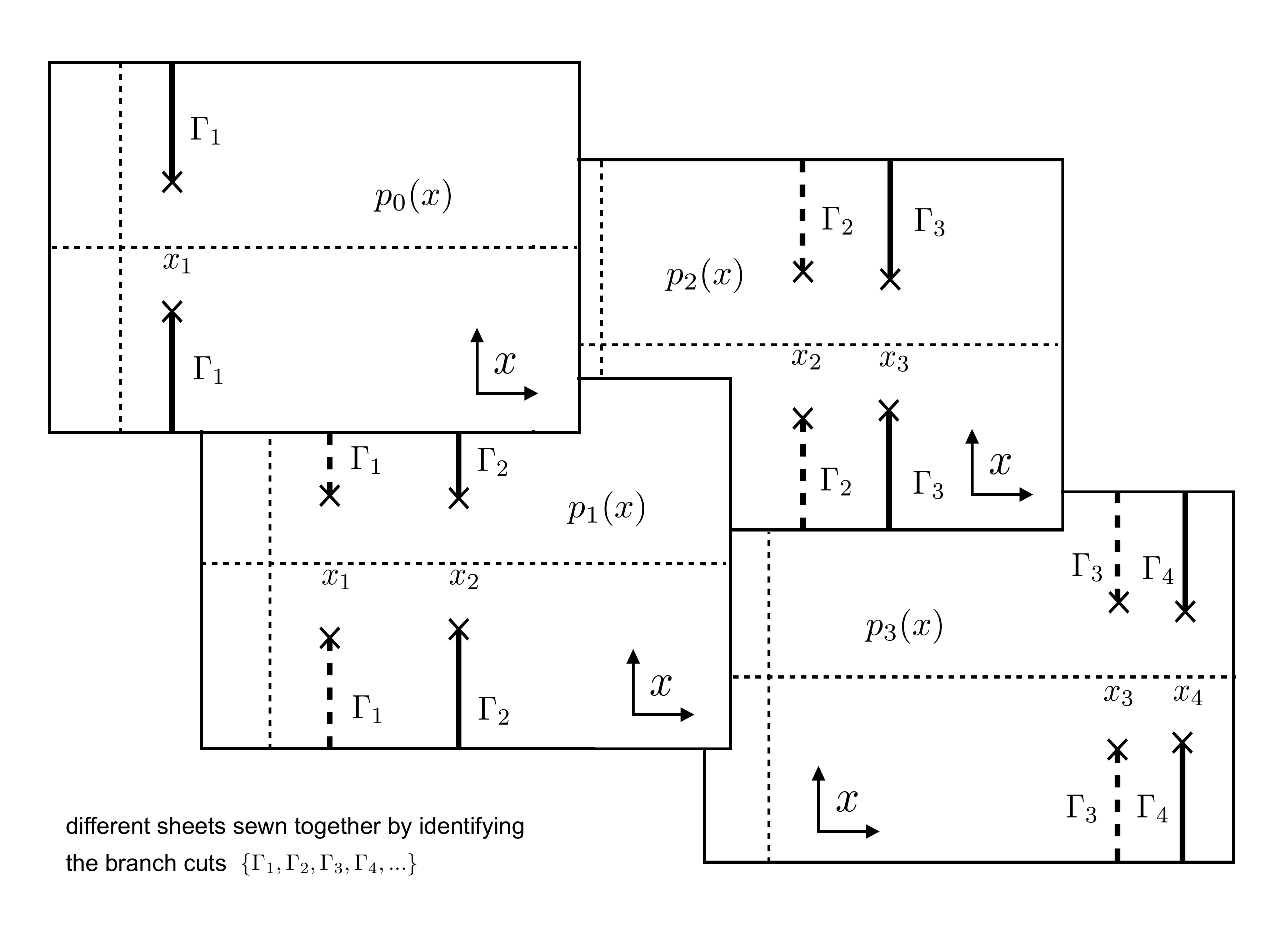}
\caption{\small{After the re-summation of probe corrections, the leading order forbidden thermal poles are ``resolved" into a series of branch-cuts. Through them a chain of additional saddles $p_n(x)$ are sewn together, and form an infinite Riemann surface $\mathcal{M}_p$.}}\label{fig: branch-cuts}
\end{figure}

We now comment the global structure of the semi-classical vacuum block $\mathcal{V}(x)$, or the associated accessory parameter $p(x)$, on the complex $x$-plane. The picture we draw is based solely on the ``quadratic" resolution discussed above. Admittedly there are additional global subtleties in the full solutions that come from with the higher order terms of the monodromy equation. We will not discuss them in this paper. We have seen that after summing over all $\mathcal{O}\left(\epsilon_L\right)$ corrections, each forbidden singularity is ``resolved" by a pair of branch points. Furthermore, we can identify an infinite number of additional saddles $p_n(x)$ with winding number $n$, which are sewn together across the branch cuts $\left\lbrace p_0(x)\to p_1(x)\to p_2(x)\to ...\right\rbrace$ in a way that resembles a one-sided ``chain".  All together they form an infinite-sheeted Riemann surface that we denote as $\mathcal{M}_p$ (see figure \ref{fig: branch-cuts}). 

To be more specific, let us describe in more detail the resulting analytic structure on $\mathcal{M}_p$. Starting near $x=0$ on the first sheet that corresponds to true vacuum block $p_0(x)$, we can approach the first forbidden singularity by taking $x\to x_1$. As one gets closer we to $x_1$ we will ``discover" its resolution into two branch points. If we choose to go beyond $x_1$ by staying close to the real axis, we remain on the same sheet, and there are no more forbidden singularities to be resolved on that sheet beyond $x_1$; however, if we choose to go around one of the branch points (say by moving above the branch point in the upper-half-plane), we enter the second sheet that corresponds to the additional saddle $p_1(x)$.\footnote{We caution the readers that the branch-cuts and multi-sheeted structure discussed here are related to the Euclidean region. They arise as artifacts of the semi-classical limit $c\to\infty$ and one should not confuse them with analytic continuation into late Lorentzian time.} Similar to what happens on the sheet $p_0(x)$, we can approach the second forbidden singularity $x_2$ on this sheet, and there are no other forbidden singularities to be resolved beyond $x_2$ if we choose to stay close the real axis and remain on the same sheet; otherwise we enter the third sheet $p_2(x)$, and so on. In other words, only two consecutive forbidden singularities (or more precisely, their resolutions) are visible on each particular sheet (except the first sheet $p_0(x)$, where only one is visible). Naively one might have hoped that the resolution of forbidden singularities would result in an array of smoothened ``bumps" while still keeping $p(x)\approx p_{\text{therm}}(x)$ along the way. After re-summing the probe effects, we see that this corresponds to crossing all the branch cuts, and the additional saddles play important roles in reproducing this.   

\section{Probe effects in micro-canonical ensembles}\label{sec: inverse_laplace}
In the last section, we studied the probe corrections to the leading order ETH results: 
\begin{equation}\label{eq: leading_ETH}
\langle E | \mathcal{O}_L(\tau)\mathcal{O}_L(0)| E\rangle \approx \langle \mathcal{O}_L(\tau)\mathcal{O}_L(0)\rangle_{\beta_E}
\end{equation}
We found that re-summing probe corrections transforms each forbidden ``thermal" poles at $\tau_n = n\beta_E$ into a pair of branch-point singularities at $\tau = \tau^{\pm}_n$. 
 
In this section, we show that such modification is respected by ETH. By this we mean that similar modification emerges from probe corrections to the RHS of ETH. Naively the finite temperature two-point functions are characterized by images of the OPE singularity along imaginary time, due to thermal periodicity. However, let us recall that it is actually the micro-canonical ensemble that ETH predicts to approximate the high-energy pure state. We will see that the probe corrections cause the micro-canonical result to differ form the canonical result in a way that parallels what we found in the previous section.
\subsection{Canonical to micro-canonical ensemble}
In the thermodynamic limit, the distinction between the canonical and micro-canonical ensemble vanishes, and one can approximate the micro-canonical ensemble by a canonical ensemble with a effective temperature $\beta = \beta_E$.

Let us first recall how this happens. The micro-canonical observable $\langle \mathcal{O}\rangle^{\text{micro}}_E$ is computed by summing over eigen-states $\psi$ whose energies lie in a thin energy shell $(E, E+\delta E)$: 
\begin{equation}
\langle \mathcal{O}\rangle^{\text{micro}}_E= N(E)^{-1} \sum_{\psi}\langle \mathcal{O}\rangle_\psi, \;\;E\leq H(\psi)\leq E+\delta E
\end{equation}
The counter-part in the canonical ensemble at temperature $\beta$ is computed by a weighted sum over all states: 
\begin{equation}
\langle \mathcal{O}\rangle_\beta = Z(\beta)^{-1}\sum_{\psi} e^{-\beta H(\psi)}\langle \mathcal{O}\rangle_\psi,\;\;Z(\beta) = \sum_{\psi} e^{-\beta H(\psi)}
\end{equation}
In the thermodynamic limit we have $E = c \mathcal{E}, c\to \infty$ with $\mathcal{E}$ finite. One can replace the discrete sum over states by an integral over a continuous distribution of states, with a density of states $\rho(\mathcal{E})\propto e^{cs(\mathcal{E})}$:
\begin{equation}\label{eq: cano-micro}
\langle \mathcal{O}\rangle_\beta = Z(\beta)^{-1}\int d\mathcal{E} e^{-c\left[\beta \mathcal{E}-s(\mathcal{E})\right]}\langle \mathcal{O}\rangle ^{\text{micro}}_{\mathcal{E}}
\end{equation} 
The ``probe limit" in this case corresponds to $\langle \mathcal{O}\rangle^{\text{micro}}_{\mathcal{E}} \propto e^{\mathcal{O}(1)}$, which then simply factors out in the saddle-point approximation: 
\begin{eqnarray}
\langle \mathcal{O}\rangle_\beta &\sim & e^{-c\left[\beta\mathcal{E}^*-s(\mathcal{E}^*)\right]}Z(\beta)^{-1}\langle \mathcal{O}\rangle ^{\text{micro}}_{\mathcal{E}^*},\;\;s'(\mathcal{E}^*)=\beta
\end{eqnarray}
 where $e^{-c\left[\beta\mathcal{E}^*-s(\mathcal{E}^*)\right]}$ is precisely the saddle point approximation for $Z(\beta)$, therefore we arrive at the equivalence between the canonical and micro-canonical ensembles: 
\begin{equation}
\langle \mathcal{O}\rangle_\beta \approx \langle \mathcal{O}\rangle^{\text{micro}}_{\mathcal{E}^*}
\end{equation}
\subsection{finite probe corrections}
\begin{figure}[h!]
\centering
\includegraphics[width=0.4\textwidth]{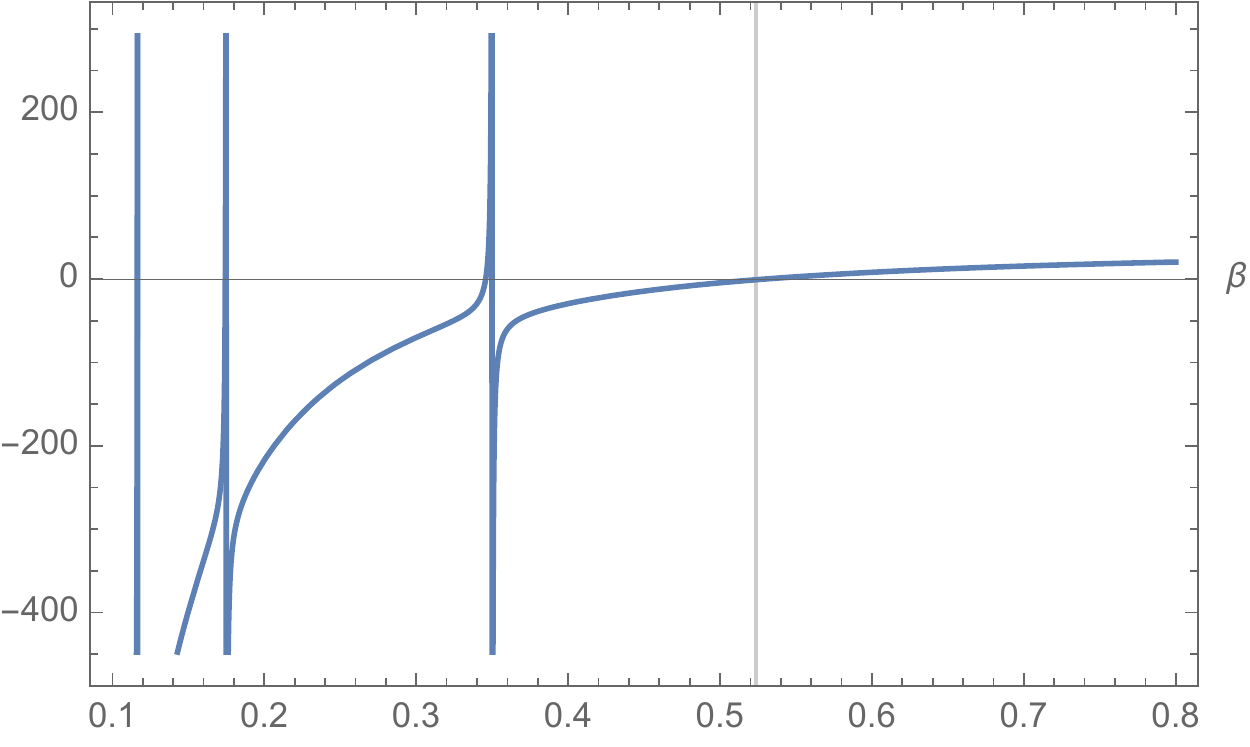}
\includegraphics[width=0.4\textwidth]{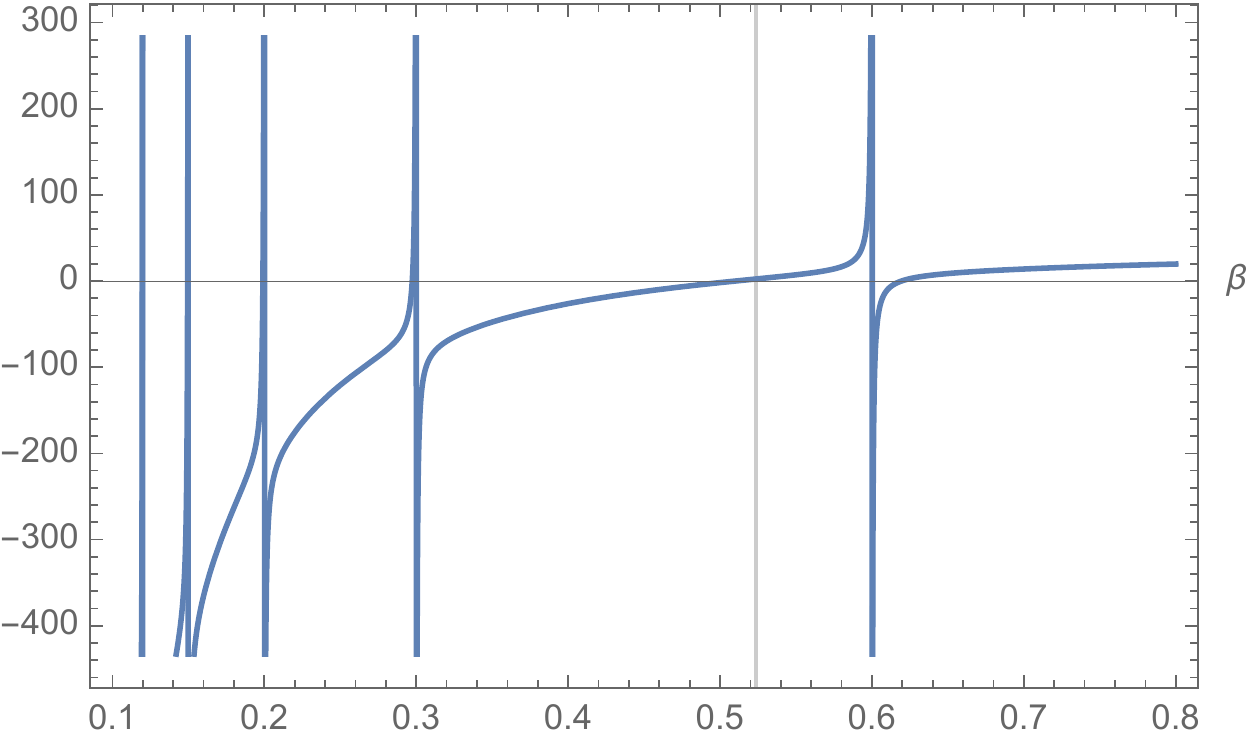}
\caption{\small{plots of the saddle-point equation $\;\epsilon_H-\frac{\pi^2}{\beta^2}-\frac{2\epsilon_L}{\beta}+2\epsilon_L\cot{\left(\frac{\pi\tau}{\beta}\right)}\frac{\pi\tau}{\beta^2}\;$ as a function of $\beta$, for $\;\epsilon_L=10^{-1}, \epsilon_H=6\mathcal{E}=36$. Left: for $\tau<\beta_{\text{thermal}}(\mathcal{E})$, the dominant saddle agrees well with $\beta_{\text{thermal}}(\mathcal{E})$ (grey line); Right: for $\tau>\beta_{\text{thermal}}$, the dominant saddle is replaced by a $\tau$-dependent new one.}}\label{fig:microcanonical_saddle}
\end{figure}

To go beyond the probe limit, we take the observable to scale exponentially with c: $\langle \mathcal{O}\rangle^{\text{micro}}_{\mathcal{E}}\sim e^{c f(\mathcal{E})},\;f(\mathcal{E})\ll 1$. The saddle point will be corrected by solving instead
\begin{equation}
s'\left(\mathcal{E}^*\right) + f'\left(\mathcal{E}^*\right) =\beta
\end{equation}
We study such corrections for the case of observable being the composite operator: $\mathcal{O}=\mathcal{O}_L(\tau)\mathcal{O}_L(0)$ with $h_L=\frac{c}{6}\epsilon_L$. The canonical ensemble results are given by two-point functions on a torus. Again in the high temperature limit we can approximate by those on infinite lines, which are fixed by conformal symmetries:
\begin{equation}
\langle \mathcal{O}_L(\tau)\mathcal{O}_L(0)\rangle_\beta  = \left(\frac{\beta}{\pi}\sin{\left(\frac{\pi\tau}{\beta}\right)}\right)^{-2h_L}
\end{equation} 
From this we can compute the micro-canonical two-point function by an inverse-Laplace transform:
\begin{equation}
\rho(\mathcal{E})\langle \mathcal{O}_L(\tau)\mathcal{O}_L(0)\rangle^{\text{micro}}_{\mathcal{E}} = \int^{\Gamma+i\infty}_{\Gamma-i\infty} d\beta\; e^{c\beta\mathcal{E}}\;Z(\beta)\langle \mathcal{O}(\tau)\mathcal{O}(0)\rangle_\beta
\end{equation}
where the vertical contour $\Gamma$ should be placed to the right of any singularity of the integrand. For illustration we work in CFTs with a gravity dual, where in the high temperature phase of Hawking-page transition we have:
\begin{equation}
Z(\beta)=e^{\frac{\pi^2 c}{6\beta}},\;\rho(\mathcal{E})=e^{2\pi c\sqrt{\frac{\mathcal{E}}{6}}}
\end{equation}
Therefore the goal is to evaluate: 
\beqn\label{eq: saddle-approx}
\langle \mathcal{O}(\tau)\mathcal{O}(0)\rangle^{\text{micro}}_{\mathcal{E}} = \int^{\Gamma+i\infty}_{\Gamma+i\infty} d\beta\; \exp{\left\lbrace c\left(\beta \mathcal{E} + \frac{\pi^2}{6\beta}-\frac{\epsilon_L}{3} \log{\left(\sin{\left(\frac{\pi\tau}{\beta}\right)}\frac{\beta}{\pi}\right)}\right)\right\rbrace}
\eeqn
We can proceed with the saddle point approximation as before, and solve for 
\beq\label{eq:beta_saddle}
\mathcal{E}-\frac{\pi^2}{6\beta_*^2}-\frac{\epsilon_L}{3\beta_*}+\frac{\epsilon_L}{3}\cot{\left(\frac{\pi\tau}{\beta_*}\right)}\frac{\pi\tau}{\beta_*^2}=0
\eeq
In the probe limit $\epsilon_L\to 0$, the saddle point is given by $\beta_{\text{thermal}}(\mathcal{E})=\pi/\sqrt{6\mathcal{E}}$, and we recover the equivalence between micro-canonical and thermal two-point functions:
\begin{equation} 
\langle \mathcal{O}_L(\tau)\mathcal{O}_L(0)\rangle^{\text{micro}}_{\mathcal{E}}=\langle \mathcal{O}_L(\tau)\mathcal{O}_L(0)\rangle_{\beta_{\text{thermal}}(\mathcal{E})}
\end{equation}  
Beyond the probe limit we need to include the $\epsilon_L$ ``back-reaction" to the saddle-point calculation. As shown in figure \ref{fig:microcanonical_saddle}, the probe corrections introduce infinitely many pairs of additional saddles as well as singularities, located near $\beta_n=\frac{\tau}{n}, n\in \mathbb{Z}$. Recall that to extract the dominant contribution, the contour $\Gamma$ needs to be positioned to the right of all singularities, this fixes the dominant saddle to always be the right-most one. As we vary $\tau$, the position of the additional-saddles move. When $\tau<\beta_{\text{thermal}}(\mathcal{E})$, the dominant saddle is approximately $\beta_{\text{thermal}}(\mathcal{E})+\mathcal{O}(\epsilon_L)$, receiving only perturbative corrections. In particular the micro-canonical two-point function is still approximated by the thermal two-point function. However, as $\tau$ crosses $\beta_{\text{thermal}}(\mathcal{E})$, the dominant saddle is replaced by a new one that is completely driven by the probe correction terms, and therefore is strongly $\tau$-dependent. From this point on the micro-canonical two-point function ceases to be approximated by the thermal one.
 
\begin{figure}[h!]
\centering
\includegraphics[width=0.4\textwidth]{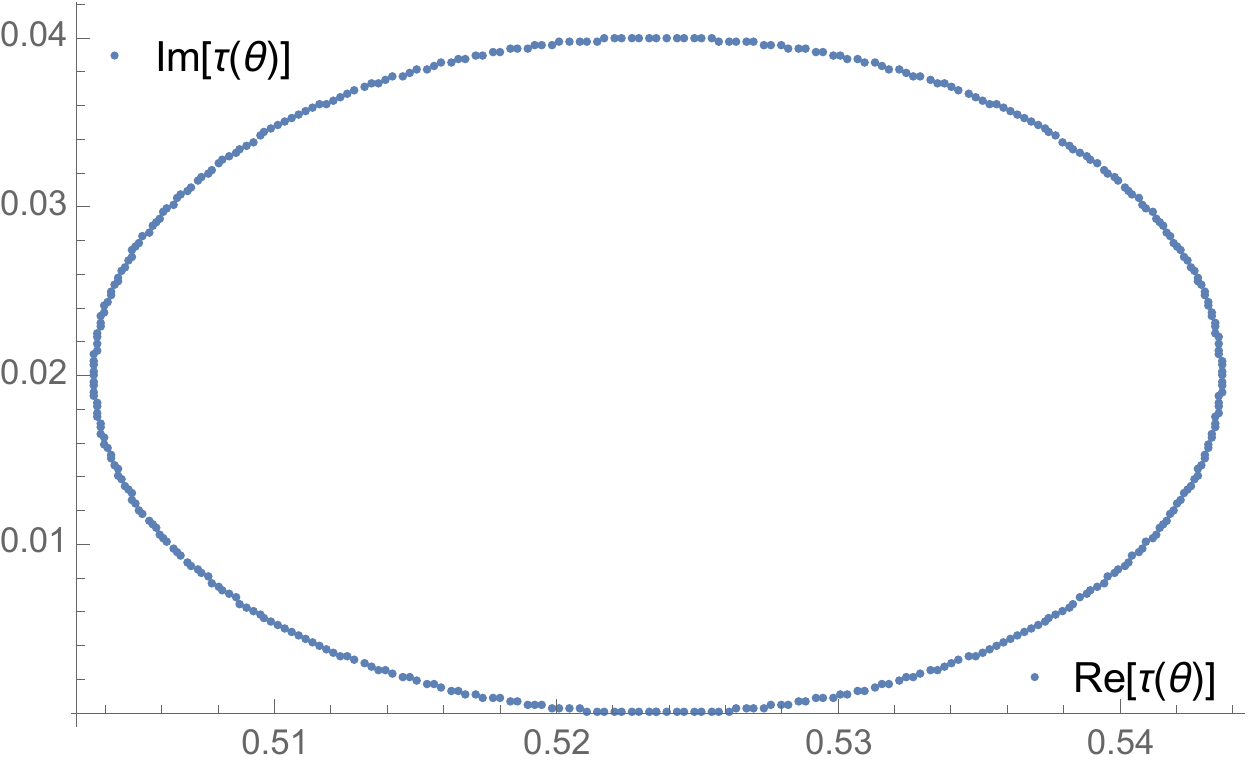}
\includegraphics[width=0.4\textwidth]{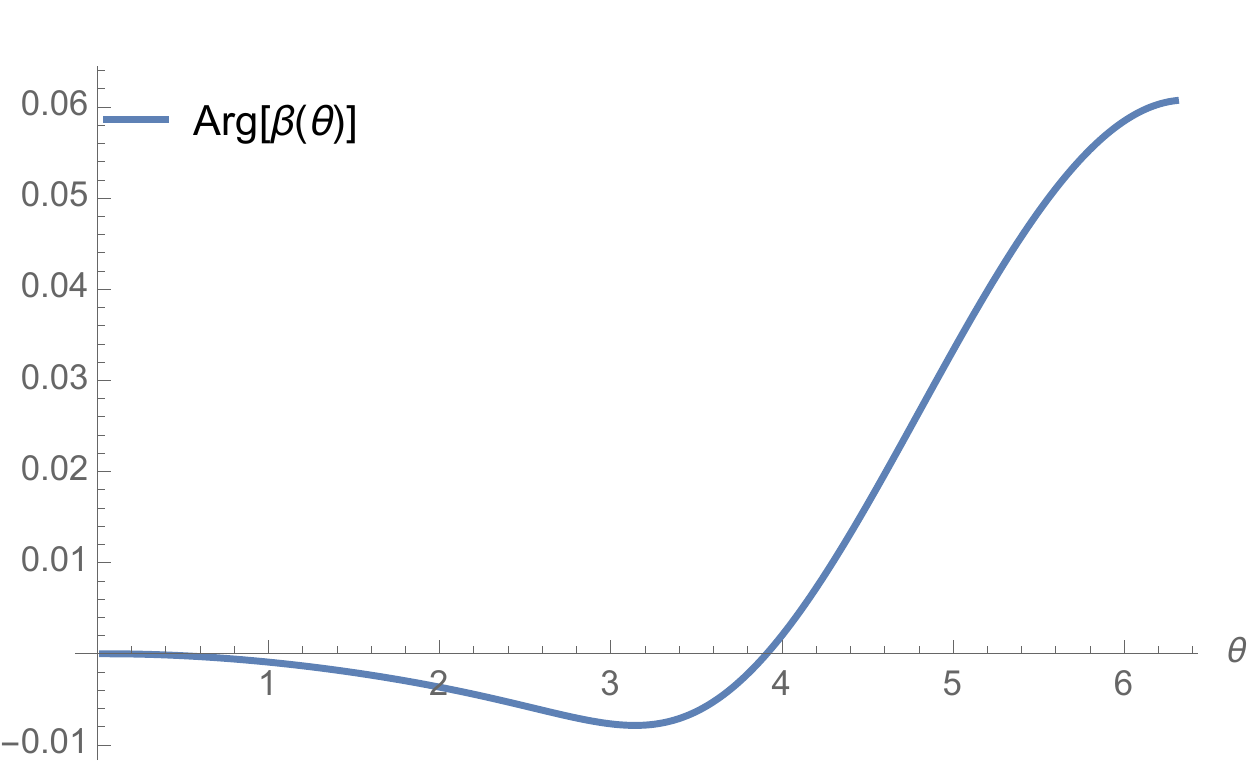}
\caption{\small{Monodromy for the dominant saddle $\beta\left(\tau(\theta)\right)$ (right) as we trace a contour (left) surrounding the a branch point near the thermal pole $\tau=\frac{\pi}{\sqrt{\epsilon_H}}\approx 0.52$ for $\epsilon_H=36, \epsilon_L=10^{-2}$. A symmetric branch point exists in the lower half complex $\tau$-plane.}}\label{fig: micro-monodromy}
\end{figure}

Effectively, the probe back-reaction modifies the saddle point in such a way that the divergence at the thermal pole $\tau=\beta_{\text{thermal}}$ in the canonical two-point function is rendered finite. Technically, this is achieved by always having the new saddle (now strongly $\tau$ dependent) satisfy $\beta_* > \tau$. In addition, a branch-cut structure analogous to what we found in the last section from the monodromy problem also emerges (see figure \ref{fig: micro-monodromy}), connecting different saddles that the probe term introduces. 

Next we present some numerical calculations of $\langle\mathcal{O}(\tau)\mathcal{O}(0)\rangle^{\text{micro}}_{\epsilon_H}$ in the saddle-point approximation $c\to \infty$ by tracing through the $\tau$-dependent saddles $\beta_*(\epsilon_H,\epsilon_L,\tau)$. To place the results in the context of ETH, we want to compare them with the excited-state calculations done in the last section using the monodromy method. For this reason we translate the results into the form of an ``accessory parameter" $p(\epsilon_H, \epsilon_L, x)$, where
\beq
p(\epsilon_H,\epsilon_L,x)=\frac{\partial {f(\epsilon_H,\epsilon_L,x)}}{\partial {x}},\;\langle\mathcal{O}(x)\mathcal{O}(0)\rangle^{\text{micro}}_{\epsilon_H}=e^{-\frac{c}{6}f(\epsilon_H,\epsilon_L,x)},\;\tau=-\log{(1-x)}
\eeq

\begin{figure}[h!]
\centering
\includegraphics[width=0.6\textwidth]{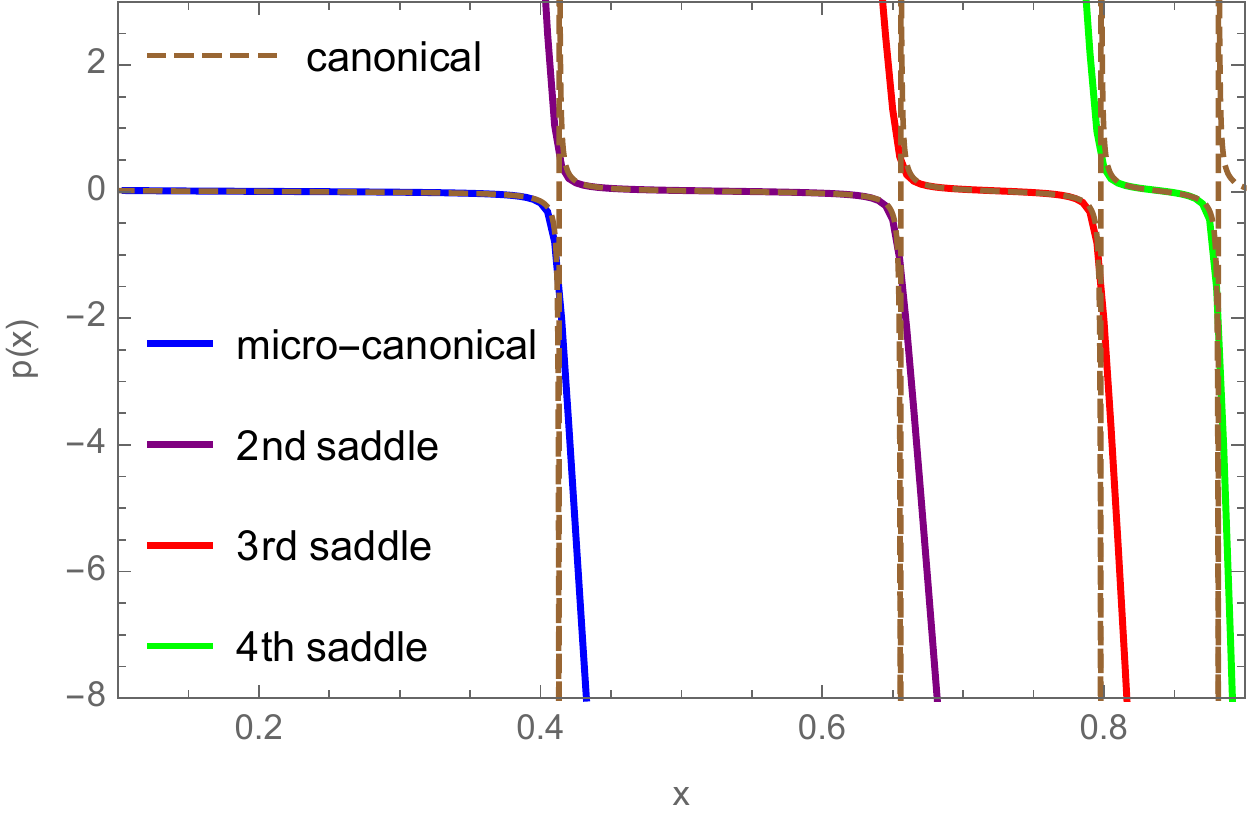}
\caption{\small{Plots of $p(\epsilon_H=36, \epsilon_L=10^{-3}, x)$ from various saddles (solid) of the inverse-Laplace transformationt; compared against the canonical result (dashed).}}\label{fig: micro-saddles}
\end{figure}

In figure \ref{fig: micro-saddles} we plot the corresponding $p(x)$ from tracing through different saddles in performing the inverse-Laplace transformation. Along imaginary time (real $\tau$), the dominant saddle gives the micro-canonical result. The sub-dominant saddles are analogous to the additional saddles that arise in solving the monodromy problem. In particular, together they form a piece-wise resolution of the thermal singularities in a way that mimic the monodromy result figure \ref{fig:accessory_real}. Into the complex $\tau$ plane, these saddles switch dominances and are connected via the branch-cuts. 

\begin{figure}[h!]
\centering
\includegraphics[width=0.6\textwidth]{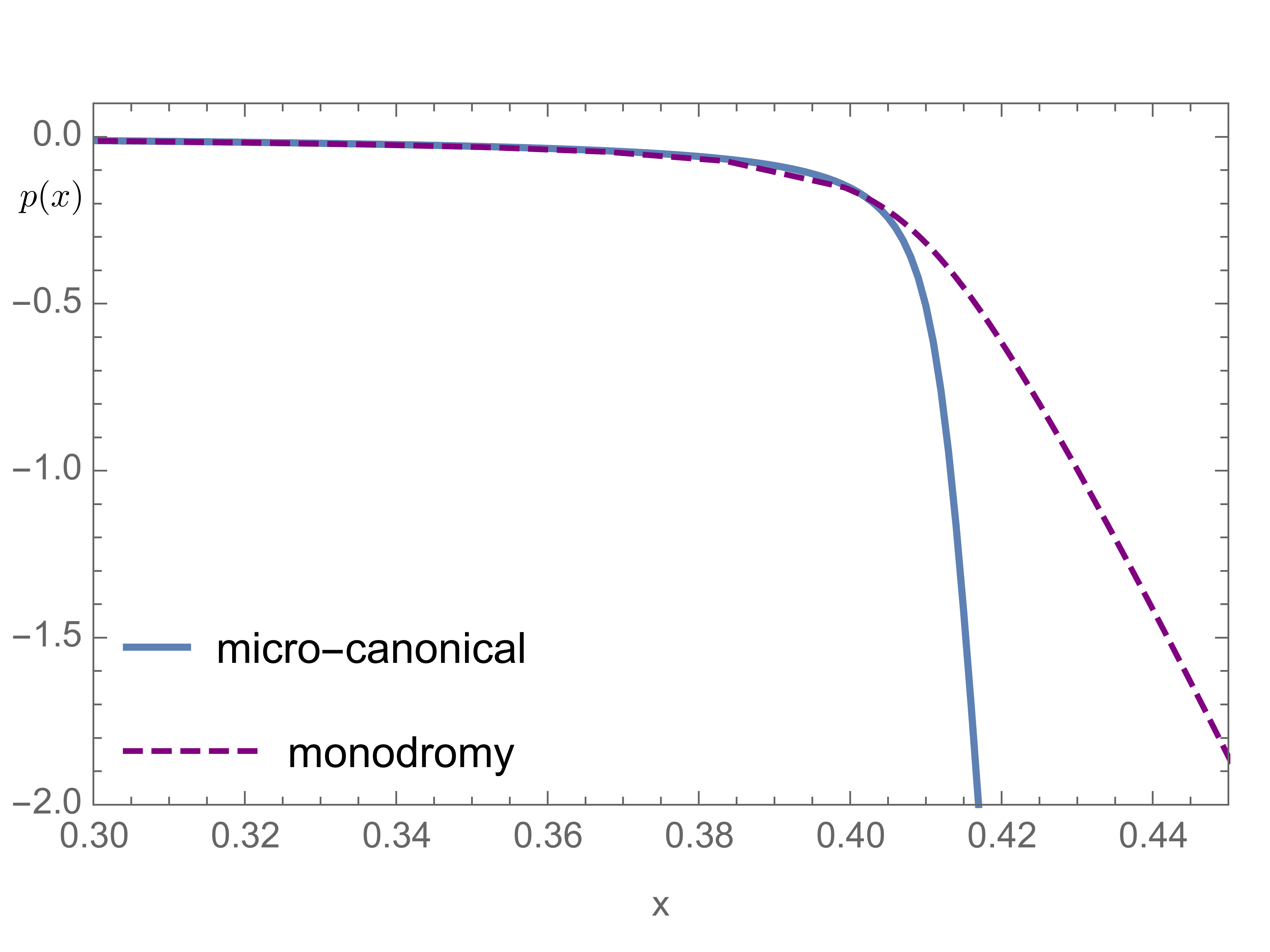}
\caption{\small{Plots of $p(\epsilon_H=36, \epsilon_L=10^{-3}, x)$. Blue is the micro-canonical ensemble results; purple is the result from the monodromy calculation.}}\label{fig: micro-vs-block}
\end{figure}

\begin{figure}[h!]
\centering
\includegraphics[width=0.47\textwidth]{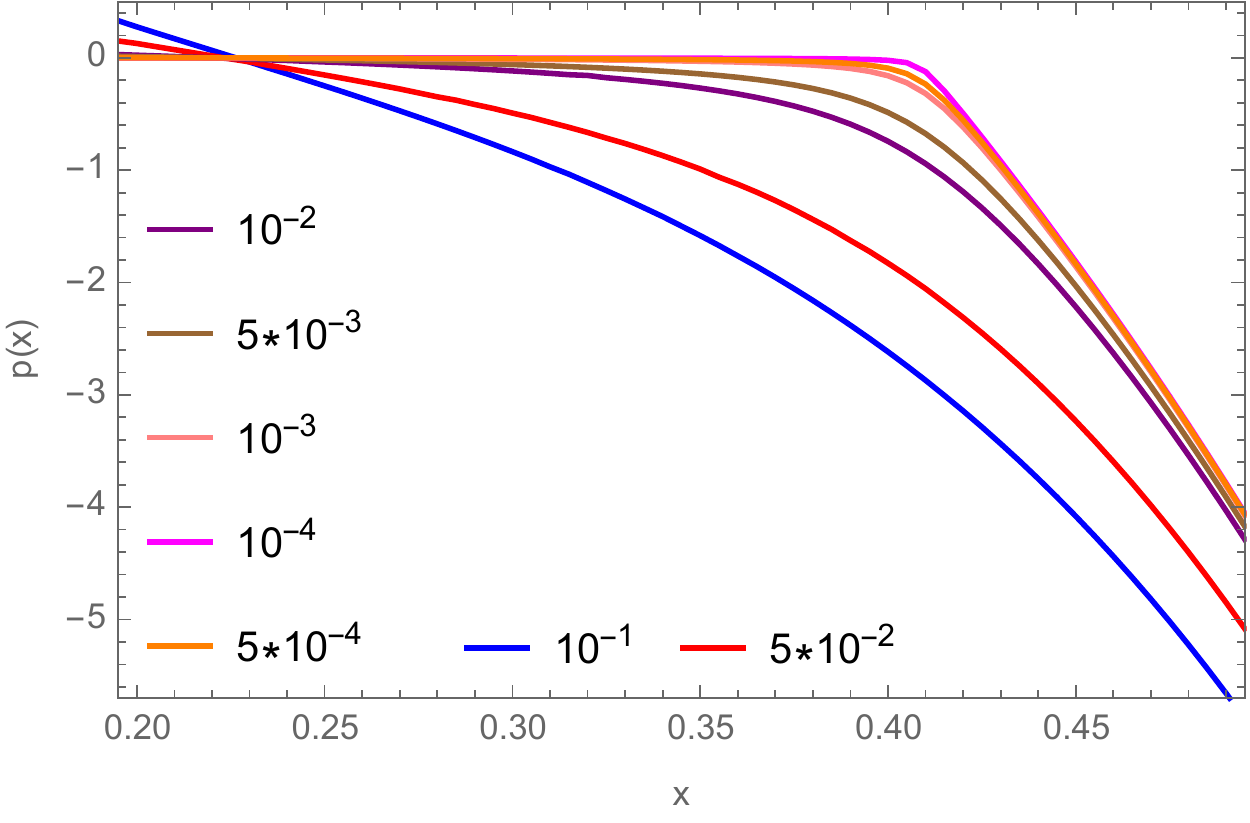}
\includegraphics[width=0.47\textwidth]{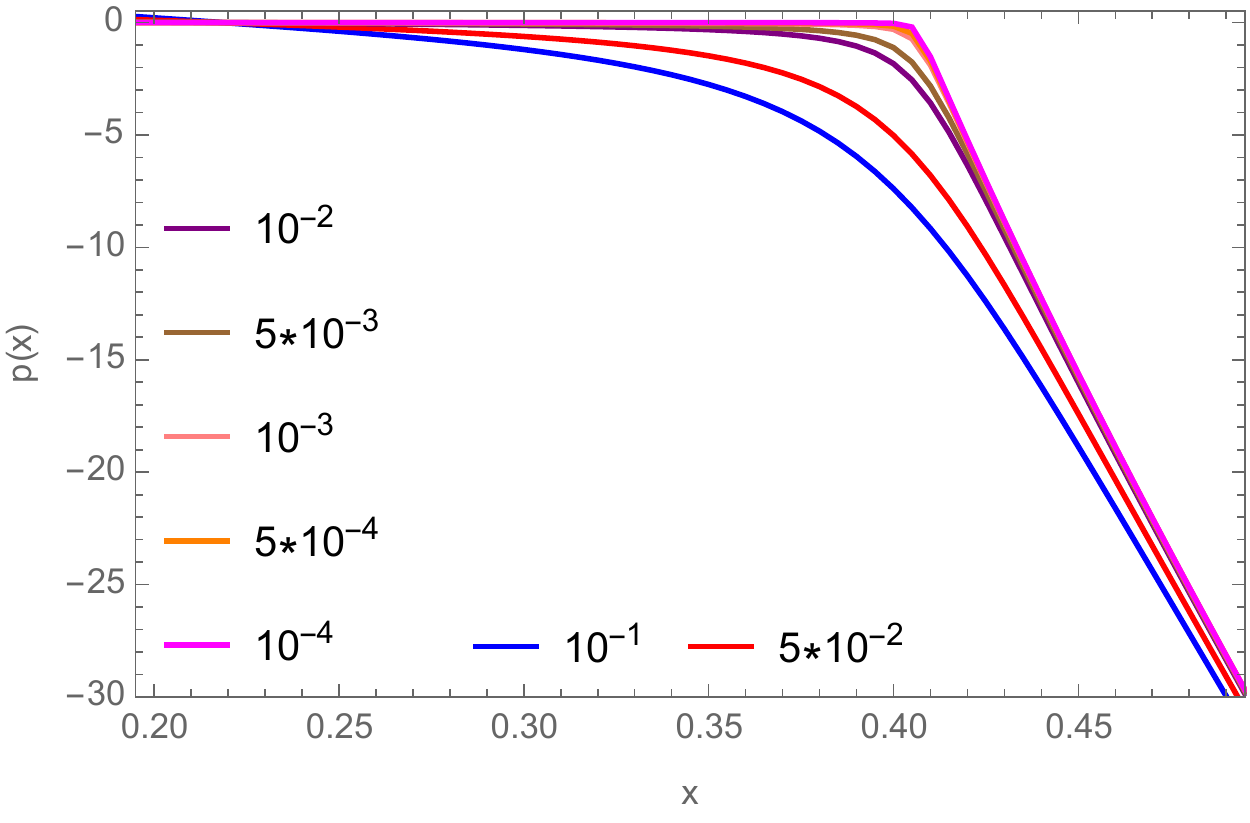}
\caption{\small{Limiting curves for $p(x)$ from the excited state (left) and micro-canonical ensemble (right), as one $\epsilon_L$ varies from $10^{-1}$ down to $10^{-4}$.}}\label{fig: limiting-micro-vs-block}
\end{figure}

We see that the drastic transformation in the analytic structure of the vacuum Virasoro block introduced by the probe corrections are reproduced in the micro-canonical ensemble. In this sense, ETH does not suffer from a qualitative breakdown away from the probe limit. Notice that the change in the analytic structure takes place for arbitrarily small but finite $h_L/c$. The fact that both sides of the ETH equality undergo the same qualitative change makes it plausible that the mismatch between $\langle \mathcal{O}_{obs}\rangle_H$ and $\langle \mathcal{O}_{obs}\rangle_{\text{micro}}$, if any, should vanish smoothly as the probe $h_L/c\to 0$. We should check this expectation with a quantitative comparison. In figure \ref{fig: micro-vs-block} we plot the accessory parameter from the micro-canonical result against that of the excited state result, picking for the same set of parameters. Surprisingly, even for $\epsilon_L$ as small as $10^{-3}$ the two develop a significant deviation from each other after the thermal length scale. In fact, we make the interesting observation (see figure {\ref{fig: limiting-micro-vs-block}}) that the excited state accessory parameters approach a limit curve beyond the thermal length scale as $\epsilon_L$ decreases; and the same is true for the micro-canonical ensemble. Therefore, the deviation shown in figure \ref{fig: micro-vs-block} does not diminish as one decrease $\epsilon_L$ further. This is puzzling and we do not have a good explanation for it. One possible issue here is that the monodromy calculation computes the block on a circle. However, in computing the micro-canonical result, we have used the universal $\langle \mathcal{O}(\tau)\mathcal{O}(0)\rangle_\beta$ on an infinite line, though in the high energy/temperature limit, one would not expect the distinction between circle and infinite lines to enter. 
We leave checking the corresponding calculations on a circle to future work.

\section{Renyi entropy for finite subsystem}\label{sec: Renyi}
In this section we switch gear and study Renyi-entropies in CFTs on a $2\pi$ circle.\footnote{We thank Tom Hartman and Tarun Grover for early collaboration on some of the results in this section.} Via the replica trick we can compute the excited state $n$-th Renyi entropy for a single interval $[0,\theta]$ as a two-point function of twist operators $\sigma_n$ in the orbifolded CFT with central charge $nc$:
\begin{equation}
S_n = \frac{1}{1-n}\ln{\text{tr}\rho_H^n},\;\;\text{tr}\rho_H^n=\langle \sigma_n(\theta)\sigma_n(0)\rangle_{h_H}\sim \langle \sigma_n(0)\sigma_n(x)\mathcal{O}_H(1)\mathcal{O}_H(\infty)\rangle
\end{equation}
which is a HL correlator with conformal ratio $x=1-e^{i\theta}$, and the twist operator is taken to be ``light" with scaling dimension $h_n = \frac{nc}{24}\left(1-\frac{1}{n^2}\right)$. Working in a large $c$ holographic like CFTs the result is approximated by the Virasoro vacuum block. For $n>1$ we move beyond the probe limit. Previous work has computed the short distance expansions of the vacuum blocks for $n>1$, and found them to differ from the thermal results \cite{Chen2013,Lin2016,He2017,Tolya1}. Having observed the interplay between probe effects and probe separations in previous sections, we extend the short distance expansions and compute Renyi-entropies for finite interval size (but not exceeding half of total system). 
In this section we will mostly compare to the more standard thermodynamic limit which is achieved by sending $\epsilon_H \rightarrow \infty$ (although our results also require taking $c \rightarrow \infty$ first.)
Analytic expressions for Renyi entropies in similar regimes were proposed in \cite{Lu2017} for more generic systems using ergodicity arguments. 

\begin{figure}[h!]
\centering
\includegraphics[width=0.4\textwidth]{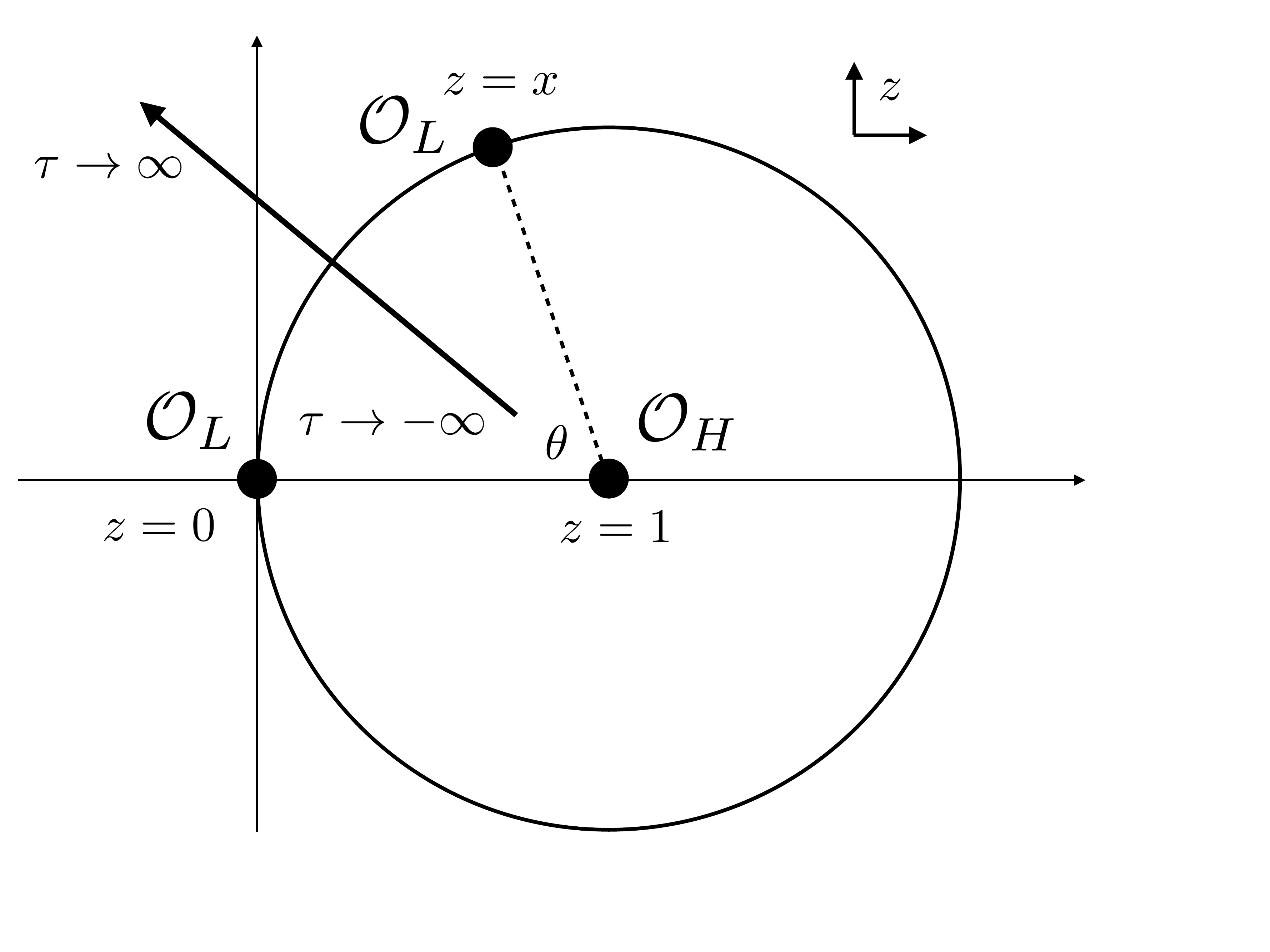}
\includegraphics[width=0.4\textwidth]{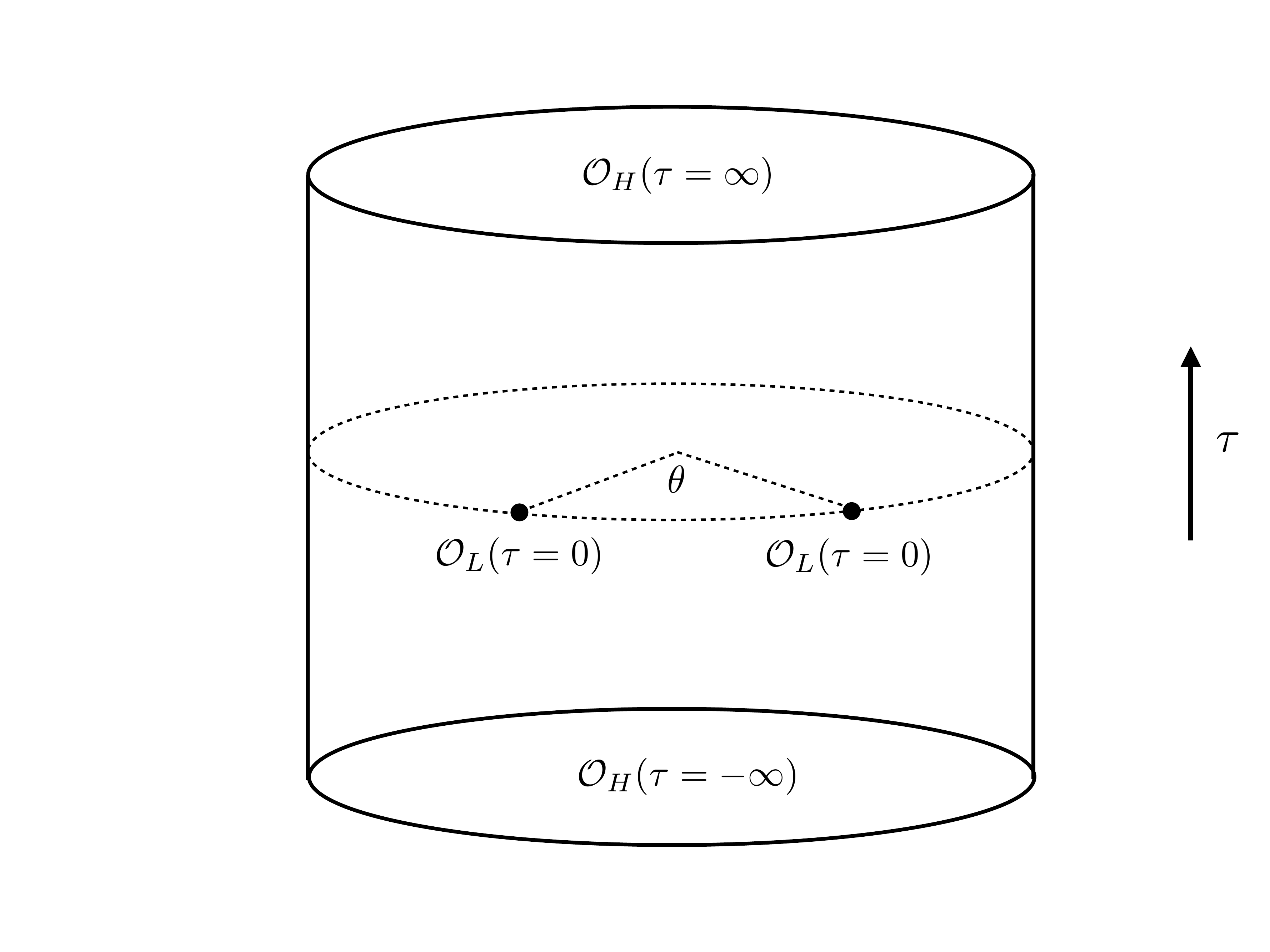}
\caption{\small{conformal mapping from the monodrompy problem (left) to a scattering problem (right)}}\label{fig:scattering}
\end{figure}

To proceed, it is easier to perform a conformal transformation $z\to 1-e^{i\theta/2+\tau}$ and at the same time re-scale $\psi(\tau)=\left(\partial_z\tau(z)\right)^{-1/2}\psi\left(z\right)$, mapping the corresponding monodromy problem onto the cylinder, see figure \ref{fig:scattering}:
\begin{eqnarray}\label{eq:Schrodinger}
&&-\psi''(\tau)+\left(V(\tau)-E\right)\psi(\tau)=0\nonumber\\
&& V(\tau)=\frac{\epsilon_L \sin{\left(\theta/2\right)}^2}{\left(\cos{\left(\theta/2\right)-\cosh{\tau}}\right)^2}+\frac{p_\theta \sin{\left(\theta/2\right)}}{\cos{\left(\theta/2\right)}-\cosh{\tau}},\;\; E=\epsilon_H-\frac{1}{4}
\end{eqnarray}
where the new accessory parameter $p_\theta$ is related to $p_x$ by $p_\theta = i p_x e^{i\theta}$. The monodromy problem (\ref{eq:Schrodinger}) takes the form of a Schrodinger equation along the real $\tau$ axis (corresponding to the original Euclidean time).
\subsection{Monodromy and reflectionless scattering}
For the vacuum block, trivial monodromy is to be imposed around a contour encircling either $z=(0,x)$ or $z=(1,\infty)$, which are mapped to either $\tau = (-i \theta/2, i\theta/2)$ or $\tau=(-\infty, \infty)$. In (\ref{eq:Schrodinger}),  trivial monodromy around $\tau=\pm\infty$ is equivalent to imposing no-reflection condition for a wave with energy $E=\epsilon_H-\frac{1}{4}$ scattering over the potential $V(\tau)$. 

To see this, expand the solution $\psi(\tau)$ near $\tau=\pm\infty$ (since $V(\pm \infty)= 0$) into plane-waves: 
\begin{eqnarray}
\psi(\tau)&\approx & e^{ik\tau} + R e^{-ik\tau},\;\;\tau\to -\infty\nonumber\\
\psi(\tau)&\approx & T e^{ik\tau},\;\;\;\;\tau\to\infty\nonumber
\end{eqnarray}
where $k=\sqrt{E}$. From these we can write the connection matrix between $\tau=\pm\infty$ in the plane-waves basis of each point as
\begin{equation}
C_\infty=\left[
\begin{array}{cc}
T^{-1} & RT^{-1}\\
\left(RT^{-1}\right)^*&\left(T^{-1}\right)^*\\
\end{array}
\right]
\end{equation}
Going around infinity picks up a phase for each plane wave: 
\begin{equation}
D_k=\left(\begin{array}{cc}e^{-2\pi k}&0\\0&e^{2\pi k}\\
\end{array}\right)
\end{equation}
The monodromy is thus given by 
\begin{equation}
M=C_\infty D_k C^{-1}_\infty D^{-1}_k = \left[
\begin{array}{cc}
\frac{1-|R|^2e^{4\pi k}}{1-|R|^2}&\frac{R}{T^2}\left(1-e^{-4\pi k}\right)\\
\left(\frac{R}{T^2}\right)^*\left(1-e^{-4\pi k}\right)&\frac{1-|R|^2e^{4\pi k}}{1-|R|^2}\\
\end{array}
\right]
\end{equation}
From this it is evident that trivial monodromy is attained by forcing zero-reflection $R=0$. We can therefore solve the monodromy problem by determining the condition under which the scattering coefficients of the Schrodinger equation (\ref{eq:Schrodinger}) has zero reflection. 
\subsection{WKB analysis}
For high energy micro-states with $\epsilon_H\gg 1$, $E>V(\tau)$ for all real $\tau$, therefore classically there is no reflection. Naively one might thus expect that $R=0$ for all choice of the accessory parameter $p_\theta$. However, through the Stoke's phenomena (see Appendix~\ref{app: stokes} for a brief summary), quantum mechanically there could be an exponentially small reflection coefficient, analogous to the tunneling rate in the case of under-scattering $E<V(\tau)_{\text{max}}$. It is this exponentially small $R$ that we aim to identify and tune to zero. 

In the limit of $E\gg 1$, we can do a WKB analysis of  Schrodinger's equation by identifying the Stoke's and anti-Stoke's curves. In the over-scattering case ($E>V(\tau)_{\text{max}}$), there are 4 turning points $\lbrace\tau_i\rbrace$ on the imaginary $\tau$ axis, defined by $V(\tau_i)=E$. 

\begin{figure}[h!]
\centering
\includegraphics[width=0.6\textwidth]{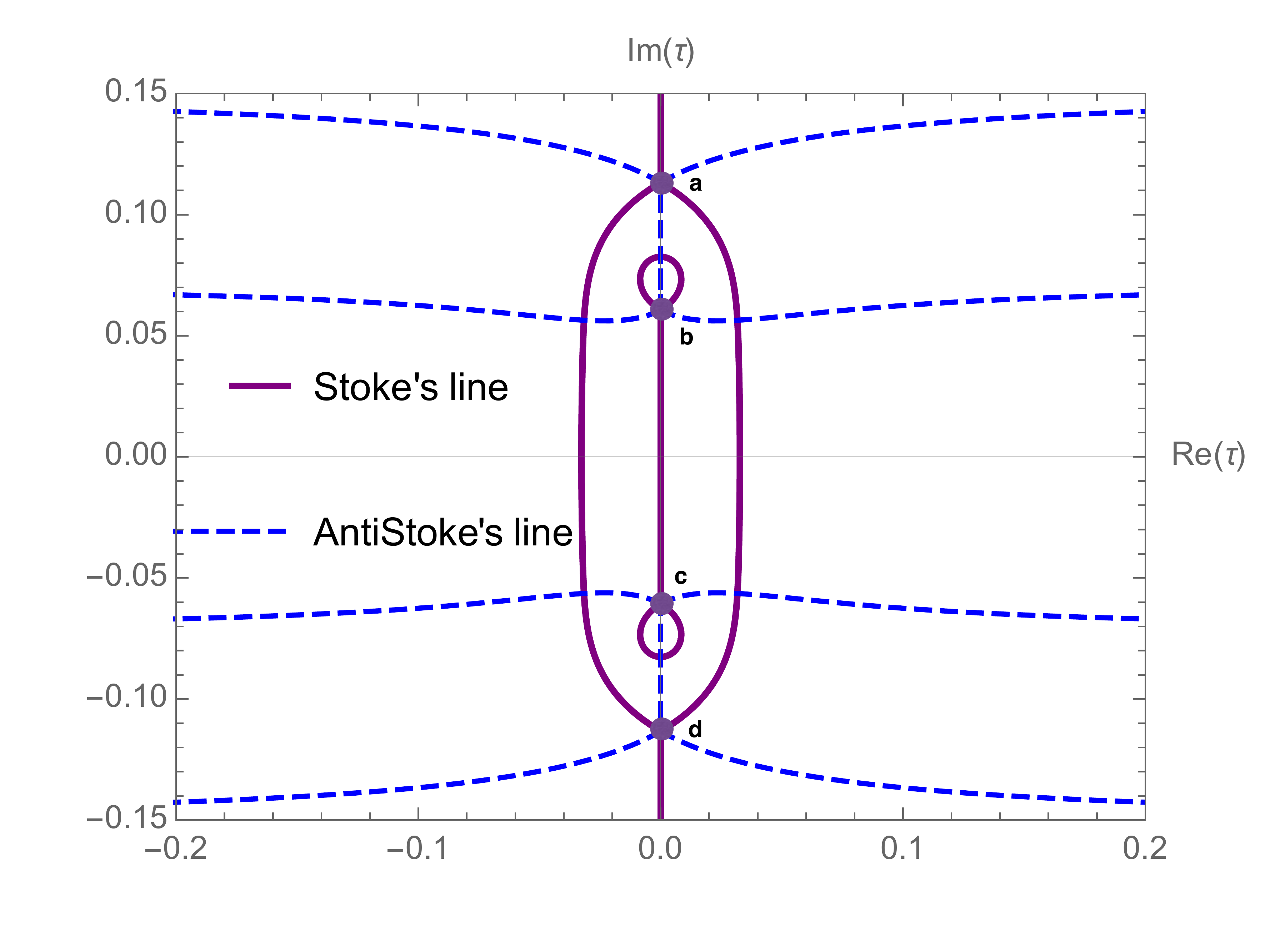}
\caption{\small{Stoke's and anti-Stoke's curves}}\label{fig:Stokes1}
\end{figure}
Let's denote the right-moving and left-moving asymptotic solutions by 
\begin{equation}
\psi_{\pm} \sim e^{\pm i k\tau}\sim e^{\pm i\int^\tau \sqrt{E-V(\tau')}d\tau'}
\end{equation} 
To compute the reflection coefficient $R$, we start from the right-moving solution $\psi_+$ in the region $\tau\to\infty$. We then continue the solution to the region $\tau\to-\infty$. Through Stoke's phenomena, each time we cross a Stoke's curve $\gamma_i$, a discontinuity is generated \cite{Heading1}: 
\begin{equation}
\psi^i_{d}\to \psi^i_{d} + i e^{-2W} \psi^i_{s},\;\; \psi^i_{s}\to\psi^i_{s}
\end{equation} 
where $\psi^i_{d/s}$ denotes the solution that increases/decreases exponentially along the Stoke's curve $\gamma_i$. The weight $W=i\int^{\tau_i}_{t_0} \sqrt{E-V(t)}dt $ is given by the integral from the corresponding turning point $\tau_i$ to the crossing point $t_0$ on the real axis.

In the case where the two turning points $\tau_{i,j}$ are connected by an anti-Stoke's curve (so that $\phi_{ij}=\int^{\tau_i}_{\tau_j}\sqrt{E-V(t)}dt$ is real), there is a relative phase $e^{2i\phi_{ij}}$ between the Stoke's phenomena at $\tau_i$ and $\tau_j$. $\phi_{ij}$ in general depends on the accessory parameter $p_\theta$. By tuning the interference such that the left-moving solution $\psi_-(\tau)$ generated from crossing all Stoke's curves cancel out, we solve the accessory parameter. 

\begin{figure}[h!]
\centering
\includegraphics[width=0.6\textwidth]{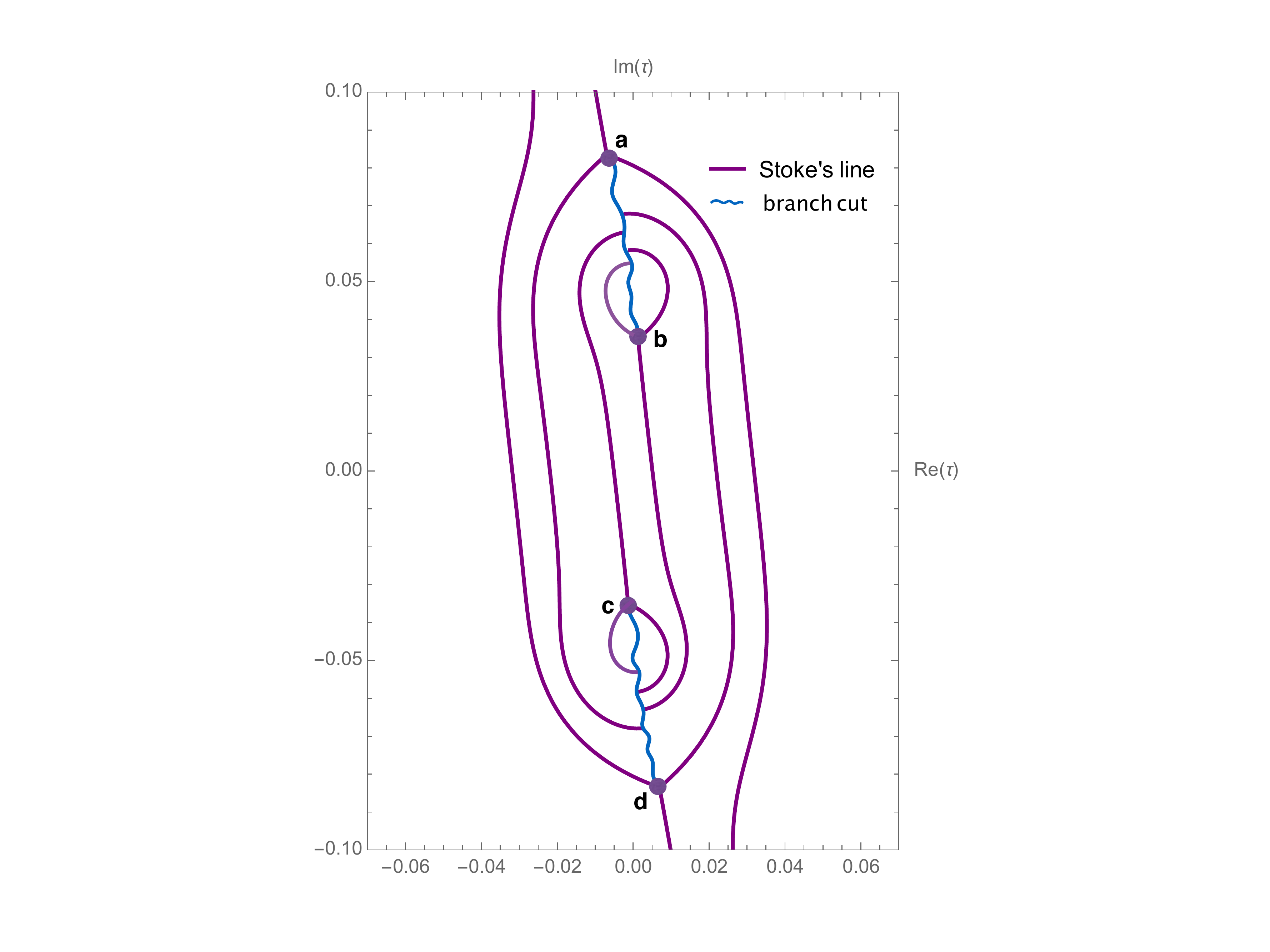}
\caption{\small{Resolved Stoke's curves (on the first sheet)}}\label{fig:Stokes2}
\end{figure}

From figure \ref{fig:Stokes1} we identify two pairs of turning points: ($a,b$) and ($c,d$), that are connected by anti-Stoke's curves. However, it turns out that the Stoke's phenomena is dominated by only ($a,b$). To see this explicitly, we need to first resolve the degeneracy of the Stoke's curves (degeneracy refers to the fact that some turning points are connected by Stoke's curves, which is due to the symmetry of the configuration) by giving the energy $E$ a tiny negative imaginary part: $E\to E-i\epsilon$. This tilts the Stoke's curves, which are now all semi-infinite (see figure \ref{fig:Stokes2}). We can therefore follow the standard procedure to cross each Stoke's curve. Carefully tracing through all of them, one can check that: 
\begin{equation}
\psi^{\infty}_+(\tau) \to \psi^{-\infty}_+(\tau) + e^{-2W}\left(1+e^{2i\phi_{ab}}\right)\psi^{-\infty}_-(\tau) + \mathcal{O}\left(e^{-4 W}\right),\;\; W\sim \mathcal{O}(H)
\end{equation}

It implies that the dominant Stoke's phenomena takes place between $a$ and $b$, while the effects from the other two turning points are further suppressed by $\mathcal{O}\left(e^{-2W}\right)$ and thus negligible. Notice that due to the pole singularity of $V$ at $\tau=i\theta/2$, which we did not take into account in the above analysis, the zero-reflection condition $R=0$ is not simply given by $1+e^{2i\phi_{ab}}=0$. A more refined analysis is needed to obtain the correct answer, which we give in the next subsection. 

\subsection{Stoke's phenomenon and Whittaker's functions}
Based on the above analysis, we can compute the reflection coefficient $R$ by zooming near the cluster of turning points and poles ($a,b,i\theta/2$). To do this, define the coordinate $\tau=i\theta/2+iy/\sqrt{4\epsilon_H}$, and re-scale $\hat{p}_\theta = 2 p_\theta/\sqrt{\epsilon_H}$. When $\epsilon_H\gg 1$, one can approximate the Schrodinger's equation for $y\ll \sqrt{\epsilon_H}$: 
\begin{equation}\label{eq:Schrodinger_zoom}
\partial^2_y \psi + \left(-\frac{1}{4}+\frac{\hat{p}_\theta}{y}+\frac{\epsilon_L}{y^2}\right)\psi = 0
\end{equation}
Corrections to (\ref{eq:Schrodinger_zoom}) are controlled by $\mathcal{O}\left(y/\sqrt{\epsilon_H}\right)$. Since the cluster of turning points and pole in $y$ coordinate are $\left(a,b,i\theta/2\right)\propto \left(2\hat{p}_\theta \pm 2\sqrt{\hat{p}_\theta^2+\epsilon_L},0\right)$, which are within $\mathcal{O}(1)\ll \sqrt{\epsilon_H}$, we can trust that (\ref{eq:Schrodinger_zoom}) captures the full Stoke's phenomena from $\left(a,b,i\theta/2\right)$ accurately. 
A generic solution to (\ref{eq:Schrodinger_zoom}) can be expressed explicitly in terms of the Whittaker's functions: 
\begin{equation}
\psi(y)=C_1 M\left(\hat{p}_\theta, -\frac{1}{2}\sqrt{1-4\epsilon_L},y\right)+C_2 W\left(\hat{p}_\theta, -\frac{1}{2}\sqrt{1-4\epsilon_L},y\right)
\end{equation}
Equivalent we can rearrange $\psi(y)$ into a sum of two functions $\lbrace P(y), Q(y)\rbrace$ having definite asymptotics: 
\begin{equation}
\psi(y)=C_+ P(y) + C_- Q(y),\;\; P(y)\sim e^{-y/2}y^{\hat{p}_\theta},\;Q(y)\sim e^{y/2}y^{-\hat{p}_\theta},\;\;\text{Re}(y)>0
\end{equation}

\begin{figure}[h!]
\centering
\includegraphics[width=0.5\textwidth]{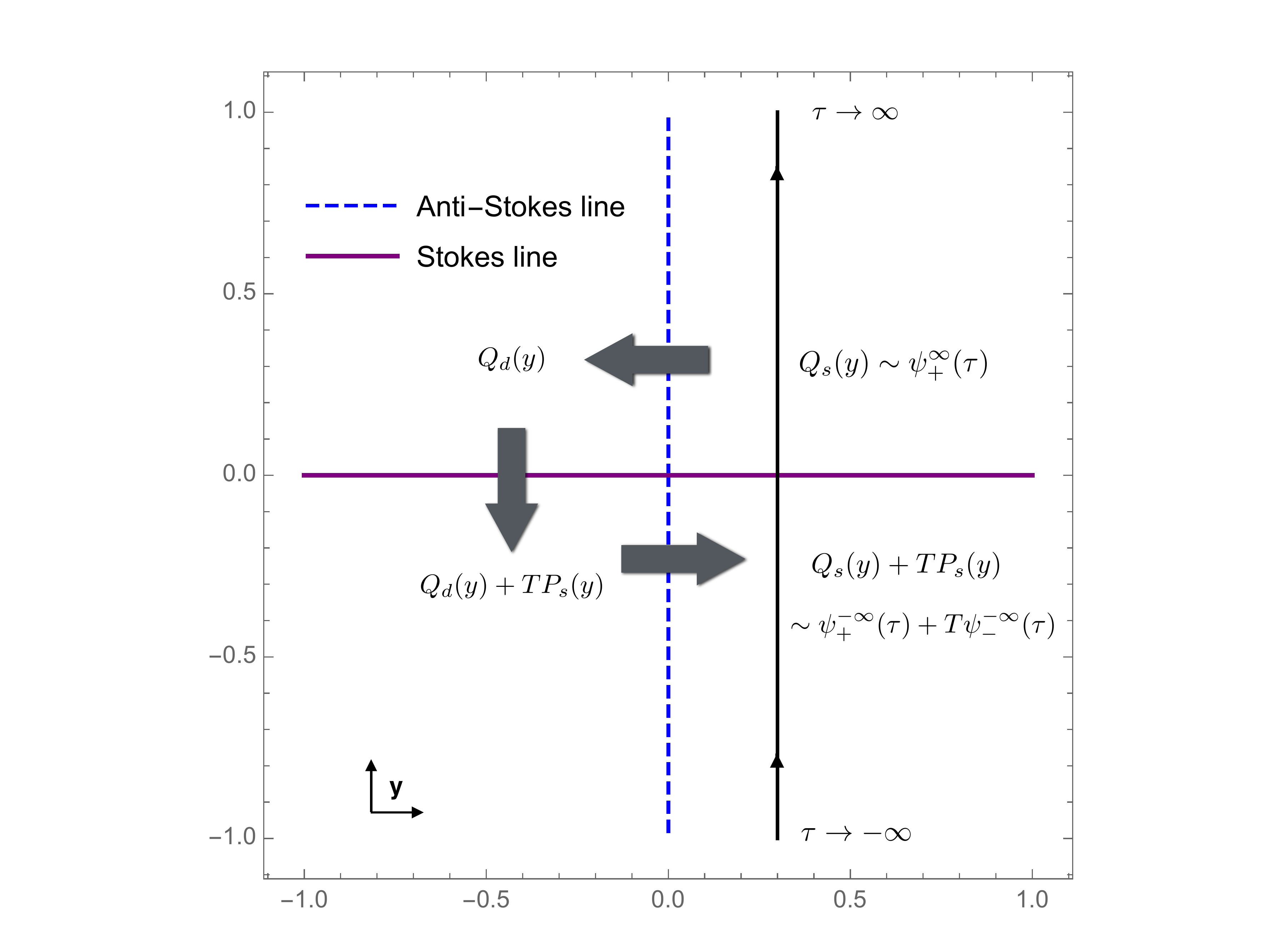}
\caption{\small{Stoke's phenomenon for the Whittaker functions}}\label{fig:Whittaker}
\end{figure}
The Stoke's phenomenon for the Whittaker function has been investigated in \cite{Heading1}. The corresponding Stokes or anti-Stokes curves are $\text{Arg}(y)=n\pi$ or $\left(n+1/2\right)\pi$ respectively (see figure \ref{fig:Whittaker}). The right-moving solution $\psi^{\infty}_+(\tau)\sim e^{i k\tau}, \tau\to \infty$ corresponds to $P(y)$ in the region $0<\text{Arg}(y)<\pi/2$, which is the sub-dominant mode. To obtain the reflected wave $\psi^{-\infty}_-(\tau)\sim Q(y)$, analytically continue $P(y)$ anti-clockwise all the way into $3\pi/2<\text{Arg(y)}$. Crossing the $\text{Arg}(y)=\pi/2$ anti-Stokes curve makes $P(y)$ the dominant mode; crossing the Stokes curve $\text{Arg}(y)=\pi$ generates the reflected mode proportional to the Stoke's constant, which has been worked out in \cite{Heading1}:
\begin{equation}
P(y)\to P(y)+T Q(y),\;\;T=\frac{2\pi i e^{2\pi i \hat{p}_\theta}}{\Gamma\left(\frac{1}{2}-	\frac{1}{2}\sqrt{1-4\epsilon_L}-\hat{p}_\theta\right)\Gamma\left(\frac{1}{2}+\frac{1}{2}\sqrt{1-4\epsilon_L}-\hat{p}_\theta\right)}
\end{equation}
crossing the $\text{Arg}(y)=3\pi/2$ anti-Stokes curve switches dominance between $P(y)$ and $Q(y)$. We therefore conclude that to achieve the reflection-less condition $T\sim R=0$, $\hat{p}_\theta$ should be tuned to hit one of the poles in the Gamma functions:\footnote{Surprisingly these solutions are very similar to the late time solutions found in \cite{Liam2016_2}.}
\begin{equation}\label{eq: Renyi-monodromy}
\hat{p}_\theta = \frac{1}{2}\pm \frac{1}{2}\sqrt{1-4\epsilon_L}+k,\;\;k=0,-1,-2...
\end{equation}
For the computation of Renyi entropy, $\epsilon_L=\frac{1}{4}\left(1-\frac{1}{n^2}\right)$, matching the correct $n\to 1$ behavior fixes $k=0$ and picks the minus sign in (\ref{eq: Renyi-monodromy}). We have thus obtained the WKB solution to the associated monodromy problem: 
\begin{equation}\label{eq:WKB}
p_\theta = \sqrt{\epsilon_H}\left(1-\frac{1}{n}\right)
\end{equation}  
\subsection{Entanglement spectrum}
The WKB solution (\ref{eq:WKB}) to the monodromy problem implies that the (vacuum subtracted) excited Renyi entropy for an arc of extension $\theta$ on a $2\pi$ circle takes the form: 
\begin{eqnarray}\label{eq: Renyi_excited}
&&\text{Tr}\rho_H^n(\theta) = \exp{\left(-\frac{cn}{6}\int p_\theta d\theta \right)}=\exp{\left[\frac{\pi c}{6\beta_H}(1-n)\theta\right]}\nonumber\\
&& S_n (\theta) = \frac{1}{1-n}\ln{\text{Tr}\rho_H^n(\theta)}=\frac{\pi c}{6\beta_H}\theta
\end{eqnarray}
where $\beta_H = \pi/\sqrt{\epsilon_H}$ is the effective temperature of the excited state $|H\rangle$. This is different from the high temperature (i.e. effectively on an infinite line) thermal result for $n>1$: 
\begin{eqnarray}\label{eq: Renyi_thermal}
\text{Tr}\rho_\beta^n(\theta) &=& \exp{\left[-\frac{cn}{12}\left(1-\frac{1}{n^2}\right)\ln{\left(\frac{\beta}{\pi}\sinh{\left(\frac{\pi \theta}{\beta}\right)}\right)}\right]}\nonumber\\
&\approx &\exp{\left[\frac{\pi c}{6\beta}\left(\frac{1}{n}-n\right)\theta\right]},\;\;\;\theta\gg \beta
\end{eqnarray} 

\begin{figure}[h!]
\centering
\includegraphics[width=0.55\textwidth]{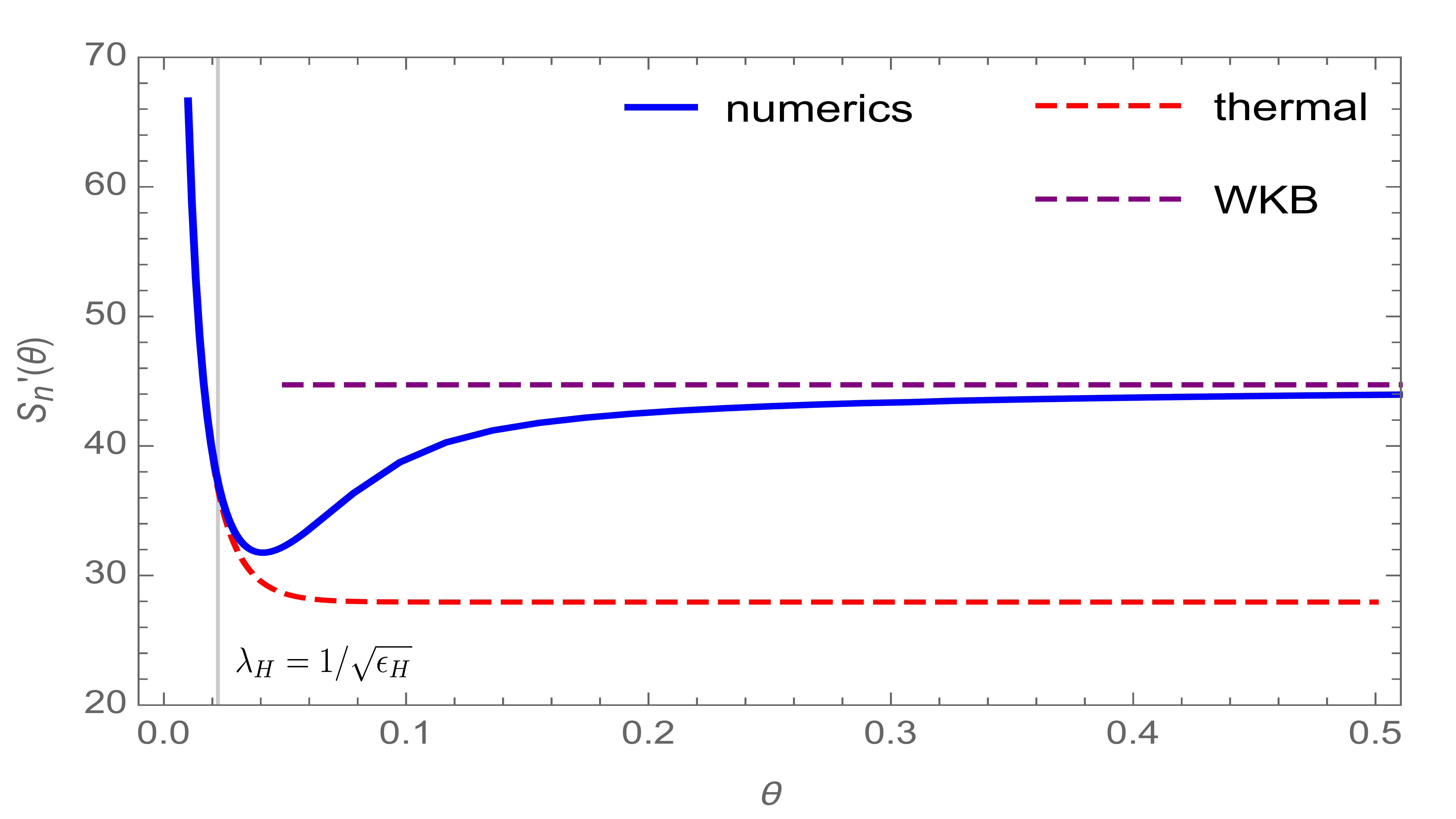}
\caption{\small{Comparison between the WKB prediction, high temperature thermal behavior and numerical result for $\lim_{c\to\infty}\frac{6}{c} S_n'(\theta)$, with $n=4$, and $\epsilon_H=2000$. The gray verticle line marks the thermal scale $\lambda_H=1/\sqrt{\epsilon_H}$.}}\label{fig:WKB_check}
\end{figure}

For $\theta\ll \beta_H$, the short distance expansion was found to be different between the excited state and thermal Renyi-entropies at sub-leading orders in the limit of $c\to \infty$. By numerically solving the relevant monodromy problem, we indeed obtain an interpolation between the short distance behavior for $\theta\ll \beta_H$ (which is close to the thermal result) and the WKB prediction for $\theta\gg \beta_H$ (figure \ref{fig:WKB_check}).

We can extract from (\ref{eq: Renyi_excited}) the spectrum of modular energies: 
\begin{eqnarray}\label{eq: entanglement_spectrum}
&&\text{Tr}\rho_H^n(\theta) =\sum_i d_i e^{-n \mathcal{E}_i}\approx \int d\mathcal{E} e^{S(\mathcal{E})} e^{-n\mathcal{E}}=\exp{\left[\frac{\pi c}{6\beta_H}(1-n)\theta\right]}
\end{eqnarray}
where $\lbrace\mathcal{E}_i\rbrace$ is the set of eigenvalues for the modular Hamiltonian $\hat{K}_H(\theta)\propto -\log{\rho_H(\theta)}$ and $\lbrace d_i \rbrace$ are the degeneracies. In the thermodynamic limit one can bin it into a continuous distribution $e^{S(\mathcal{E})}$. Our result is consistent with an entanglement spectrum that is strongly peaked at $\mathcal{E}^*\sim \frac{\pi c \theta}{6\beta_H}$ with density of states $\exp{\left(\mathcal{E}^*\right)}$. We conclude that to good accuracy in this thermodynamic limit the entanglement spectrum is flat. 

To extract the range of validity for the WKB result (\ref{eq:WKB}), we simply plug it back to (\ref{eq:Schrodinger}), and require that the resulting Schrodinger problem be of the over-scattering type, i.e. $V(\tau)\ll E$ for all $-\infty<\tau<\infty$. It is easy to derive from this:
\begin{eqnarray}
\frac{p_{\theta}+\sqrt{p_{\theta}^2+4\epsilon_L E}}{2\epsilon_L}\gg \frac{\sin{\left(\frac{\theta}{2}\right)}}{1-\cos{\left(\frac{\theta}{2}\right)}}
\end{eqnarray}
In the limit $\epsilon_H\gg 1$, this reduces to $\theta \gg 1/\sqrt{\epsilon_H}\sim\lambda_T$, where $\lambda_T$ is the thermal wavelength. It complements the regime $\theta \ll \lambda_T$ where the short distance expansion is valid. 

One possible caveat is that for the high energy micro-state Renyi entropies with $n>1$, the corresponding ``all-heavy" correlation functions are not necessarily dominated by the universal virasoro vacuum block we computed here.\footnote{We thank Alex Belin for pointing this out.} See for example \cite{Belin}, which corresponds to replacing both $\mathcal{O}_H$ and $\mathcal{O}_L$ by the heavy twist operator $\sigma_{n=3}$. In principle there could be a theory-specific critical $n^*$ above which the vacuum block approximation is no longer valid. We cannot rigorously rule out such possibilities. However, the flatness of the entanglement spectrum, the main feature of our result, is relevant for the high temperature regime and valid for $n$ not far from $n=1$. We expect both limits to be away from the possible low temperature (large $n$) instabilities that yield the non-universalities.
 
\section{Finite $c$ resolution}\label{sec: finite_c}
Finally we discuss the vacuum block for the HL correlator at finite $c$. We have seen that re-summing the $\epsilon_L$ corrections to all orders regularizes the accessory parameter $p(x)$ at the forbidden singularities, while giving rise to a pair of branch-cuts close to the thermal poles. Via the branch-cuts, infinitely many saddles that solve the monodromy problem with different winding numbers are stitched together, they form a Riemann surface $\mathcal{M}_p$. Of course at finite $c$, the block is analytic away from the OPE singularities, and the branch-cuts should eventually disappear after summing back all finite $c$ corrections. 

There are in general two ingredients for the resolution of branch-cuts. At the local level, the finite $c$ corrections smoothen-out the discontinuities across the branch cuts; at the global level, the finite $c$ corrections single out a particular way that $\mathcal{M}_p$ pinch off, and become disconnected. We will look at both aspects in this section. In particular, we use the Zamolodchikov recursion relation to perform a high-order $q$ expansion, which re-sums all finite $c$ corrections. Using the numerical results, we first investigate the local aspects of the resolution. After that, we study the global aspect of the resolution. We will see interesting roles played by the non-perturbative effects in $c$, manifested by Stoke's phenomena. Numerical work of similar nature was done in \cite{Hongbin}, which also explored the late time behavior and found interesting power-law tails universal in all blocks. 

\subsection{Local resolution}
We begin with the local resolution. To get some intuition, we ask the reverse question: how would branch-cuts in the accessory parameter $p(x)\sim\frac{ \partial_x\mathcal{V}(x,c)}{\mathcal{V}(x,c)}$ emerge as the $c\to \infty$ limit of the analytic vacuum block $\mathcal{V}(x,c)$.  The most natural possibility is that at finite $c$, the accessory parameter $p(x)$ possesses a series of poles that become more and more densely packed as we increase $c$. In the limit of $c\to\infty$, they condense and form a branch-cut. For analytic $\mathcal{V}(x,c)$, the only possible poles for the accessory parameter $p(x)$ are those with integer residues, they correspond to zeros of the block $\mathcal{V}(x,c)$. 

Following this, we propose that at finite $c$, there are series of zeros $\lbrace x_i\rbrace$ for the vacuum block $\mathcal{V}(x_i, c)=0$. Furthermore, these zeros become increasingly dense as we increase $c$, and in the limit $c\to\infty$ coalesce into lines that match the branch-cuts in the semi-classical picture. 

We confirm such a picture numerically. Effective computations for generic finite $c$ blocks have been limited apart from the Zamolodchikov's recursion relations \cite{Zamolodchikov1984, Zamolodchikov1987}, which is briefly summarized in Appendix \ref{app: Zamo}. We adopt a brute-force approach by solving Zamolodchikov's recursion relation to high enough order in $q$-expansion, whose coefficients contain all finite $c$ corrections. The goal is to make the domain of convergence large enough to reveal the ``forbidden branch-cut" singularities. A formal solution to the recursion relation was worked out in \cite{Perlmutter2015}. Here we simply use the recursion relation and write a mathematica code to generate the coefficient list to a few hundred orders.\footnote{During the work, a code with very similar approach has been developed and published in \cite{Hongbin}. }
 
\subsubsection{Numerical results}
We compute the vacuum block by generating the q-series coefficients for $c=1000,\epsilon_H=36,\epsilon_L=5*10^{-2}$ to 800 order. To investigate the local resolution we start by focusing on the neighborhood of the first forbidden singularity $x_0=1-e^{-\frac{2\pi}{\sqrt{4\epsilon_H-1}}}\approx 0.41$.

\begin{figure}[h!]
\centering
\includegraphics[width=0.6\textwidth]{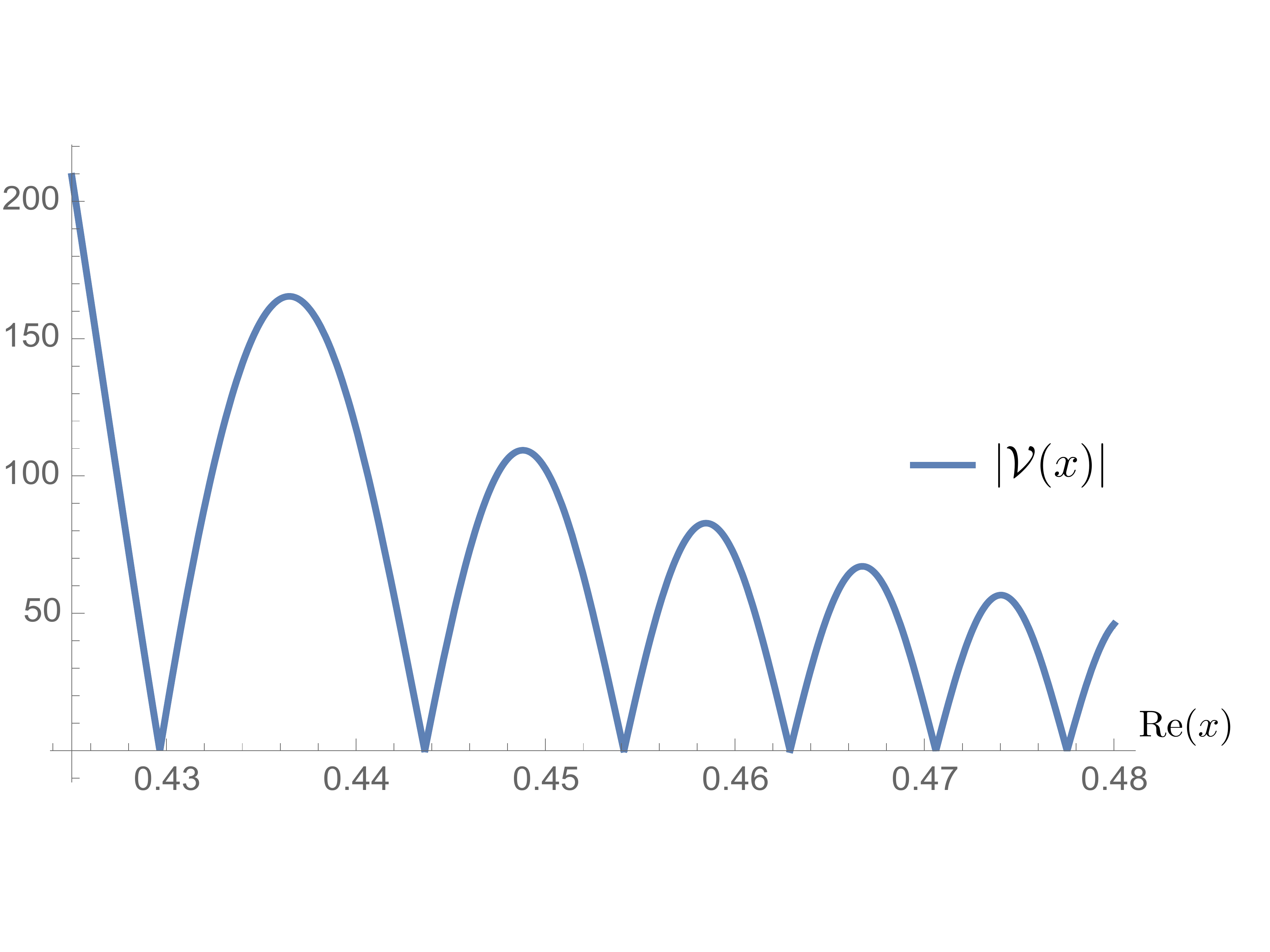}
\caption{\small{modulus of the finite $c$ vacuum block $|\mathcal{V}(x,c)|$ along a path on the complex $x$ plane that connects the zeros}}\label{fig:zeros_of_block}
\end{figure}

We found that in agreement with the prediction, the vacuum block at finite $c$ has a series of zeros along a path that coincides with the semi-classical branch-cuts. In figure \ref{fig:zeros_of_block} we plot the modulus of the block $|\mathcal{V}(x,c)|$ along such a path.  

\begin{figure}[h!]
\centering
\includegraphics[width=0.85\textwidth]{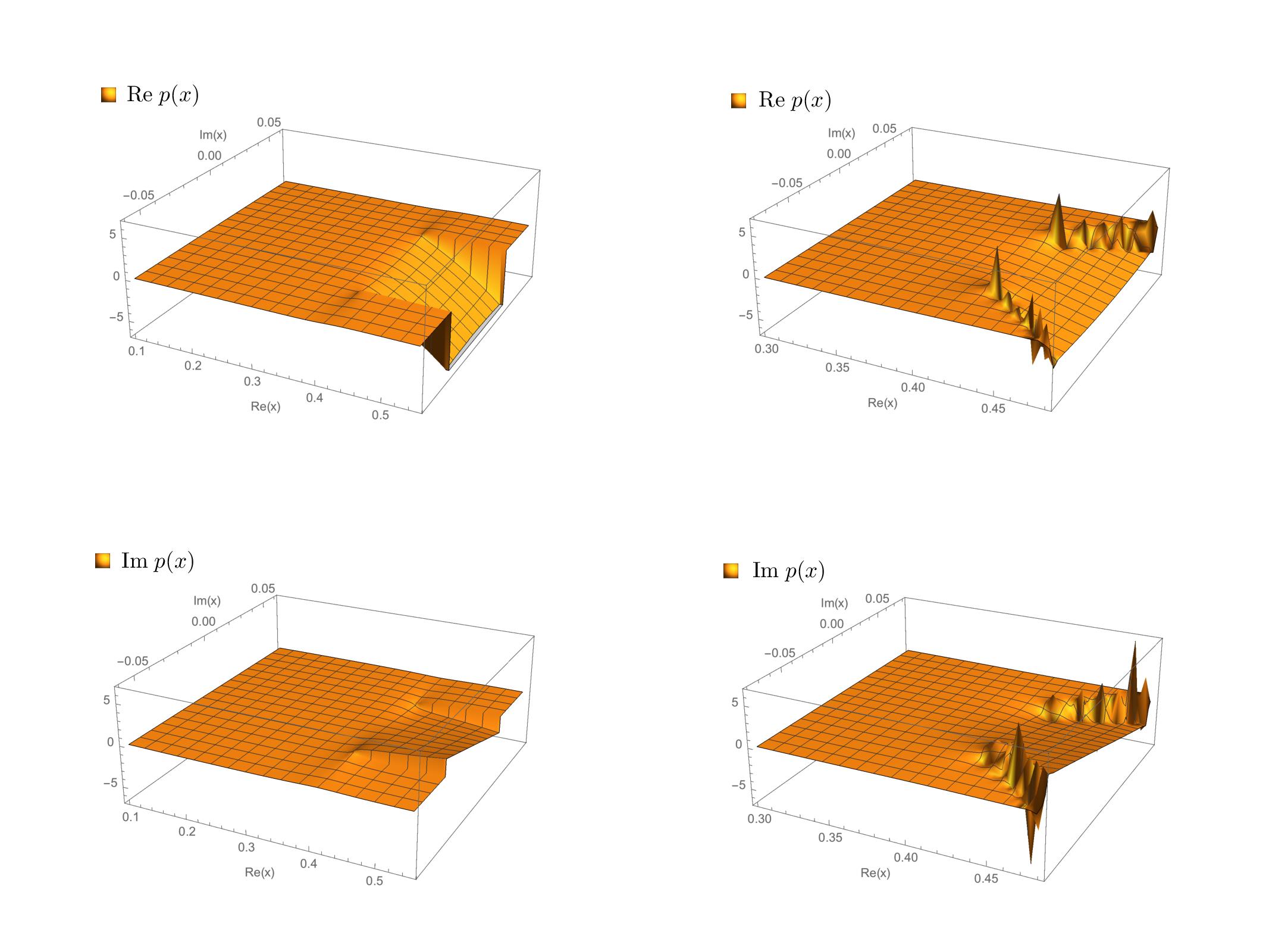}
\caption{\small{real (upper left) and imaginary (lower left) part of $p(x)$ computed semi-classically using the monodromy method; real (upper right) and imaginary (lower right) part of $p(x)$ computed using the recursive series expansion}}\label{fig: resolving_c}
\end{figure}

To visualize the resolution taking place, we also plot the finite $c$ accessory parameter $p(x)\sim \frac{\partial_x \mathcal{V}(x,c)}{\mathcal{V}(x,c)}$, and compare it with semi-classical result obtained via the monodromy method. To see the analytic structure, we plot both the real and imaginary parts of $p(x)$ (figure \ref{fig: resolving_c}) on a region close to the first forbidden singularity in the complex $x$-plane. 

We see that the finite $c$ corrections fix a particular direction for the branch-cuts, along which they are resolved into series of poles. One can also numerically integrate around the poles to compute the residues, which are found to be all unity: $\frac{1}{2\pi i}\oint dx\; p(x)=1$, implying that the zeros in the vacuum block are of order one.  

\subsection{Global resolution}
Next we discuss what happens globally to the infinite-sheeted Riemann surface $\mathcal{M}_p$. This manifold arose from solving the monodromy problem, so let us take a step back and re-consider the monodromy equation. In fact we would like to draw an analogy between the monodromy equation and the WKB solution to a linear differential equation. This is explicitly true for blocks involving degenerate operators, so we begin with them. These blocks satisfy linear differential equations whose order $\ell$ depend on the indices of the degenerate operators:
\beq
\sum^\ell_{k=0} h_k(c,x)\partial^k_x\mathcal{V}(x)=0
\eeq
At each order $k$, the coefficient scales as $h_k(c,x)\sim \left(\frac{c}{6}\right)^{1-k}g_k(x),\;g_k(x)\sim \mathcal{O}(1)$. Substitute the ansatz $\mathcal{V}(x)=\exp{\left[-\frac{c}{6}f(x)\right]}$, at leading order in large $c$ the equation becomes algebraic in terms of the ``accessory parameter" $p(x)\approx \partial_x f(x)$: 
\beq
\sum^n_{k=0} g_k(x) p(x)^k =0 
\eeq
This is a polynomial equation with $\ell$ branches. They correspond to the (finitely many) different fusion channels of the degenerate operators. The WKB solution to the differential equation is then given in terms of the accessory parameter:  
\beq
\mathcal{V}(x)\approx \exp{\left[-\frac{c}{6}\int^x\; dx'\; p(x') \right]}
\eeq
One can imagine that for a physical block, instead of a finite order differential equation, it satisfies an infinite order differential equation whose exact nature is unknown to us at the moment. The polynomial equation for the accessory parameter $c_x$ is then replaced by the monodromy equation, in the form of a ``transcendental" equation with infinitely many branches. The branches correspond to the additional ``saddles" we have seen. 

At this point, it is very tempting to associate the branch-points we have identified with the turning points of the WKB solutions, just as in section~\ref{sec: Renyi}. From them one can locate the Stokes and anti-Stokes curves. In terms of the accessory parameters, they correspond to trajectories of $x$ such that: 
\beqn
\text{Im} \int^x p_n(x') dx' = \text{Im} \int^x p_m(x') dx',\;\;  \text{Re} \int^x p_n(x') dx' = \text{Re} \int^x p_m(x') dx'
\eeqn 
for distinct branches $n\neq m$.

The prediction for the global resolution of $\mathcal{M}_p$, based on the analogy proposed between the monodromy problem and the WKB solution, is as follows. The way in which adjacent sheets of $\mathcal{M}_p$ pinch off near a forbidden branch cut is determined by working out the Stokes phenomena for the participating  WKB solutions near the corresponding turning point. In particular, this implies a concrete prediction for the locations of the resolved branch-cuts, or the poles/zeros in the accessory parameters/blocks: the anti-Stokes curves. The semi-classical discontinuities of $p_n(x)$ are due to exchange of the dominant WKB saddles for $\mathcal{V}_n(x)$ across the anti-Stokes curves. Along the curves, participating WKB saddles become oscillatory and produce the series of zeros we observe. 

\subsubsection{Stoke's phenomena}\label{subsec: local_stokes}
By studying blocks involving degenerate operators, the authors in \cite{Liam2016} discussed a form of ``universal resolution" near the forbidden singularity $x_n$:  
\beq
\mathcal{V}(x)\propto \int^\infty_0 dp\; p^{2c\epsilon_L-1}e^{-p(x-x_n)-\frac{\sigma^2}{4 c\alpha_H}p^2} 
\eeq
which solves the second order differential equation: 
\beq\label{eq: universal_res}
\frac{\sigma^2}{2c}\mathcal{V}''(x)+\alpha_H (x-x_n)\mathcal{V}'(x)-2\alpha_H c\epsilon_L\mathcal{V}(x)=0
\eeq
Based on the analogy proposed, this effectively performs the quadratic resolution near $x_n$ for the accessory parameter. As discussed in section \ref{sec: semi-resol}, there are two branch-points $ x^{\pm}_n=x_n\pm 2i\sigma\sqrt{\frac{\epsilon_L}{\alpha_H}}$, we could view them as turning points for the Stoke's phenomena.\footnote{The ``forbidden" branch-cuts discussed here are somewhat obscure in \cite{Liam2016}, as they remained in the probe limit: $h_L\sim \mathcal{O}(1)$ while organizing finite $c$ corrections.} The quadratic resolution allows us to identify the Stokes and anti-Stokes curves near each point. Label a ray from $x^{\pm}_n$ by $\theta^\pm$: $x^{\pm}(r)=x^\pm_n + r e^{i\theta^\pm}$, the Stokes and anti-Stokes curves emerge from $x^\pm_n$ as rays with: 
\beqn\label{eq:anti-stokes}
\theta^+_{anti-stokes} &=& \left\lbrace \frac{\pi}{6},\frac{5\pi}{6},\frac{3\pi}{2}\right\rbrace,\;\;\theta^+_{stokes}=\left\lbrace-\frac{\pi}{6},\frac{\pi}{2},\frac{7\pi}{6}\right\rbrace\nonumber\\
\theta^-_{anti-stokes} &=& \left\lbrace-\frac{\pi}{6},\frac{7\pi}{6},\frac{\pi}{2}\right\rbrace,\;\;\theta^-_{stokes}=\left\lbrace \frac{\pi}{6},\frac{3\pi}{2},\frac{5\pi}{6}\right\rbrace\nonumber
\eeqn

\begin{figure}[h!]
\centering
\includegraphics[width=0.45\textwidth]{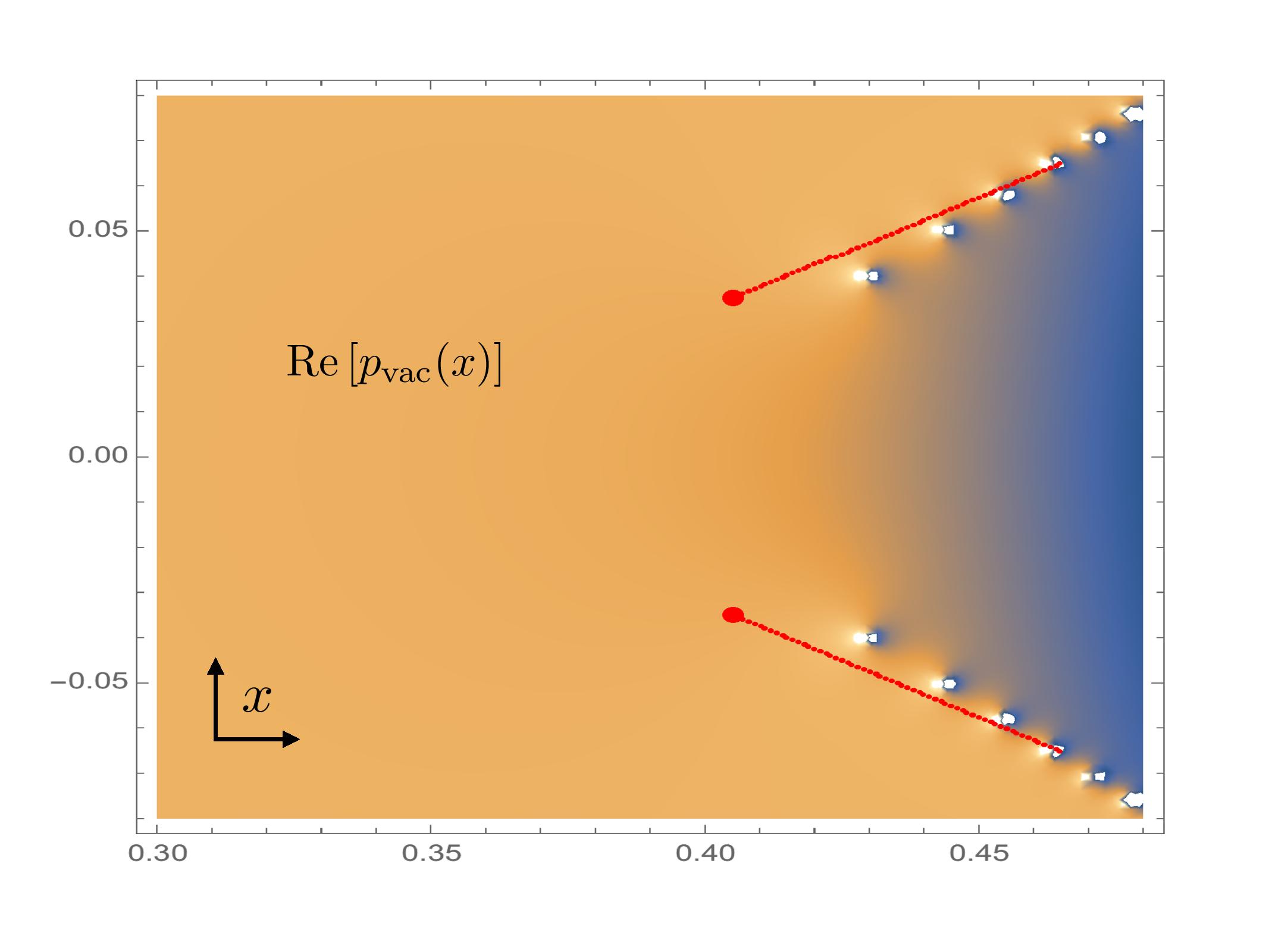}
\includegraphics[width=0.45\textwidth]{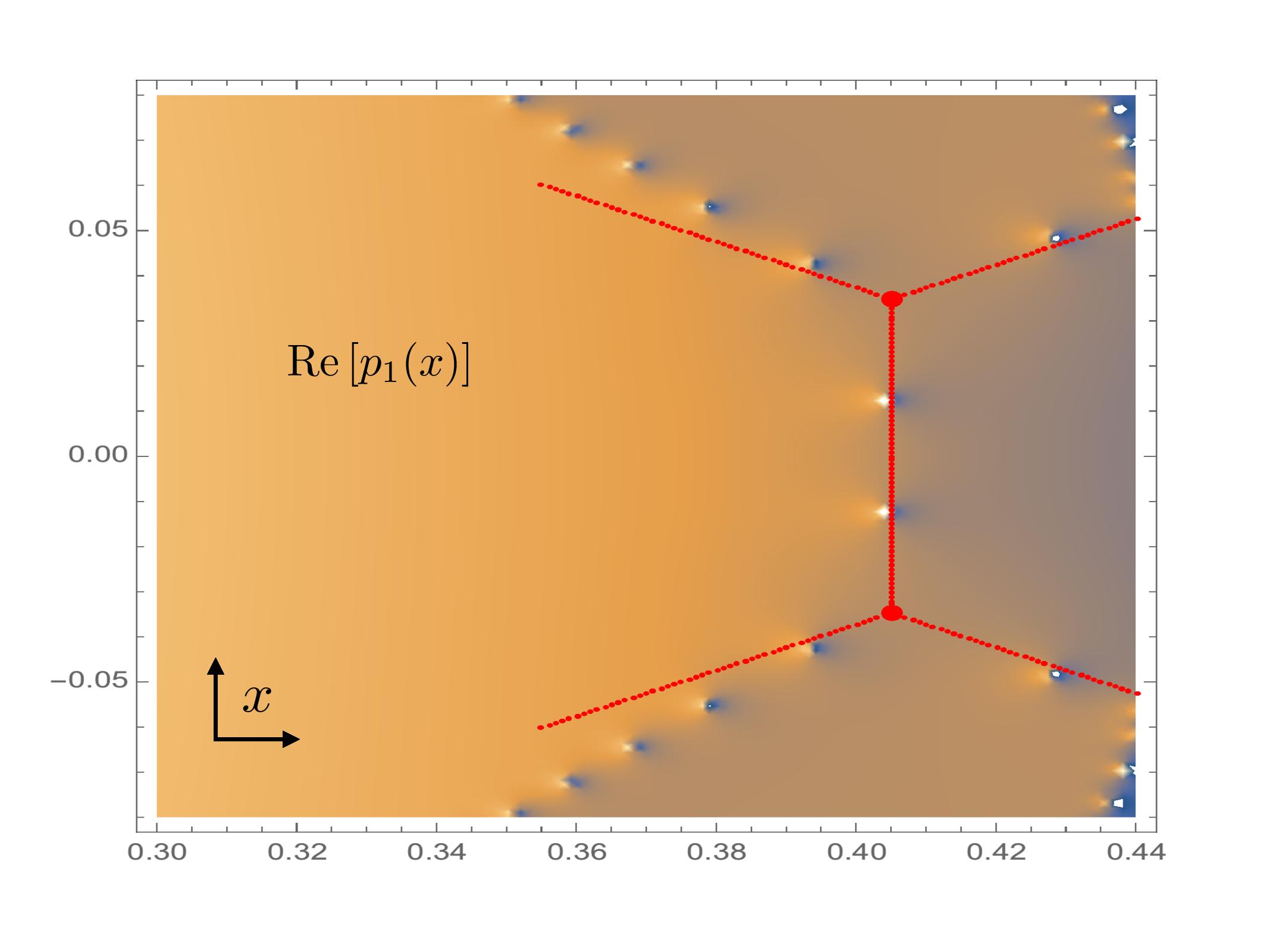}
\caption{\small{Accessory parameters associated with the exact blocks at finite $c$. Left: real part of $p_{\text{vac}}(x)$; right: real part of $p_1(x)$. Red curves are the anti-Stokes rays predicted in (\ref{eq:anti-stokes}). Parameters used: $c=1000, \epsilon_L = 1/200, \epsilon_H=36$, branch points at $x^\pm_0\approx 0.405\pm 0.035 i$}}\label{fig: stokes_curves_accessory}
\end{figure}

To check this prediction, in figure \ref{fig: stokes_curves_accessory} we plot the finite $c$ accessory parameters $p_{\text{vac}}(x)$ and $p
_1(x)$ associated with the vacuum block and the first ``unphysical" block near the first forbidden singularity $x_0$. Semi-classically they are connected by a pair of branch cuts close to $x_0$. At finite $c$, we see that the location of the poles align approximately with the predicted anti-Stokes curves in (\ref{eq:anti-stokes}). Furthermore, the pattern for which anti-Stoke's curve show up as a series of poles is also consistent with the underlying Stoke's phenomena between the two WKB solutions. 

\begin{figure}[h!]
\centering
\includegraphics[width=0.45\textwidth]{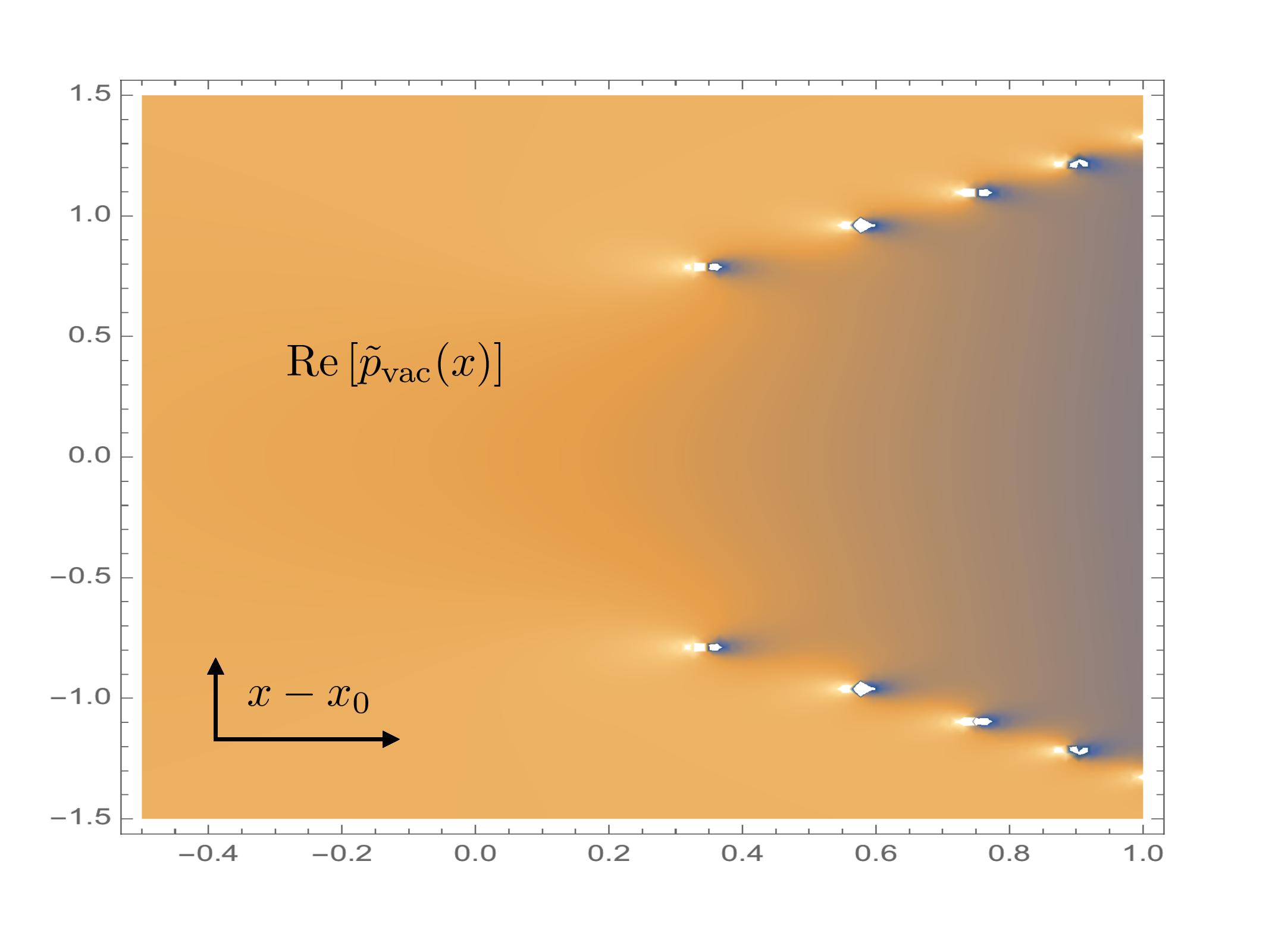}
\includegraphics[width=0.45\textwidth]{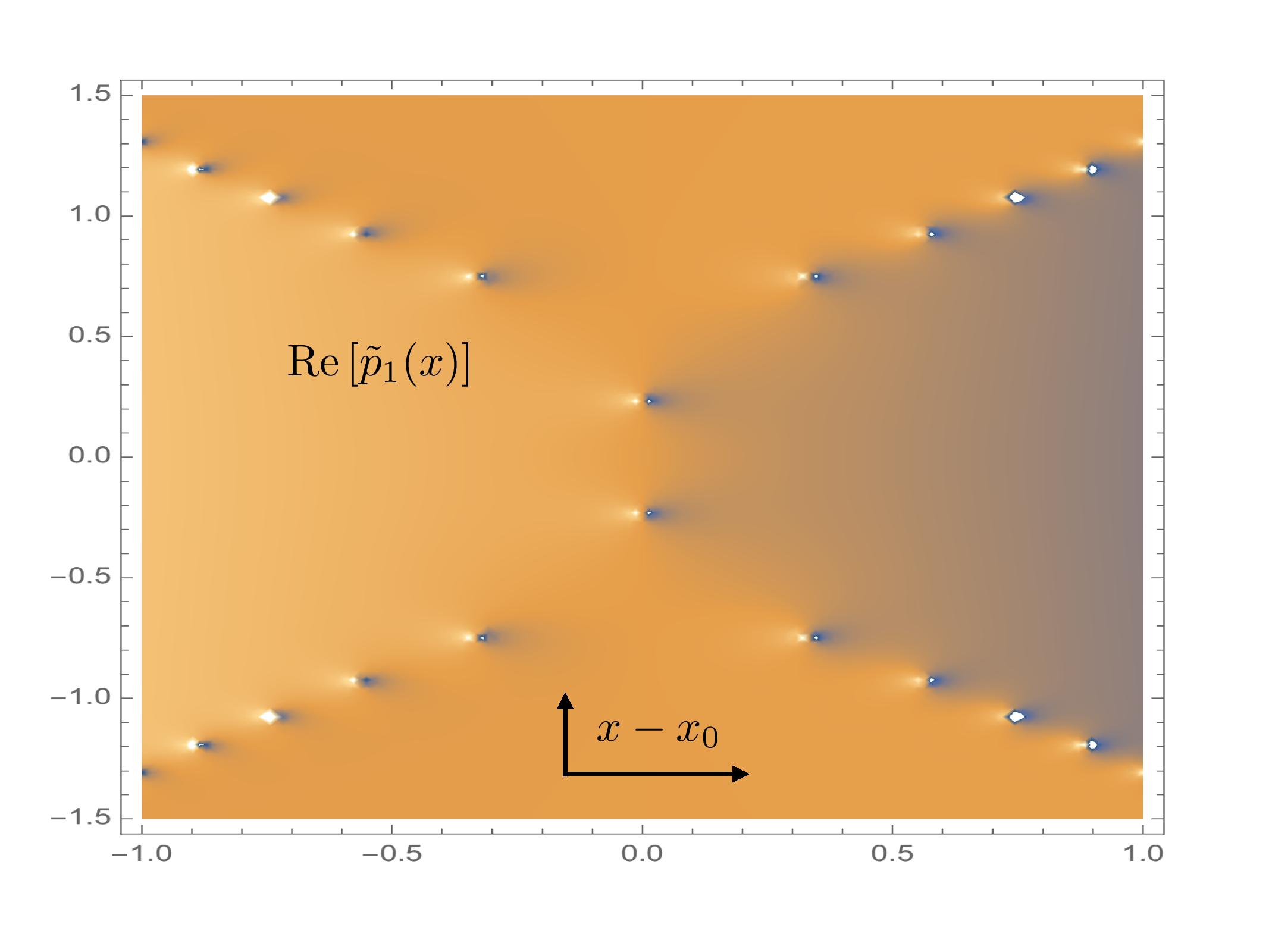}
\caption{\small{Left: real part of $\tilde{p}_{\text{vac}(x)}$; right: real part of $\tilde{p}_1(x)$. Parameters used: $c=10, \epsilon_L = 1/10, \alpha_H=1, \sigma=1$}}\label{fig: ptildes}
\end{figure}

Using (\ref{eq: universal_res}) one can analytically capture the essence of the Stoke's phenomena near $x_0$. The participating blocks are $\mathcal{V}_{\text{vac}}$ and $\mathcal{V}_1$.  There is no forbidden singularity to the left of $x_0$, where we can extract the asymptotic behaviors from the semi-classical results: 
\beq\label{eq: asymptotic}
\mathcal{V}_{\text{vac}}(x)\propto \exp{\left[-\frac{c}{6}\int^x p_-(x')dx'\right]},\;\;\mathcal{V}_1(x)\propto \exp{\left[-\frac{c}{6}\int^x p_+(x')dx'\right]}
\eeq  
Solutions to (\ref{eq: universal_res}) are confluent hypergeometric functions of the first kind. We fix the linear combinations by matching (\ref{eq: asymptotic}). In figure \ref{fig: ptildes} we plot the real parts of the corresponding accessory parameters $\tilde{p}_{\text{vac}}(x)$ and $\tilde{p}_1(x)$, they capture the essential features of the exact results in figure \ref{fig: stokes_curves_accessory}. 

\subsubsection{Numerical results}
The Stoke's phenomena involving the remaining ``unphysical" blocks $\mathcal{V}_n(x), n>1$ are more complicated. Naively one needs to take into account the interplay between more than one clusters of turning points for each pair of ``adjacent" blocks.  The ``universal" local approximation proposed in the last section \ref{subsec: local_stokes} may not be adequate for capturing all the Stoke's phenomena. Therefore we resort again to numerical works for revealing what happens there.

For this purpose we need to significantly improve the range of convergence for the numerical series expansion. It was observed in \cite{Hongbin} that optimal convergence happens near the ``boundary" of the HL kinematic limit: for moderately heavy state $\epsilon_H\sim \mathcal{O}(1)\geq 1/4$ as well as moderately large $c$. For such choices the forbidden singularities are too densely packed near $x=1$. To dilute them we plot the results in the $q$-plane, the natural variable in the Zamolodchikov's recursion relation.  

\begin{figure}[h!]
\centering
\includegraphics[width=0.32\textwidth]{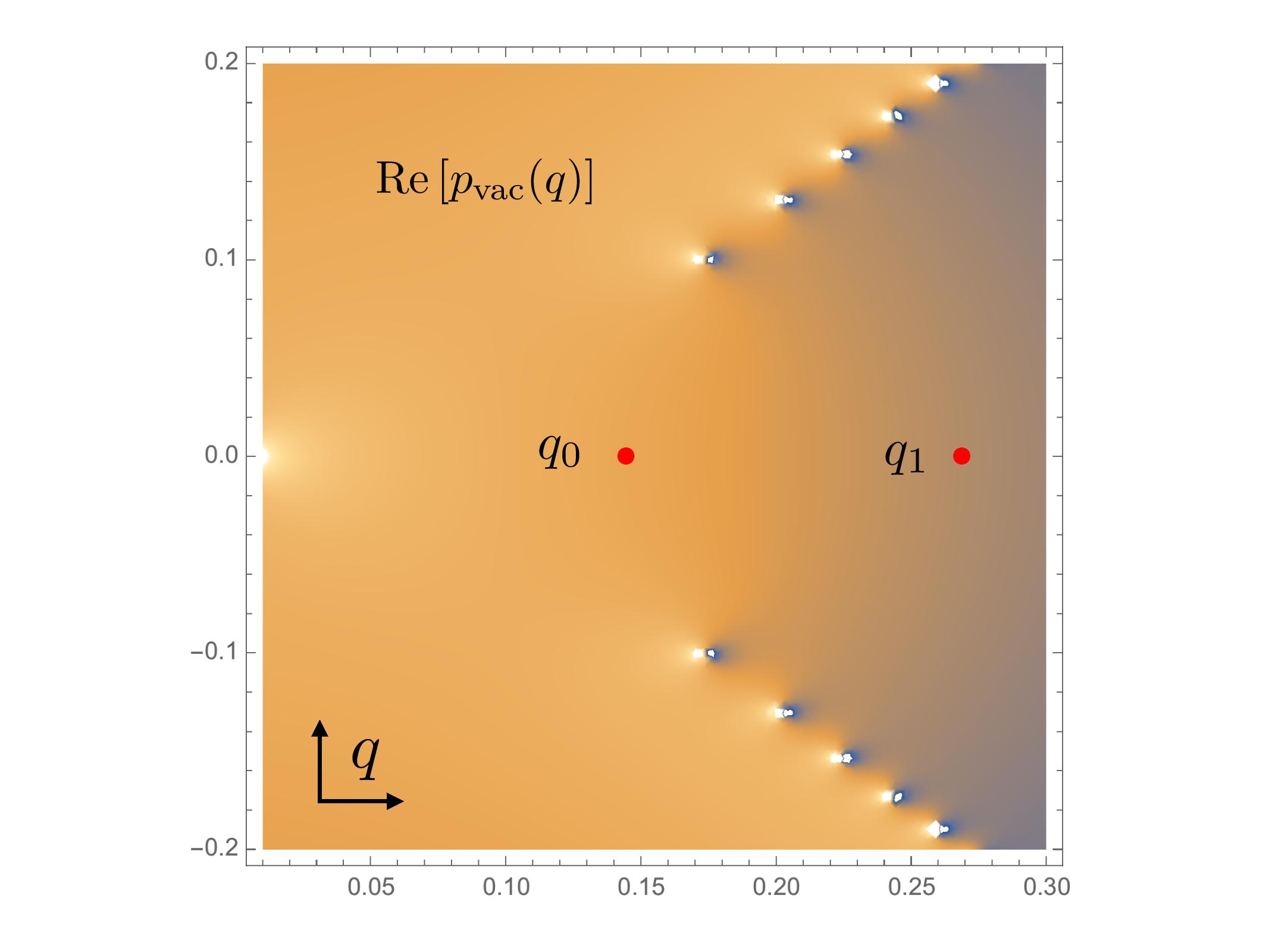}
\includegraphics[width=0.32\textwidth]{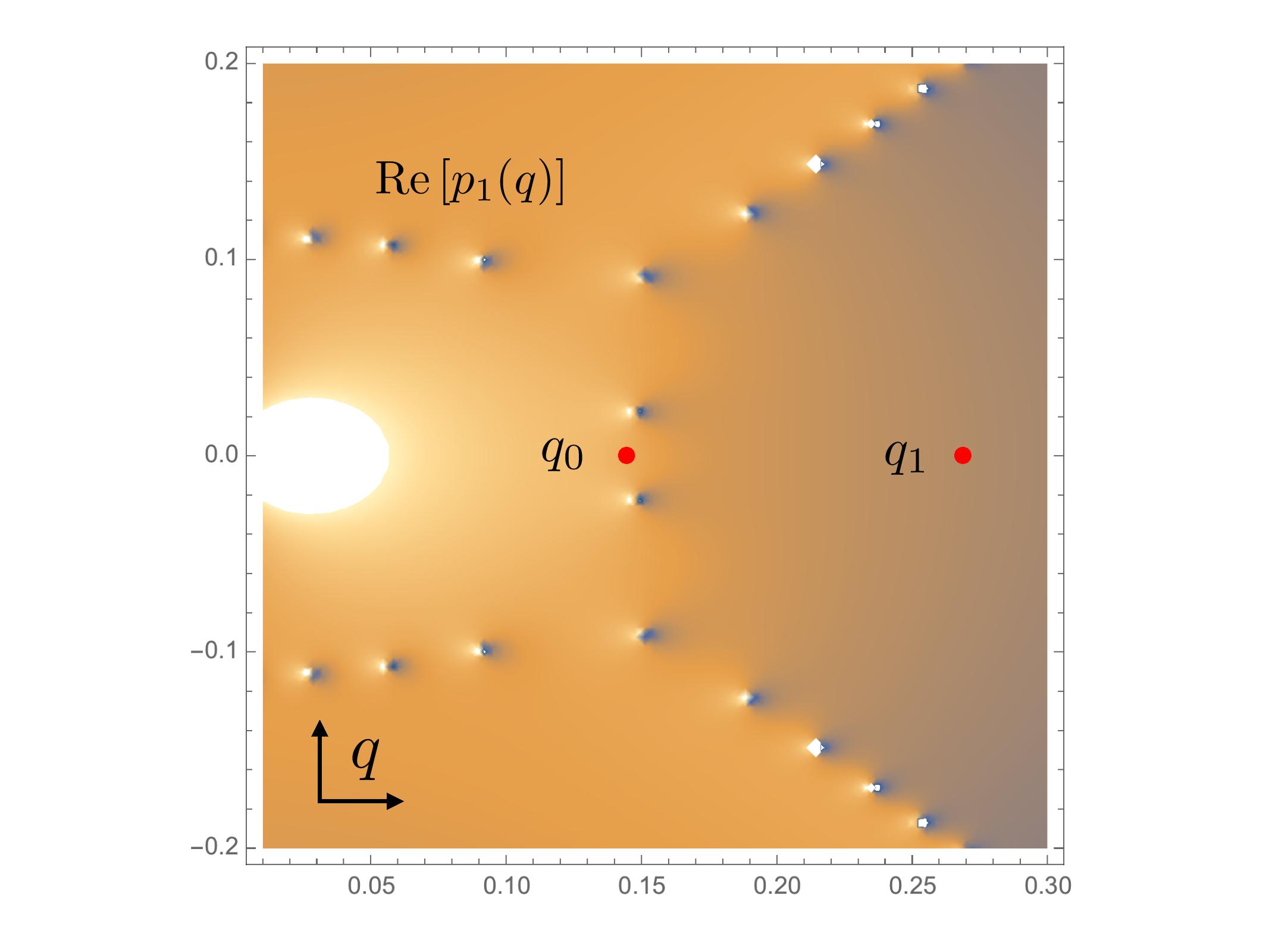}
\includegraphics[width=0.32\textwidth]{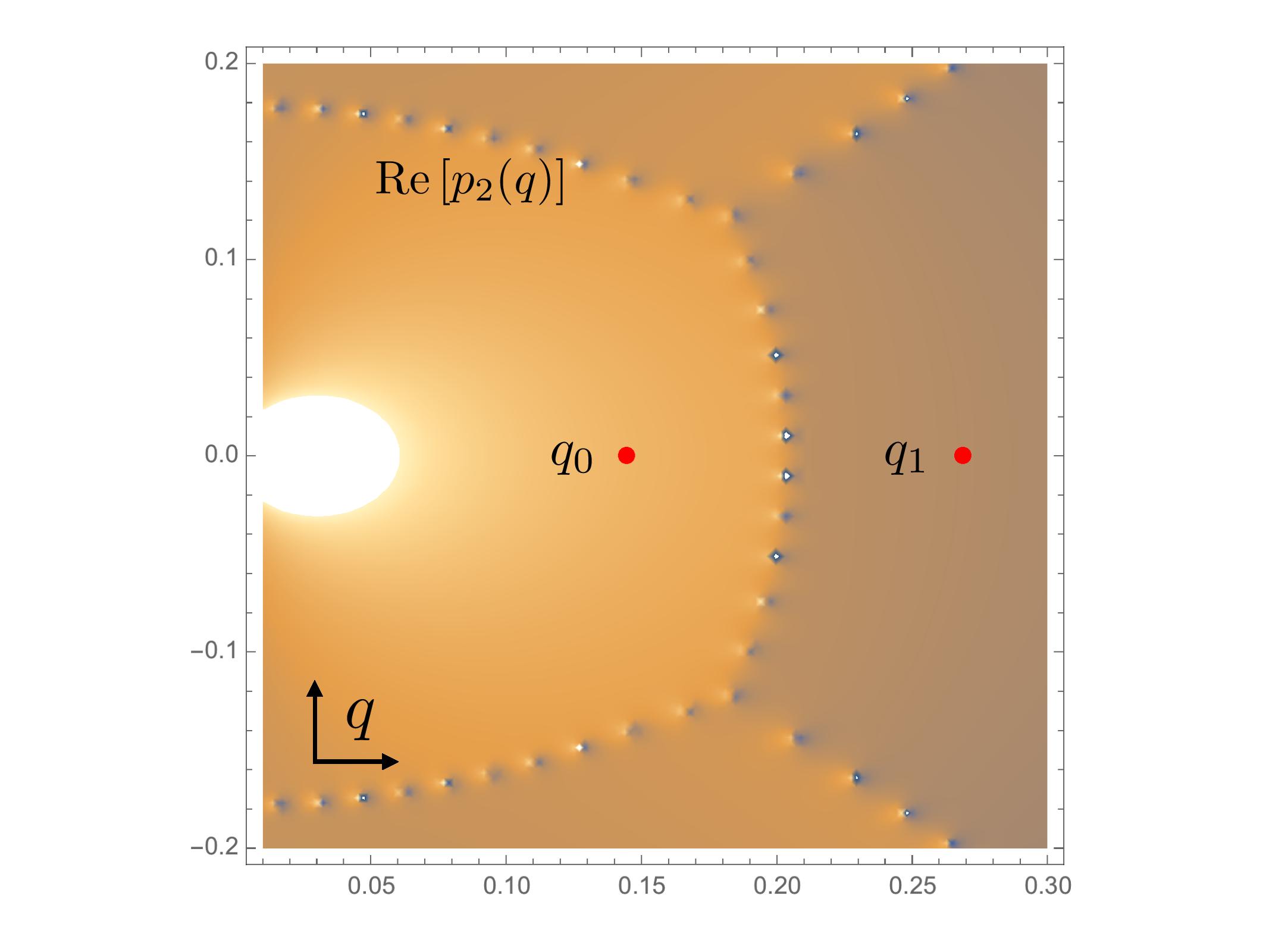}
\includegraphics[width=0.32\textwidth]{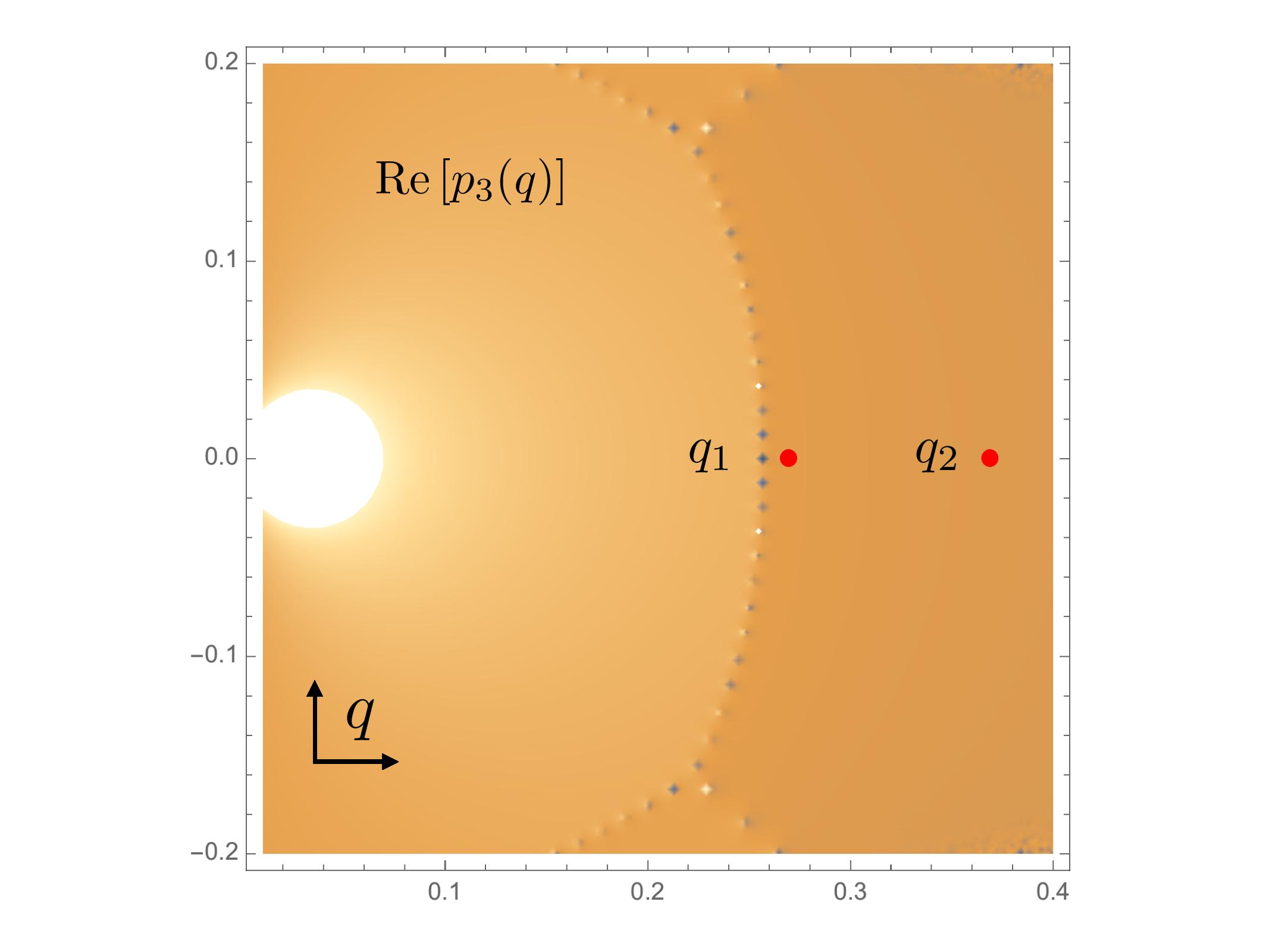}
\includegraphics[width=0.32\textwidth]{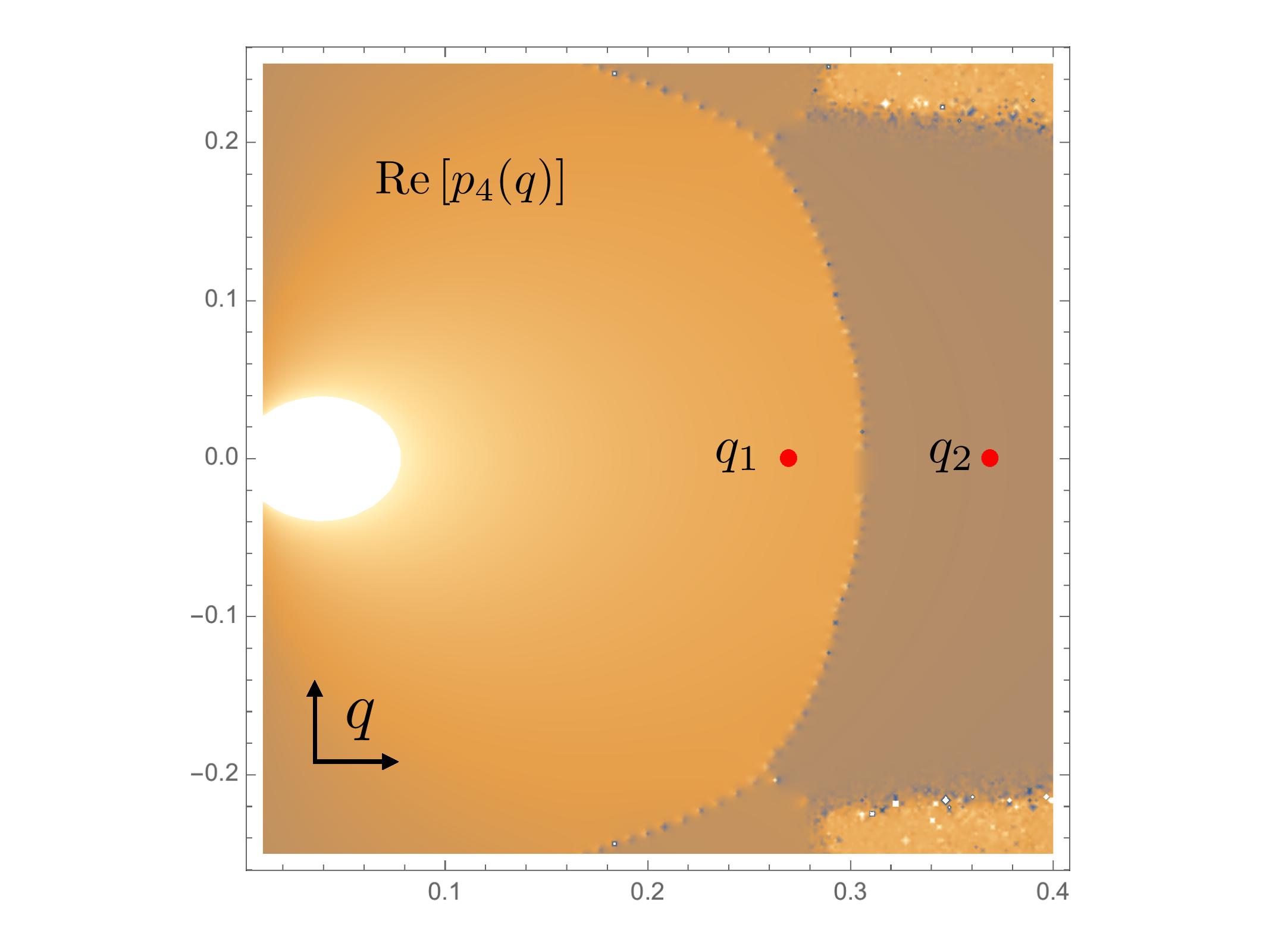}
\includegraphics[width=0.32\textwidth]{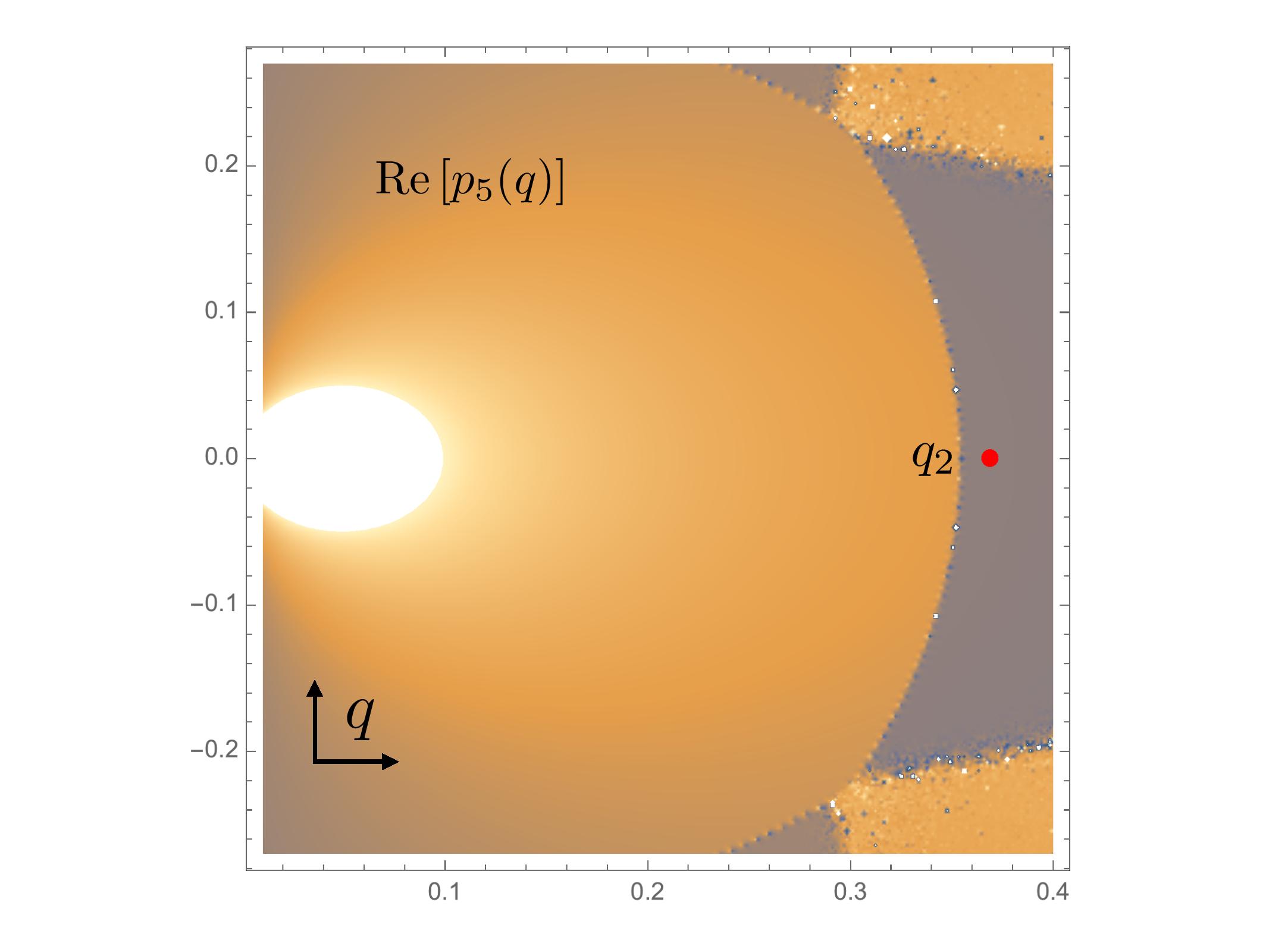}
\caption{\small{Real parts of the finite $c$ accessory parameters $p_{\text{vac}}(q)$ and $p_n(q)$ for $n=1,...,5$, on the complex $q$-plane. Parameters used: $c=60, \epsilon_L = 1/10, \epsilon_H=2$. Red dots $q_i$ are locations for the forbidden singularities in the $q$-plane.}}\label{fig: complete_stokes}
\end{figure}

In figure \ref{fig: complete_stokes} we plot on the complex $q$-plane the real parts of the finite $c$ accessory parameters $p_{\text{vac}}(q)$ and $p_n(q)$ for $n=1,2,3,4,5$. A few observations can be made regarding the Stoke's phenomena. Firstly, for each block, only one cluster of anti-Stokes curves are visible, which seems to differ from the semi-classical picture for $\mathcal{M}_p$ that two pairs of forbidden branch-cuts are present on each sheet $n\geq 1$. Secondly, there are cases (e.g. $n=2,4$) where the cluster of anti-Stokes curves are not close to any forbidden singularity $q_i$. Both observations suggest that the complete Stoke's phenomena associated with the vacuum block and its associated ``unphysical" blocks are not simply described by a chain of local ``universal" Stoke's phenomena proposed in section \ref{subsec: local_stokes}. The full Stoke's geometry could be much more complicated. In fact, it was pointed in \cite{Aoki1,Aoki2,Berk,Daalhuis} that for higher order ($>2$) differential equations, there are new complications. For example, not all Stokes curves emanate from the turning points (which could be related to our second observation), and that Stokes curves can be partially inactive (which could be related to our first observation). Clearly much more work is required to understand these patterns, we leave them for future investigations.

\begin{figure}[h!]
\centering
\includegraphics[width=0.43\textwidth]{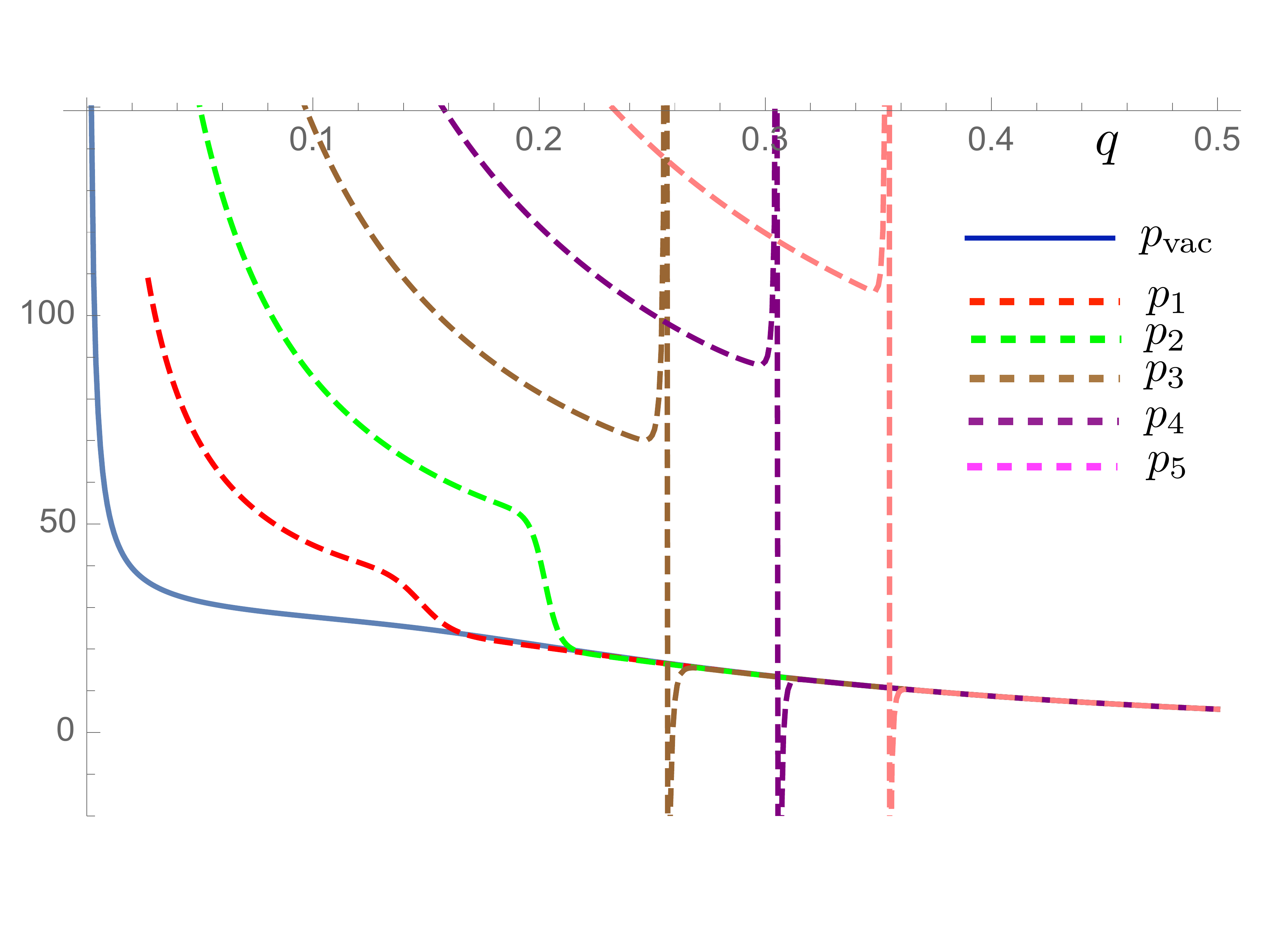}
\includegraphics[width=0.43\textwidth]{accessory_real}
\caption{\small{Left: plots of $p_{\text{vac}}(q)$ and $p_n(q)$ for $n=1,...,5$ on the real $q$-axis. Right: semi-classical results in section \ref{sec: semi_global} (different parameters picked) for qualitative comparison.}}\label{fig: blocks_collapse}
\end{figure}

A more interesting observation is the following. For all the ``unphysical" blocks we have examined, $p_n(q)$ collapses onto $p_{\text{vac}}(q)$ after going through the Stoke's phenomena (see figure \ref{fig: blocks_collapse}). This is again in conflict with the semi-classical picture that $p_n(q)$ should be connected with $p_{n\pm 1}(q)$ across via the forbidden branch-cuts on $\mathcal{M}_p$. In addition, the monodromy problems for the physical non-vacuum blocks $\mathcal{V}_h(q)$ of positive internal dimensions $h\propto c \to \infty$ (but below the BTZ black hole threshold $c\leq c/24$) should have similar features as the vacuum block: ``forbidden" singularities resolved into ``forbidden" branch-cuts connecting additional ``saddles". There should be Stoke's phenomena for them as well. We have checked a few such physical non-vaccum blocks, interestingly the collapse onto the same $p_{\text{vac}}(x)$ also happens for them. It implies that all blocks seem to have a universal outcome of the Stoke's phenomena in terms of the accessory parameters. In other words they are all dominated by a universal WKB saddle beyond the anti-Stokes curves. This observation, if true, would have very interesting and useful implications for many other computations. We will pursue these in future works. 

\section{Discussion}
In this paper, we studied probe corrections to ETH in 2D CFTs, focusing on observables in the form of non-local composite operators $\mathcal{O}_{obs}\sim \mathcal{O}_L(x)\mathcal{O}_L(0)$. In large central charge CFTs with a spectral gap, expectation values of these observables can be approximated by the Virasoro vacuum blocks. A sharp feature of ETH is the emergence of ``forbidden singularities" along the imaginary time direction of $x$. They arise already at the level of Virasoro vacuum blocks. By considering probe corrections in the monodromy problem, which computes the block in the $c\to \infty$ limit, we identified a mechanism to regulate the divergences at the forbidden singularities. The mechanism is non-perturbative in nature, and gives rise to ``forbidden branch-cuts" near the resolved singularities. We found that by crossing these branch-cuts,  the vacuum block is connected to other ``unphysical" blocks of negative conformal dimensions, which can be interpreted as additional saddles that solve the same monodromy equation but with different winding numbers. Though apparently drastic, the alterations in the analytic structure does not indicate a violent breakdown of ETH:
\beq\label{eq:dis1}
\langle \mathcal{O}_L(x)\mathcal{O}_L(0)\rangle_{\psi_H}\approx \langle \mathcal{O}_L(x)\mathcal{O}_L(0)\rangle_{\text{micro}} 
\eeq
Analogous alterations also arise on the RHS by considering probe corrections to the micro-canonical ensemble in the same limit. In the saddle point approximation for the inverse Laplace transform which relates the canonical and micro-canonical ensembles, the probe effects modify the saddles in a way that reproduces many features of the LHS of (\ref{eq:dis1}). With this said, the two sides do seem to exhibit quantitative deviations even for $\epsilon_L\ll 1$, especially for separations $x$ greater than the thermal scale. This discrepancy poses a puzzle that needs to be clarified in future work.
  
It turns out that the probe corrections are crucial for understanding what happens at finite $c$, especially for the resolution of forbidden singularities. Having obtained the partial resolution into the ``forbidden branch-cuts", one is very naturally led to the correct guess: a series of zeros for $\mathcal{V}(x)$ that become more densely packed as $c$ increases, whose condensation at $c\to \infty$ reproduce the forbidden branch-cuts in the accessory parameters. This was verified numerically.   

Such condensation of zeros also arise in other contexts. For example, phase transitions of Lee-Yang type \cite{Yang1952} are accompanied by condensations of zeros in the partition functions. A special case that might bare some connections to the present work is the modular invariant partition function of pure quantum gravity in $\text{AdS}_3$ \cite{Maloney2007}. In the future, it would be interesting to obtain a better understanding of the properties of zeros (density, distributions, etc) as well as their physical implications at the level of Virasoro blocks. 

An analogy between the monodromy problem and the WKB solutions to an infinite order differential equation was discussed. Led by this analogy, we found very strong evidence for the Stoke's phenomena taking place at the level of vacuum block and the associated ``unphysical" blocks. The locations for the series of zeros correspond to the anti-Stokes curves. It would be extremely interesting to make this analogy more concrete in the future. What is the object that plays the role of the infinite order differential equation? Speculatively, one way to achieve this is by mapping the Virasoro block calculation into a well-defined quantum mechanical problem with a path-integral representation. The object we seek could be the corresponding equation of motion (or the Schrodinger's equation), and many non-perturbative effects such as the Stoke's phenomena we have observed will then have clear interpretations. Some fruitful efforts along this direction has been initiated in \cite{Liam2017_3}. We leave these fascinating questions for future investigations. 

\begin{figure}[h!]
\centering
\includegraphics[width=0.47\textwidth]{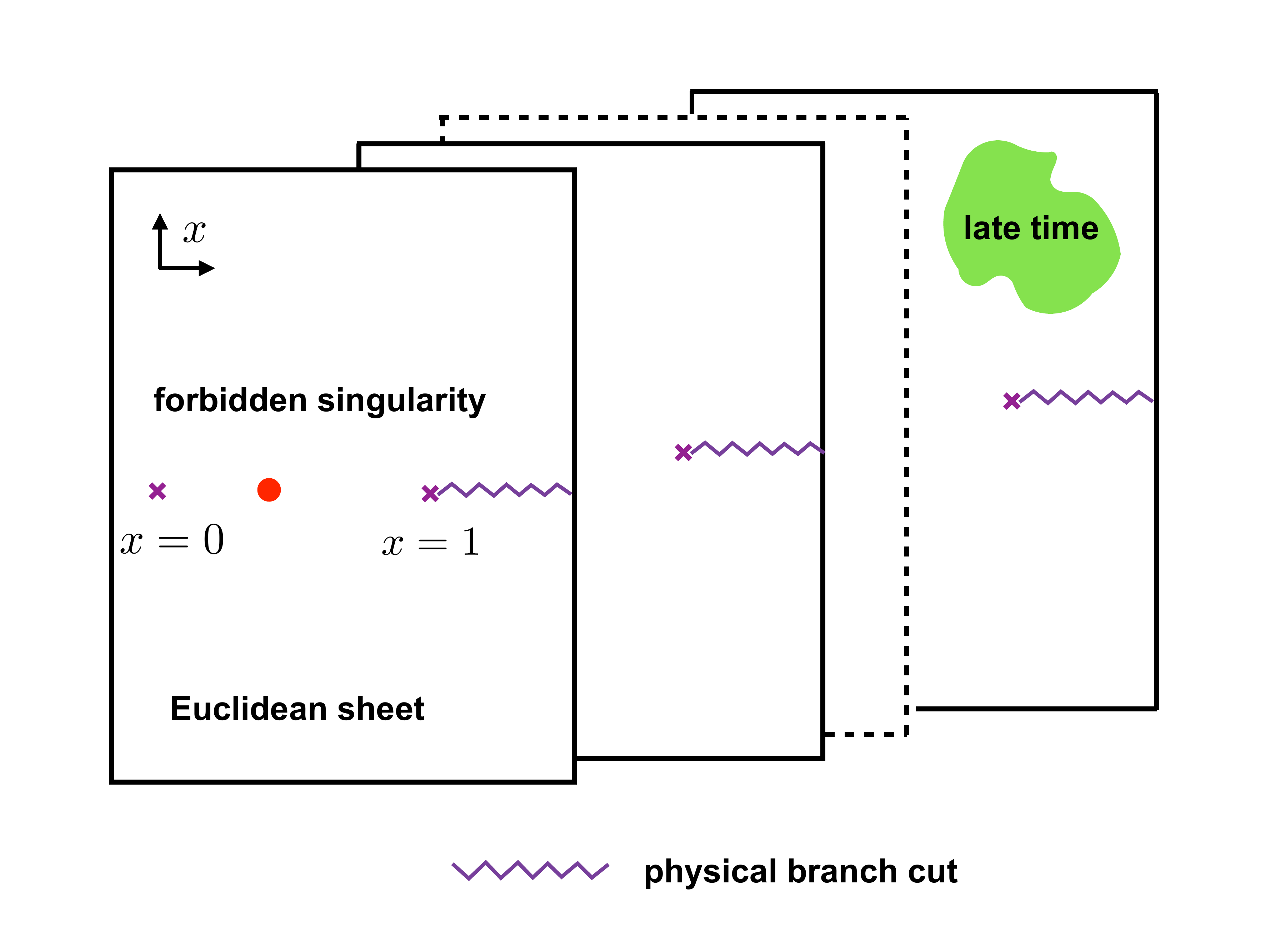}
\includegraphics[width=0.47\textwidth]{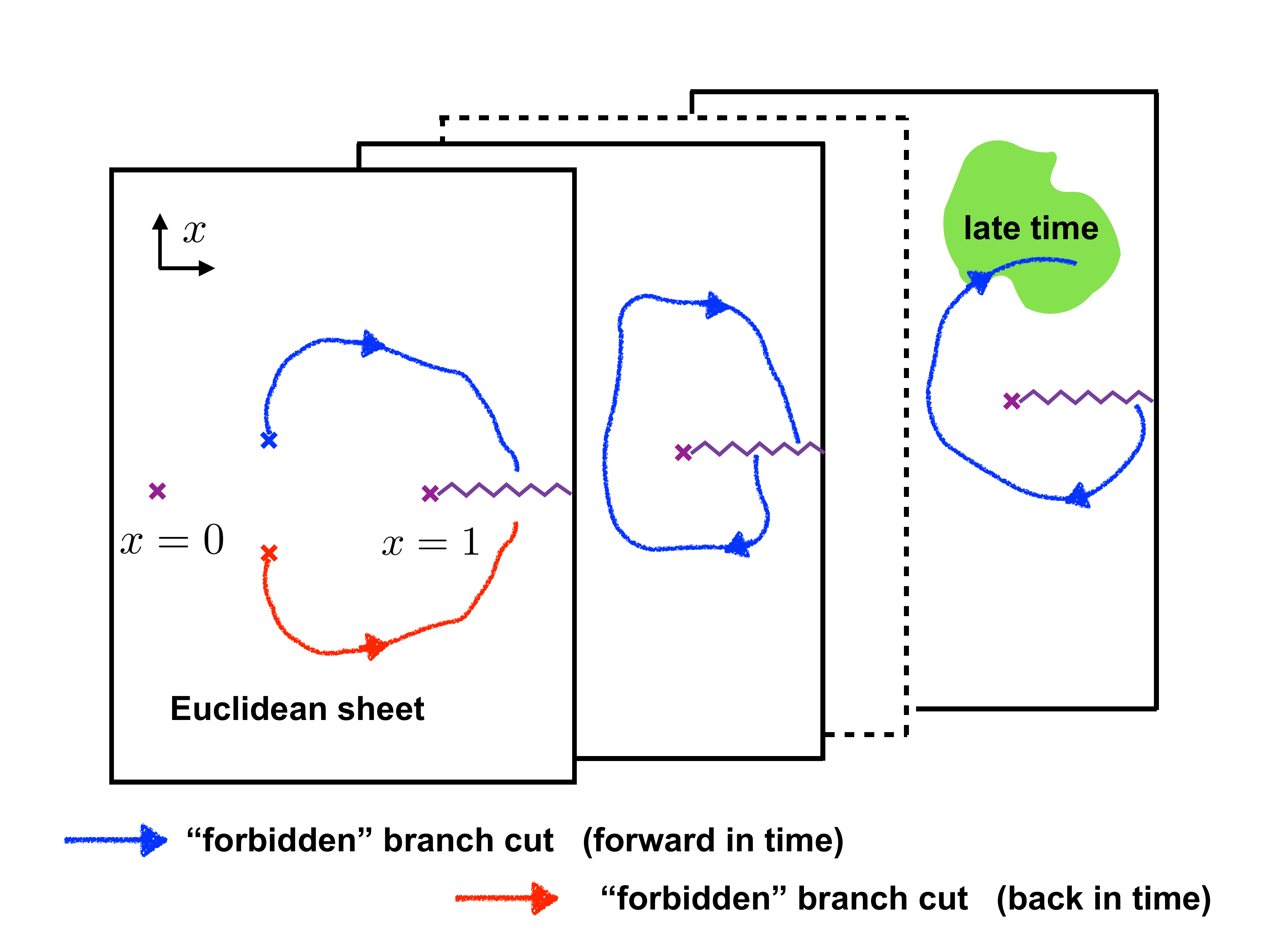}
\caption{\small{Left: forbidden singularity separated from the late time. Right: ``forbidden" branch-cut can potentially reach the late time.}}\label{fig: late_time_1}
\end{figure}

The numerical studies in \cite{Hongbin} identified universal late time behaviors $\mathcal{V}_h(t)\propto t^{-3/2} $ for all Virasoro blocks. In terms of the accessory parameters, it implies that $p_h(t)$ become all identical beyond some onset time. This is analogous to what we are finding for arbitrary blocks: they all collapse onto the same $p_{\text{vac}}(x)$ beyond the anti-Stokes curves. In this aspect, it seems to suggest some connection between the two phenomena. Recall that via radial quantization, the real time trajectories are given by $x(t)\propto 1-r e^{it}$. Technically one can smear the operator $\mathcal{O}_L(x)$ to make $r<1$, so as to regulate the periodic OPE singularity. Naively, one might argue that the late time behaviors of the blocks are obtained by going through the physical branch cut starting from $x=1$ many times, reaching out to some distant Lorentzian sheet of $x$; while the forbidden singularities take place only on the first/Euclidean sheet. It is not clear how the two can be related. However, once we partially resolve the thermal poles in $p(x)$ into extended ``forbidden branch-cuts", in principle they can cross the physical branch-cut and extend into the late time Lorentzian sheets (see figure \ref{fig: late_time_1}). 

\begin{figure}[h!]
\centering
\includegraphics[width=0.43\textwidth]{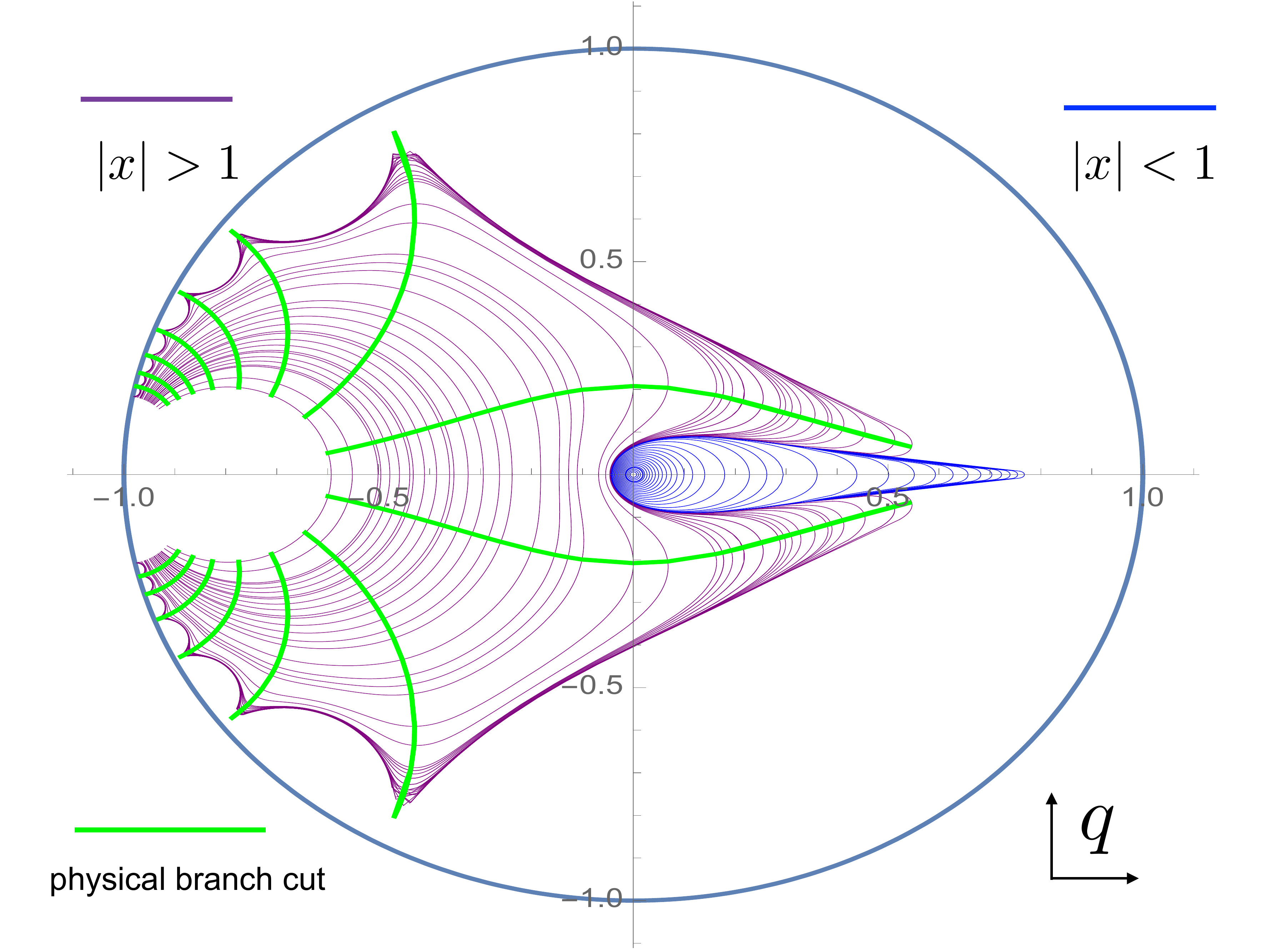}
\includegraphics[width=0.43\textwidth]{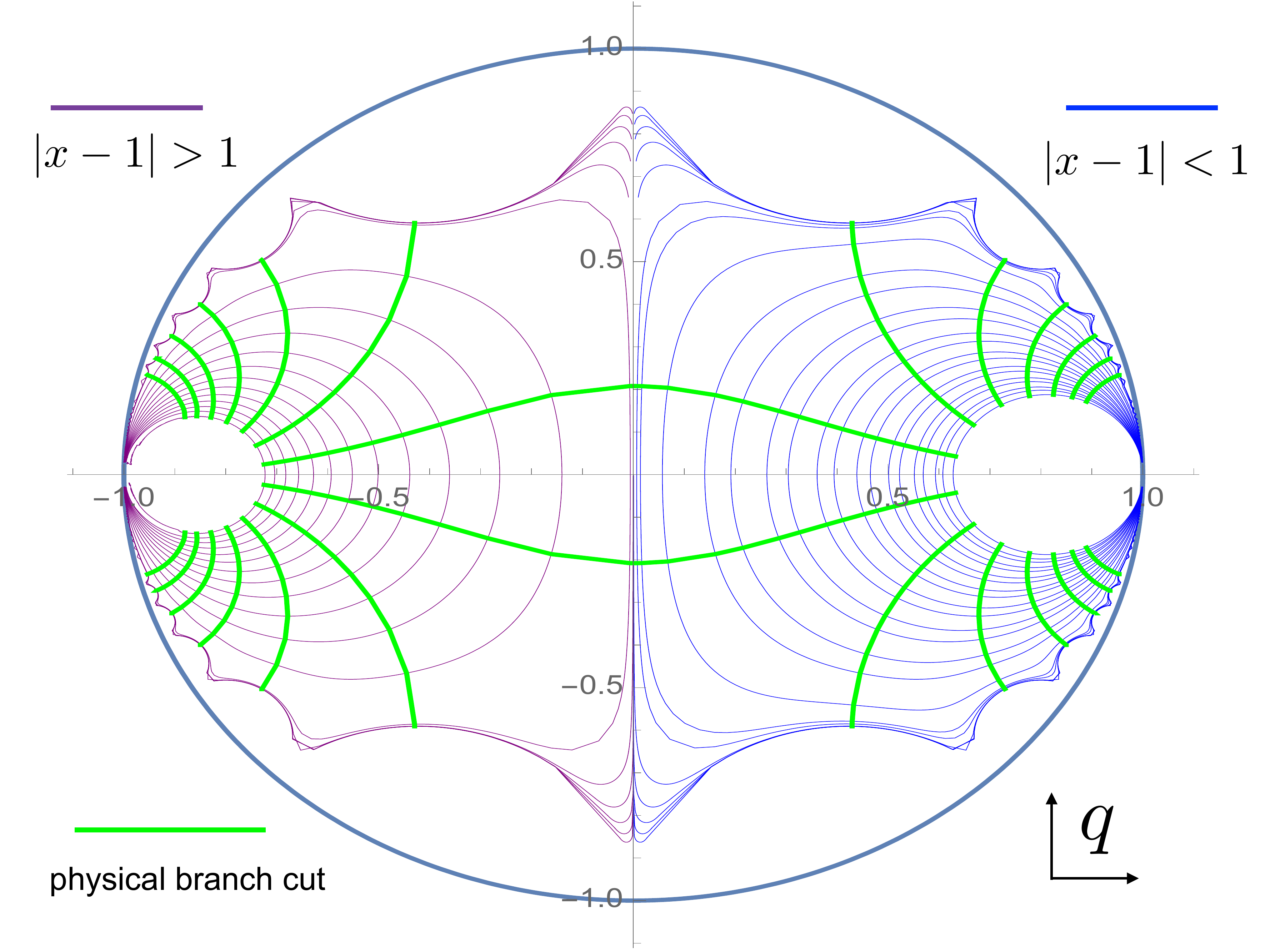}
\caption{\small{Mapping between the $q$-plane and the (Lorentzian sheets of) $x$-plane. Left: contours of constant $|x|$ circles on the $q$ plane; right: contours of constant $|x-1|$ circles on the $q$ plane. Green lines correspond to the physical branch-cuts from $x=1$ to $x=\infty$ in the original $x$-plane, distinct Lorentzian sheets in the $x$-plane are mapped to disjoint regions in the $q$-plane bounded by the green lines. }}\label{fig: qmap}
\end{figure}

As discussed before, the trajectory of the ``forbidden branch cut"s are fixed at finite $c$ along the anti-Stokes curves. Indeed we demonstrate explicitly that they do extend into the late time. To see this we present results on the $q$-plane, whose unit-disc contains all Lorentzian-sheets of the $x$-plane. For reference in figure \ref{fig: qmap} we provide a visual map between the $x$-sheets and the $q$-plane. 

\begin{figure}[h!]
\centering
\includegraphics[width=0.45\textwidth]{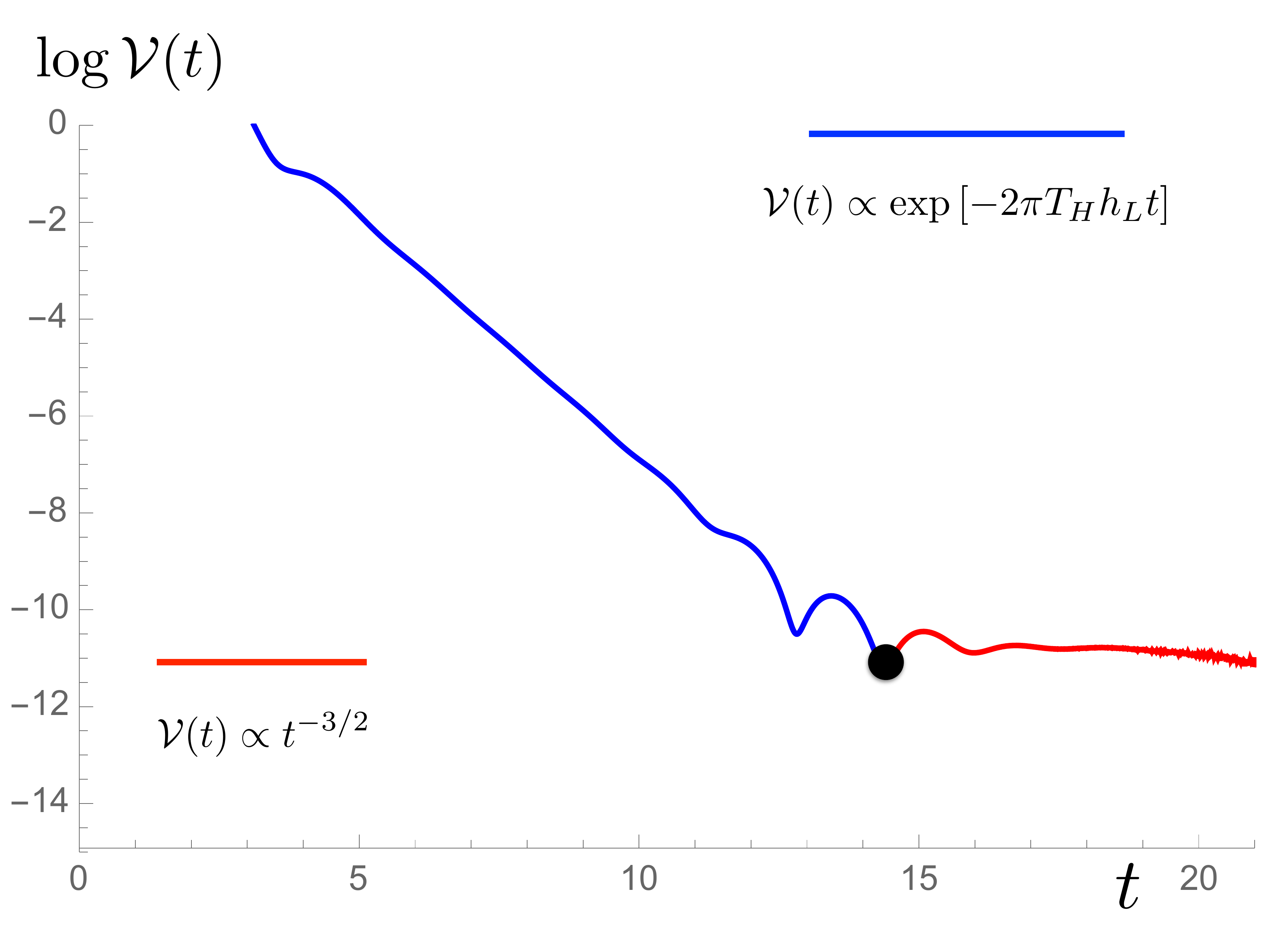}
\includegraphics[width=0.45\textwidth]{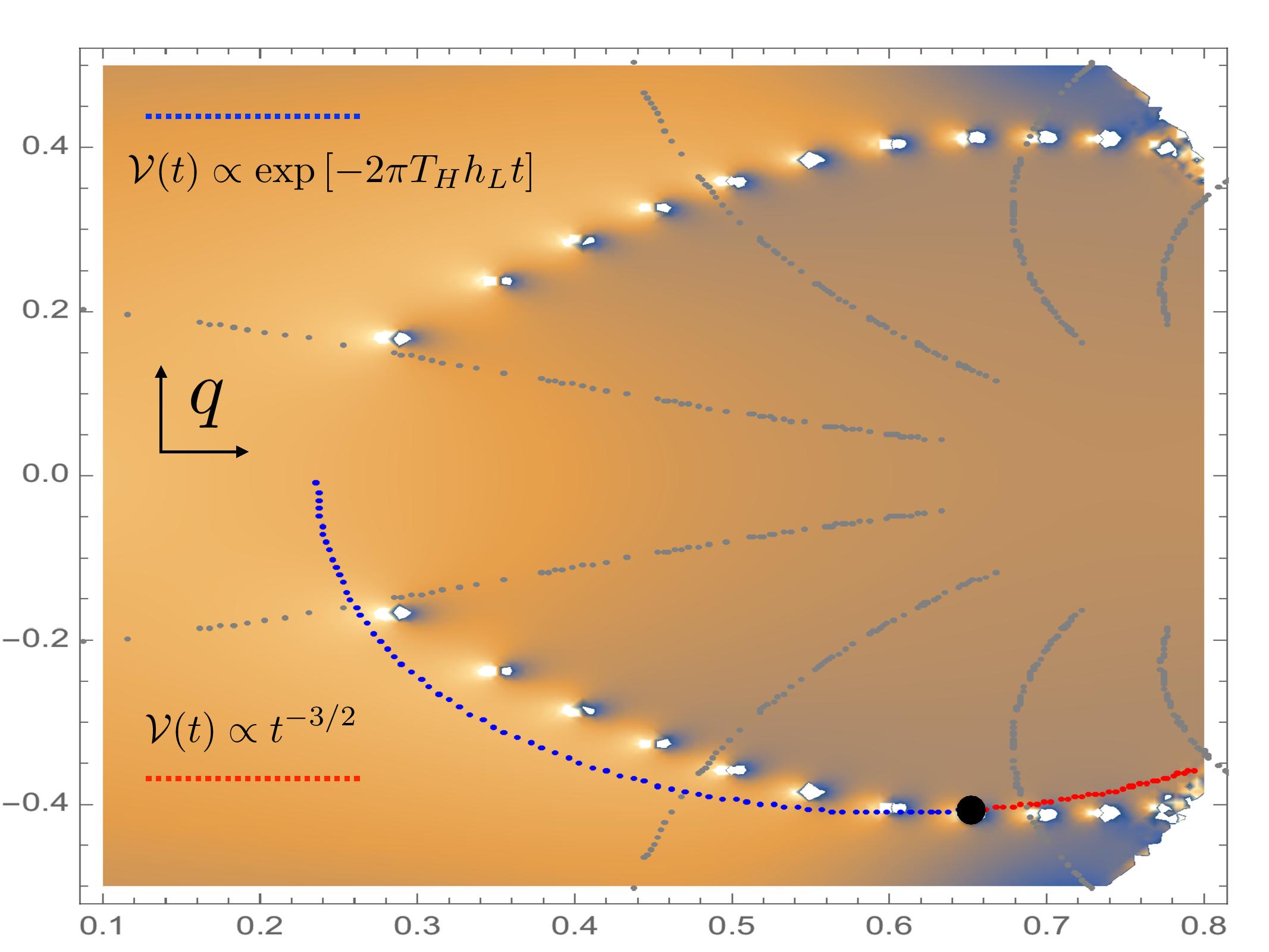}
\caption{\small{Left: real-time behavior of the vacuum block, with a marked late-time transition from thermal exponential decay to a power-law behavior. Right: corresponding trajectory $q(t)$ on $q$-plane (blue and red dashed lines) with transition point marked, against the density plot of the accessory parameter $p_{\text{vac}}(q)$, gray lines are the physical branch-cuts at $x=1$. Transition point coincides with the intersction between $q(t)$ and the resolved branch-cut, which is now a string of poles (zeros) in $p_{\text{vac}}(q)(\mathcal{V}_{\text{vac}}(q))$. Parameters used: $c=30, h_H=5, h_L=1/2$.}}\label{fig: late_time_2}
\end{figure}

The results are plotted in figure \ref{fig: late_time_2}. We see that on the $q$-plane, the resolved branch-cuts (anti-Stokes curves) consisting of poles (zeros) for $p(q)$($\mathcal{V}(q)$) keep crossing the physical branch-cuts, indicating their extensions into the late times. Furthermore, the transition from the exponential decay $\mathcal{V}(t)\propto e^{-2\pi T_H h_L t}$ to the late time behavior $\mathcal{V}(t)\propto t^{-3/2}$ found in \cite{Hongbin} is precisely due to the real-time trajectory $q(t)$ crossing the resolved branch-cut. From a physical point of view, both the exponential decay and the forbidden singularities are consequences/manifestations of the emergent thermal behavior, and thus are related by the underlying ETH dynamics. It is natural that the late time exit from exponential decay and the resolutions of the forbidden singularities are connected.\footnote{However, the ultra-late time transitions at $t\sim e^{S}$ may not be accessible at the level of individual blocks, as observed in \cite{Hongbin}. In principle they are related to the discreteness of the full spectrum, see also \cite{Cotler2016,Dyer2017}.} The intermediate step of re-summing all probe corrections is crucial for revealing such a connection. 

\begin{figure}[h!]
\centering
\includegraphics[width=0.45\textwidth]{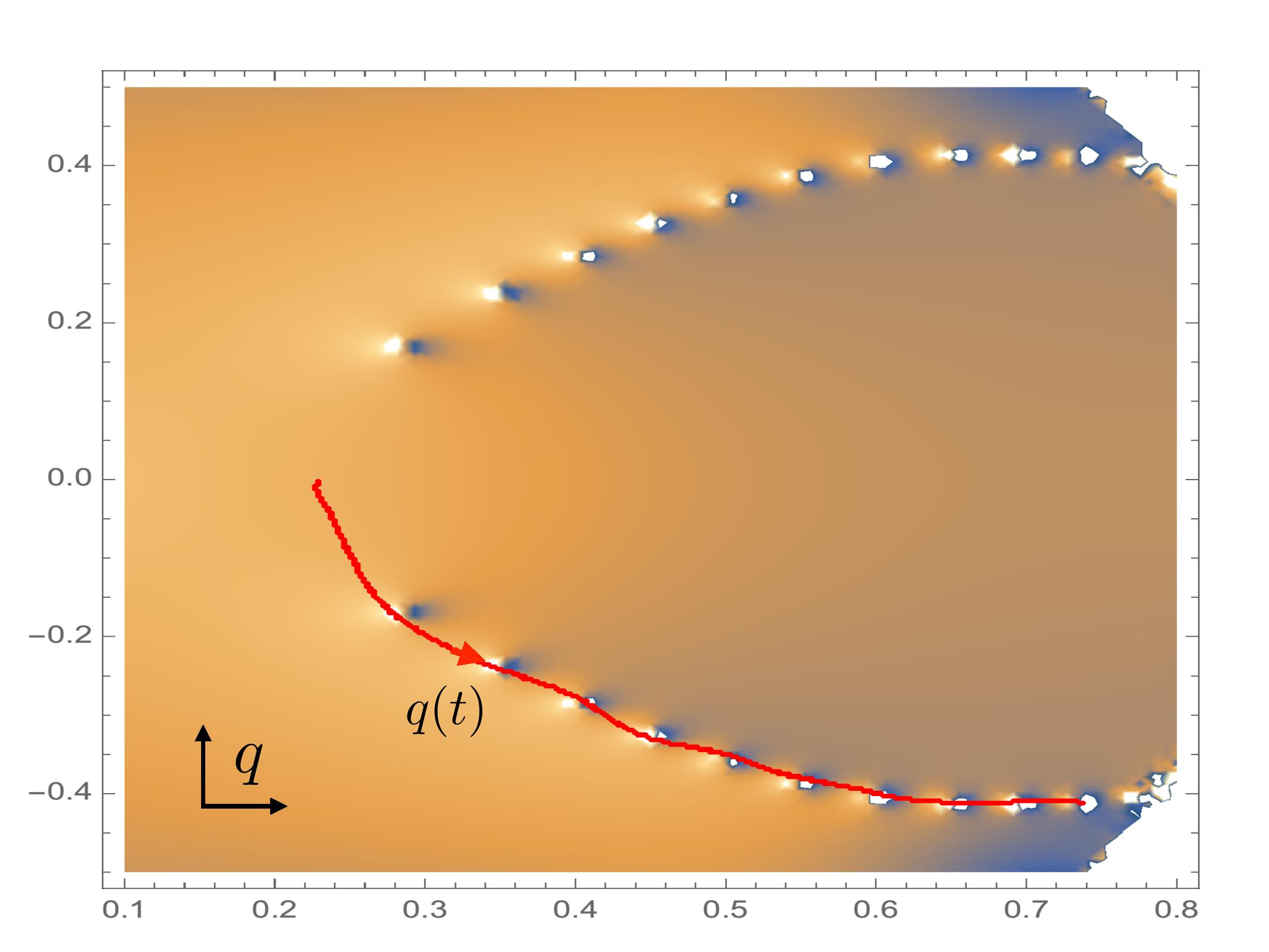}
\includegraphics[width=0.45\textwidth]{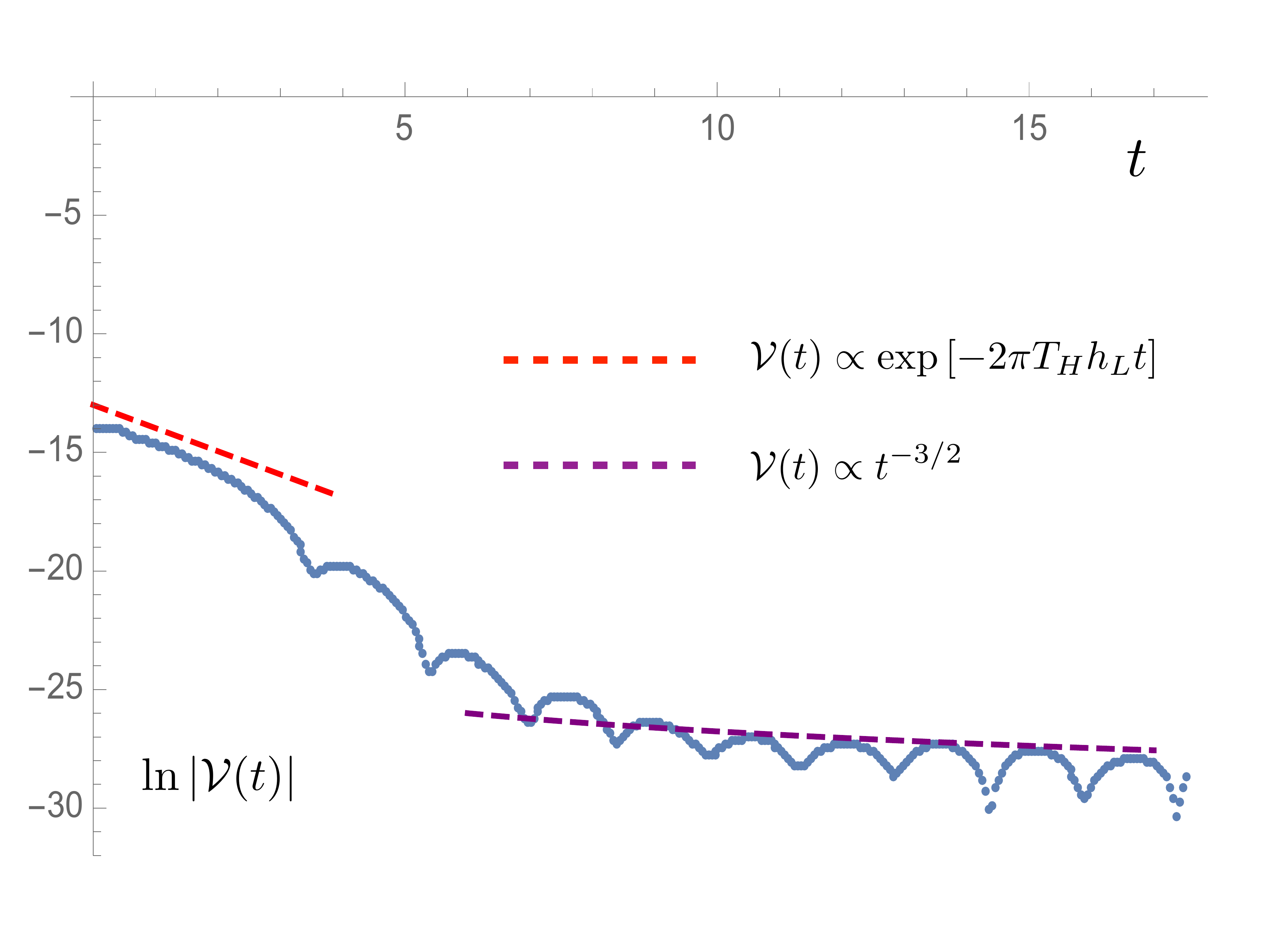}
\caption{\small{Left: trajectory for $x(t)=1-r(t)e^{it}$ in the $q$-plane, passing through the poles/zeros for $p(q)$/$\mathcal{V}(q)$. Right: plot of $\ln{|\mathcal{V}(x(t))|}$, with an early exponential decay followed by ``ripples" of approximately $t^{-3/2}$ envelope. Parameters used: $c=30, h_H=5, h_L=1/2$.}}\label{fig: ripples}
\end{figure}

It is worth pointing out that similar transitions into the $t^{-3/2}$ behavior following early exponential decay $e^{-\alpha t}$ were also observed in computing the spectral form factors $|Z(\beta+it)|$, both for BTZ black holes \cite{Dyer2017} and for the SYK models \cite{Cotler2016}. In particular for BTZ black holes, transitions into the power-law behavior for $|Z(\beta+it)|$ are accompanied by oscillatory ``ripples", representing the re-shuffling of the dominant modular image related to the vacuum character. In fact, it seems that similar ``ripples" with an $t^{-3/2}$ envelope can emerge for the block by going to the late time while staying on the anti-Stoke's curve (figure \ref{fig: ripples}). Technically this requires that the smearing factor for the trajectory to be time-dependent: $x(t)=1-r(t)e^{it}$. A set of universal features seem to be present in different contexts. However, the underlying mechanisms are quite different. For example, the power-law slope for the SYK model can be derived from the 1-loop Schwarzian effective action, and thus only encode perturbative $1/N$ effects; while for the blocks it is from the Stoke's phenomena and are non-perturbative in $c$. The ``ripples" in $|Z(\beta+it)|$ for BTZ black holes encode many saddle exchanges; while those for the blocks are only related to a single anti-Stokes curve. Understanding the connections and distinctions between such ubiquitous phenomena in the different contexts is definitely worth future investigations. 

In this work, we also computed the excited-state renyi-entropy $S_n(\theta)$ on a circle. This is done by studying the same monodromy problem in a slightly different kinematic setting. For sufficiently high energy and subsystem size $\lambda_T\ll\theta\leq \frac{1}{2}$, where the short-distance expansion is not useful, we obtained a WKB solution to the monodromy problem: $S_n(\theta) = \frac{\pi c}{6\beta_H}\theta$. 

It is illuminating to compare our results with those of \cite{Tolya1} and \cite{Lu2017}, which suggests based on general ergodicity argument that: 
\beq\label{eq:tarun}
S^A_n(\rho_A)=\frac{1}{1-n}\log{\left[\frac{\sum_{E_A} e^{S_A(E_A)+n S_{\bar{A}}(E-E_A)}}{\left(\sum_{E_A} e^{S_A(E_A)+ S_{\bar{A}}(E-E_A)}\right)^n}\right]}
\eeq 
where $e^{S_{A,\bar{A}}(E)}$ are the subsystem density of states. Furthermore, it is argued that $S^A_n$ is a convex(concave) function of $V_A/V$ for $n>1(n<1)$. While our results seem to suggest that $S_n(\theta)$ is linear for all $n$ (and for $V_A/V < 1/2$), taking into account the full interpolating solution at finite energy does introduce curvatures that agrees with the convexity/concavity constraints of \cite{Lu2017} (see figure \ref{fig:WKB_check}). 

In the future, it may be fruitful to zoom into the cross-over region between the short-distance limit $\theta \ll \lambda_T$ and the WKB limit $\theta \gg \lambda_T$. They are crucial for understanding the curvature as a function of $V_A/V$. Interestingly the corresponding monodromy problem, which features irregular singular points, is mathematically related to computing scattering amplitudes in black hole spacetimes \cite{Castro2013}.

On the other hand, the $n$-dependence seems to differ from the general formula (\ref{eq:tarun}). For finite subsystem in the thermodynamic limit, one can use the saddle-point approximations to evaluate both the numerator and denominator. In general, the numerator is peaked at $E^*_A(n)$ that is $n$-dependent, for example by substituting the Cardy's formula for $e^{S_{A,\bar{A}}(E)}$. However, our $n$-independent result implies that both the numerator and denominator are peaked at the same value $E^*_A(n=1)$. Intuitively one can understand the offset between $E^*_A(n\neq 1)$ and $E^*_A(n=1)$ as coming from the width of spreading near the peak of the spectrum for the reduced density matrix $\rho_A$. The fact that there is no offset in our result indicates that the subsystem has energy fluctuations that are suppressed compared to the thermal expectation \cite{Garrison}:
\beq
\Delta E_A^2\propto c_V T^2 \frac{V_A V_{\bar{A}}}{V_A+V_{\bar{A}}}\propto c T^2 (T L) \theta (2\pi -\theta)  
\eeq 
where $c_V$ is the specific heat per volume. For small subsystem $\theta\ll 1$, $\Delta E_A^2$ proportional to the volume $V_A = L\theta$. In fact, there is a clear distinction between the entanglement spectrum implied by our result and by (\ref{eq:tarun}). Although both are controlled by the same saddle point when computing $S_{n=1}$, the former suggests a strong peak in the entanglement spectrum itself; while the later features a continuous distribution in entanglement spectrum with density:
\beq 
d\left(S_{\bar{A}}\left(E_{\bar{A}}\right)\right)\propto e^{S_A(E-E_{\bar{A}})}
\eeq
The suppression in subsystem energy fluctuation can be seen explicitly by noting that for primary states $|H\rangle \propto \mathcal{O}_H |0\rangle $, which our result concerns, one can compute the fluctuations in the subsystem energy:  
\beq
E_\theta\propto \int^\theta_0 d\phi \left[T_L(\phi)+T_R(\phi) \right]\propto L_0+\bar{L}_0+\sum_{k\neq 1}\frac{2\sin{\left(\frac{k\theta}{2}\right)}}{k}\left[e^{-\frac{ik\theta}{2}}L_k+e^{\frac{ik\theta}{2}}\bar{L}_k\right]
\eeq
where we have written the unit step function $\Theta\left(0<\phi<\theta\right)$ on the circle as an infinite sum over its Fourier modes. For primary states, the energy fluctuation is 
\beq 
\langle \Delta E_\theta^2 \rangle_H \propto c T_H^2 \log{\left[\frac{1-\cos{\theta}}{1-\cos{\delta}}\right]} 
\eeq
where $T_H \propto \sqrt{H/c}$ and the cutoff $\delta$ is introduced to round off the sharp edges in the step function. One can interpret this as saying that the energy fluctuations only comes from the edges and is not extensive over the subsystem. This is consistent with the $n$-independence of our result. Primary states are special (yet generic at high energies) infinitely symmetric states for which not only total energies, but also local energy densities are conserved.\footnote{We thank Tom Hartman for this comment.} It is therefore not surprising that the subsystem energy fluctuations only come from the boundaries. This fact is then related to the infinite number of extra conserved charges that exist in any CFT, the KdV charges \cite{Bazhanov}. These charges should be properly accounted for by comparing to a Generalized Gibbs Ensemble \cite{He:2017txy,Lashkari:2017hwq} in order to account for the correct energy fluctuations.  The expectation value of the KdV charges take particular values for primary states and different values for descendent states.   For descendent states $|K, H\rangle$ at level $K$ above the primary state $|H\rangle$, the subsystem fluctuations can exhibits a rich variety of behaviors. For $K,H\gg 1$ one can compute in the two extreme case $| K_1\rangle \propto L_{-1}...L_{-1}|H\rangle$ and $| K_2\rangle \propto L_{-K}|H\rangle$:
\beqn
\langle \Delta E_\theta^2\rangle_{K_2} &=& \langle \Delta E_\theta^2\rangle_H +\mathcal{O}\left(K^0\right)\nonumber\\
\langle \Delta E_\theta^2\rangle_{K_1} &=& \langle \Delta E_\theta^2\rangle_H + \frac{K^2}{8}\theta (2\pi-\theta)
\eeqn
For example by taking $K\sim H$, one can get descendent states whose subsystem energy fluctuations range from only boundary-dependent (e.g. $| K_1\rangle $) to super-volume dependent $\propto (cTL)^2$ (e.g. $| K_2\rangle$). This is a special feature of 2D CFTs, and special extensions/generalizations of ETH may be needed to fully capture the chaotic dynamics in these theories. It would be very interesting to explore these in the future. 
 
\section*{Acknowledgements}

TF would like to thank Tom Hartman and Tarun Grover for early collaboration on the topic of Renyi entropies of highly excited states. We thank Alex Belin, Tolya Dymarsky, Liam Fitzpatrick, Jared Kaplan, Daliang Li, and Junpu Wang for discussions and comments on the draft. This research was supported by the DARPA YFA program, contract D15AP00108

\appendix 
\section{Stoke's phenomena}\label{app: stokes}
In this appendix we briefly summarize some basic ingredients of the Stoke's phenomena. For simplicity, suppose we are computing some observable in a (-1) dimensional quantum mechanical model (finite-dimensional integral): 
\beq 
Z(k)=\int dx^i\; e^{\mathcal{I}(x^i,k)}
\eeq 
In the limit where a saddle point approximation is valid, one only needs to consider small neighborhoods around the critical points $q^i_m(k)$, which depends on the external parameter $k$, and $m$ denotes the discrete number of them. For each critical point, one can identify the germs of the local ``downward" flow (i.e. $\text{Re}(\mathcal{I})$ decreases along the flow), whose number equals the number of negative eigenvalues for the second derivative matrix $K_{ij}=\frac{d^2 \text{Re}(\mathcal{I})}{dx^i dx^j}|_{q^i}$. The submanifold traced out by following all possible downward flow defines the so-called Lefshitz thimble $\mathcal{J}_q$ associated with $q^i$, provided that the ``downward" flows do not terminate on another critical point $p$. Loosely speaking $\mathcal{J}_q$ are building blocks of integration contours where the integral is convergent: 
\beq
Z(k)_q \equiv \int_{\mathcal{J}_q} dx^i e^{\mathcal{I}(x^i,k)}<\infty 
\eeq

When the above critierion fails, namely for values of external parameter $k$ such that there exists pairs of critical points $q$ and $p$ that are connected by some ``downward" flow, then the integral is ambiguous up to Stoke's phenomena. Assuming that the action $\mathcal{I}(x)$ is a holomorphic function of complex $x$, one can show that the imaginary part of $\mathcal{I}(x)$ is conserved along the flow. Therefore Stoke's phenomena happens when there are critical points $q$ and $p$ with equal imaginary part of the action: 
\beq
\text{Im}\left[ \mathcal{I}(q,k)\right]=\text{Im} \left[\mathcal{I}(p,k)\right] 
\eeq
Trajectories of $k$ where this happen constitute the so-called Stokes curves. In particular, assume that $\text{Re}\left[\mathcal{I}(p,k)\right]>\text{Re}\left[\mathcal{I}(q,k)\right]$, by crossing such a line, the Lefshitz thimble associated with $\mathcal{J}_p$ undergoes a shift, while that of $\mathcal{J}_q$ remains the same: 
\beq 
\mathcal{J}_p\to \mathcal{J}_p\pm \mathcal{J}_q,\;\;\mathcal{J}_q\to \mathcal{J}_q
\eeq
In other words, by crossing the Stokes curves, the dominant saddle receives a sub-dominant correction, while the sub-dominant saddle remains the same. One can also define anti-Stokes curves as trajectories of $k$ such that 
\beq 
\text{Re}\left[ \mathcal{I}(q,k)\right]=\text{Re} \left[\mathcal{I}(p,k)\right] 
\eeq
Along the anti-Stokes curves both saddles become comparable. Stokes and anti-Stokes curves intersect at points where $\mathcal{I}(q,k)=\mathcal{I}(p,k)$, and we call these the turning points. 
\section{Zamolodchikov's recursion relation}\label{app: Zamo}
We briefly summarize Zamolodchikov's recursion relation. A convenient representation of the conformal block at central charge $c$ with external dimension $\lbrace h_i\rbrace$ and internal dimension $h_p$ is given by: 
\beqn 
&&\mathcal{V}(c,h_i,h_p,x)=\left(16q\right)^{h_p-\frac{c-1}{24}}x^{\frac{c-1}{24}}(1-x)^{\frac{c-1}{24}-h_2-h_3}\theta_3(q)^{\frac{c-1}{2}-4\sum_i h_i}H(c,h_i,h_p,q)\nonumber\\
&& q=e^{i\pi\tau},\;\tau=i\frac{K(1-x)}{K(x)},\;\theta_3(q)=\sum^\infty_{n=-\infty}q^{n^2}
\eeqn
, where $K(x)$ is the complete elliptic integral of the first kind, and $\theta_3(q)$ is the Jacobi theta function. Zamolodchikov proposed the following recursion relation: 
\beqn
 H(c,h_i,h_p,q)&=& 1+\sum^\infty_{m\geq 1,n\geq 1}\frac{\left(16q\right)^{mn}\hat{R}_{mn}(c,h_i)}{h_p-h_{p,mn}(c)}H(c,h_i,h_{p,mn}+mn,q)\nonumber\\
 h_{p,mn}(c)&=&\frac{1}{4}(n^2-1)t(c)+\frac{1}{4}(m^2-1)\frac{1}{t(c)}-\frac{1}{2}(mn-1)\nonumber\\
 t(c) &=& 1+\frac{1}{12}\left(1-c\pm\sqrt{(1-c)(25-c)}\right)\nonumber\\
 \hat{R}_{mn}(c,h_i)&=&-\frac{1}{2}\frac{\prod_{j,k}\left(\lambda_2+\lambda_1-\frac{\lambda_{jk}}{2}\right)\left(\lambda_2-\lambda_1-\frac{\lambda_{jk}}{2}\right)\left(\lambda_3+\lambda_4-\frac{\lambda_{jk}}{2}\right)\left(\lambda_3-\lambda_4-\frac{\lambda_{jk}}{2}\right)}{\prod_{a,b}\lambda_{ab}}\nonumber\\
 i &=& -m+1,-m+3, ... ,m-3, m-1;\;\;j= -n+1,-n+3, ... ,n-3, n-1\nonumber\\
&& -m+1\leq a\leq m,\;-n+1\leq b \leq n,\; (a,b)\neq (0,0), (a,b)\neq (m,n)\nonumber\\
\lambda_i &=& \sqrt{h_i+\frac{1-c}{24}},\;\;\lambda_{pq}=\frac{1}{\sqrt{24}}\left\lbrace\left(p+q\right)\sqrt{1-c}+\left(p-q\right)\sqrt{25-c}\right\rbrace
\eeqn
, from which one can obtain a recursion relation for the coefficients of $q$-series expansion $H(c,h_i,h_p,q)=\sum^\infty_{N=0}H_N(c,h_i,h_p)q^N$: 
\beq
H_{\ell}(c,h_i,h_p)=\sum_{mn\leq \ell-1}(16)^{mn}\hat{R}_{mn}(c,h_i)\frac{H_{\ell-mn}(c,h_i,h_{p,mn}+mn)}{h_p-h_{p,mn}},\;H_{0}(c,h_i,h_{p,mn}+mn)=1
\eeq

\bibliographystyle{JHEP}
\bibliography{ETH}

\end{document}